\documentclass[12pt,aps,amsmath,amssymb,prb,reprint,preprintnumbers, nofootinbib, showkeys, showpacs, ]{revtex4-1}
\usepackage{graphicx, textcomp, wasysym, amssymb, epsfig}

\begin{document}

\preprint{APS/123-QED}            

\title{Electronic correlations in FeGa$_3$ and the effect of hole doping on its magnetic properties}

\author{M. B. Gam\.za$^{1,a}$, J. M. Tomczak$^{2,b}$, C. Brown$^{3,4}$, A. Puri$^{1,5}$,  G. Kotliar$^{2}$ and M. C. Aronson$^{1,5}$}

\affiliation{$^{1}$Condensed Matter Physics and Materials Science Department, Brookhaven National Laboratory, Upton, NY 11973-5000, USA}
\affiliation{$^{a}$Current address: Department of Physics, Royal Holloway, University of London, Egham, TW20 0EX, UK}
\affiliation{$^{2}$Department of Physics and Astronomy, Rutgers University, Piscataway, New Jersey 08854, USA}
\affiliation{$^{b}$Current address: Institute of Solid State Physics, Vienna University of Technology, 1040 Wien, Austria }
\affiliation{$^{3}$NIST Center for Neutron Research, National Institute of Standards and Technology; Gaithersburg, MD 20899, USA}
\affiliation{$^{4}$Department of Chemical and Biomolecular Engineering, University of Delaware, DE 19716, USA}
\affiliation{$^{5}$Department of Physics and Astronomy, State University of New York, Stony Brook, NY 11794-3800, USA}

\date{\today}

\begin{abstract}

We investigate signatures of electronic correlations in the narrow-gap semiconductor FeGa$_3$ by means of electrical resistivity and thermodynamic measurements performed on single crystals of FeGa$_3$, Fe$_{1-x}$Mn$_x$Ga$_3$ and FeGa$_{3-y}$Zn$_y$, complemented by a study of the 4$d$ analog material RuGa$_3$.
We find that the inclusion of sizable amounts of Mn and Zn dopants into FeGa$_3$ does not induce an insulator-to-metal transition. Our study indicates that both substitution of Zn onto the Ga site and replacement of Fe by Mn introduces states into the semiconducting gap that remain localized even at highest doping levels.   
Most importantly, using neutron powder diffraction measurements, we establish that FeGa$_3$ orders magnetically above room temperature in a complex structure, which is almost unaffected by the doping with Mn and Zn. 
Using realistic many-body calculations within the framework of dynamical mean field theory (DMFT), we argue that while the iron atoms in FeGa$_3$ are dominantly in an \mbox{$S=1$} state, there
are strong charge and spin fluctuations on short time scales, which are independent of temperature.
Further, the low magnitude of local contributions to the spin susceptibility advocates an itinerant mechanism for the spin response in FeGa$_3$.
Our joint experimental and theoretical investigations classify FeGa$_3$ as a correlated band insulator with only small dynamical correlation effects, in which non--local exchange interactions are responsible for the spin gap of 0.4~eV and the antiferromagnetic order.
We show that hole doping of FeGa$_3$ leads, within DMFT, to a notable strengthening of many--body renormalizations. 

\end{abstract}

 

\maketitle

\section{Introduction}

The role of electron--electron correlation effects in narrow gap $d$--electron semiconductors has been a subject of extensive study for over two decades.\cite{FeSiJac, FeSb2exp1, FeSb2exp2, anisimovFeSi, FeSb2nearlyFM, Takahashi, Delaire, FeSiXCo, FeSb2th, FeSiJan, FeSiJanB, FeSb2Jan, FuDoniach, Sentef, FeSb2DMFT, Sunthermo, FeSi1, FeSiKI, FeSiGe, FeCoSi, FeSiAl, FeSb2KI, FeSb2Sn, FeCoSb2, FeCrSb2, FeSb23, FeSb2FeAs2Sun, FeAs2, Fe2VAlexp, Fe2TiSn, Arita, Hadano, Picket, Bittar, Takabatake, pressure, Neel}
For archetypal compounds of this family, FeSi and FeSb$_2$, an intriguing crossover was observed from a nonmagnetic semiconductor at low temperatures to a paramagnetic metal with a Curie--Weiss--like magnetic susceptibility at high temperatures that are, however, still small relative to the gap energy.\cite{FeSiJac, FeSb2exp1, FeSb2exp2} 
Furthermore, when passing this crossover, optical spectral weight that is suppressed upon cooling due to the opening of the gap is recovered only at energies that are very high with respect to the charge gap. These distinct similarities of FeSi and FeSb$_2$ to heavy fermion Kondo insulators\cite{Kondoinsulators} in both charge and spin degrees of freedom, has caused great interest in this class of materials.  

Models that were proposed to explain the unusual behaviors of FeSi and FeSb$_2$ include a nearly magnetic semiconductor scenario based on a spin fluctuation theory of itinerant electrons,\cite{anisimovFeSi,FeSb2nearlyFM,Takahashi} and alternatively a renormalization of the electronic structure mediated by a strong electron--phonon coupling.\cite{Delaire}
Recent studies tend to treat FeSi and FeSb$_2$ as correlated band insulators.\cite{FeSiXCo, FeSb2th, FeSiJan, FeSiJanB, FeSb2Jan, FuDoniach,Sentef,FeSb2DMFT} 
Unlike conventional semiconductors, the metallization of correlated band insulators is believed to be caused by the emergence of incoherent states in the gap, accompanied by a massive reorganization of the spin excitation spectrum.\cite{FeSiJan}  
For FeSi, this description led to a coherence--incoherence scenario in which the temperature induced metallization is associated with the unlocking of fluctuating iron moments that are almost temperature independent on short time scales.\cite{FeSiJan}
It remains to be seen whether the latter paradigm is applicable to other compounds, motivating these investigations of FeGa$_3$. 

Early reports on FeGa$_3$ classify the compound as a diamagnetic semiconductor with a narrow gap of \mbox{0.4--0.5~eV.}\cite{Hadano, Imai, Hauserman, Amagai, Arita, Tsuji} The magnetic susceptibility increases strongly at temperatures above $\sim$500~K,\cite{Tsuji} suggestive of an approaching crossover to a paramagnetic metallic state.
Electronic band structure calculations within the local density approximation (LDA) find a gap of 0.4~eV in the ground--state that is produced by the hybridization of Fe~3$d$ states with $p$ states of Ga,\cite{Hauserman, Arita, Picket, pressure} in good agreement with experimental results.\cite{Arita, Hadano} 
However, an angle--resolved photoemission spectroscopy (ARPES) study disclosed differences between the measured electronic dispersions and those obtained in the LDA calculations,\cite{Arita} 
suggesting a band--narrowing due to electronic correlations. 
With these findings, FeGa$_3$ is an appealing compound for studying the consequences of electron--electron correlation effects in a $d$--electron semiconductor, in which the band gap is about one order of magnitude larger than in FeSi\cite{FeSi1} and FeSb$_2$\cite{FeSb2exp1, FeSb2exp2}.

Similar to those latter compounds, the presence of narrow iron 3$d$ states near the Fermi level raises the question of whether there are local Fe moments and/or magnetic order in FeGa$_3$. 
So far, there is no consensus in either experimental or theoretical studies as to the overall magnetic character of FeGa$_3$.\cite{Picket, pressure, Arita, Petrovic, Singh, Moessbauer} 
LSDA+U electronic structure calculations find the existence of local Fe moments in FeGa$_3$, independent of which double counting scheme is used, with an antiferromagnetic order being lowest in energy.\cite{Picket} Further, the size of the band gap obtained assuming moderate values of the on--site Coulomb interaction for the Fe 3$d$ states, \mbox{$U\sim 2$~eV,} coincides with experimental results.\cite{Picket} However, a non--magnetic state is stabilized for \mbox{$U\apprle 1.5$ eV} and even for larger values of $U$ if screening effects are included in the LSDA+U formalism via a Yukawa ansatz.\cite{pressure} 
On the experimental side, a recent muon spin rotation study detected spectroscopic features characteristic of electron confinement into spin polarons.\cite{Petrovic} Formation of spin polarons requires the existence of Fe moments, whereas a $^{57}$Fe M$\ddot{o}$ssbauer study did not show the presence of an internal magnetic field at the Fe site, indicating a nonmagnetic state of Fe.\cite{Moessbauer}

In this work, we address the magnetic properties of FeGa$_3$ using neutron powder diffraction experiments. The diffraction patterns reveal the presence of ordered Fe moments, featuring a complex room temperature magnetic structure. We performed thermodynamic and electrical resistivity measurements on single crystals of FeGa$_3$ and of its isostructural homolog RuGa$_3$.
A comparative study of the magnetic properties and the electrical resistivities of these compounds allows us to separate the contributions originating from strongly correlated 3$d$ electrons.
Analysis of the specific heat of FeGa$_3$ gives us insight into the vibrational properties of Fe and Ga sublattices. 
The experimental investigation is complemented by electronic structure calculations based on dynamical mean--field theory (DMFT) aimed at exploring the charge and spin states of iron in FeGa$_3$ and effects of many--body renormalizations. 

A characteristic feature of correlated insulators is a strong suppression of the band gap by doping, accompanied by large enhancements of the electronic specific heat coefficient and magnetization.\cite{FeCoSi, FeSiGe, FeCoSb2, FeSb2Sn} 
Furthermore, weak doping often results in a magnetic instability.\cite{FeCoSi, FeSiGe, FeCoSb2, FeCrSb2} 
For FeGa$_3$, electron--type doping induces a crossover to a correlated metallic state at \mbox{$x\approx 0.05$} and \mbox{$y\approx 0.006$} in Fe$_{1-x}$Co$_x$Ga$_3$\cite{Bittar} and FeGa$_{3-y}$Ge$_y$\cite{Takabatake}, respectively.
Interestingly, further doping with Ge in FeGa$_{3-y}$Ge$_y$ leads to a ferromagnetic quantum critical point at $x\approx 0.016$--0.05,\cite{Takabatake, Neel} whereas for Fe$_{0.5}$Co$_{0.5}$Ga$_3$ nuclear spin--lattice relaxation measurements  revealed a very fast relaxation with temperature dependence $1/T_1\propto T^{1/2}$ being a unique feature of weakly and nearly antiferromagnetic metals.\cite{Sascha}   
A recent computational study suggested the formation of an itinerant ferromagnetic state with half--metallic properties in FeGa$_3$ in case of doping with both electrons and holes.\cite{Singh} 
Motivated by these results, we dope FeGa$_3$ with holes on both Fe and Ga sites. We choose Mn and Zn as dopants because they lie in the periodic table next to iron and gallium, respectively, and have one less valence electron, corresponding to hole doping. We investigate the evolution in the thermodynamic and transport properties of single crystals of Fe$_{1-x}$Mn$_x$Ga$_3$ and FeGa$_{3-y}$Zn$_y$ by means of magnetization, specific heat, electrical resistivity and neutron powder diffraction measurements. We interpret the experimental results in the context of electronic structure calculations based on DMFT.

\section{Methods}\label{Methods}
\subsection{Experimental\protect\footnote{Certain commercial equipment, instruments, or materials are identified in this document. Such identification does not imply recommendation or endorsement by the National Institute of Standards and Technology nor does it imply that the products identified are necessarily the best available for the purpose.} }

Single crystals of FeGa$_3$ and RuGa$_3$, as well as the series Fe$_{1-x}$Mn$_x$Ga$_3$ \mbox{($0\textless x \textless 0.12$)} and FeGa$_{3-y}$Zn$_y$ \mbox{($0\textless y\textless 0.06$),} were grown from gallium flux. Mixtures of high purity elements in the ratio Fe:Ga~=~1:6.7--11.5, Ru:Ga~=~1:15.7, \mbox{Fe:Mn:Ga~=~($1-x_{nom}$):$x_{nom}$:8.7--9.3} \mbox{($0\leq x_{nom}\textless 70$\%)} or \mbox{Fe:Zn:Ga~=~0.066:$x_{nom}$:($1-x_{nom}$)} \mbox{($0\leq x_{nom}\textless 28.3$\%)} were placed in Al$_2$O$_3$ crucibles and sealed in quartz ampoules under a low pressure of \mbox{$\sim$200~mbar} of Ar gas.  The ampoules were heated up to $1050^{\circ}$C and then slowly \mbox{($\sim$3$^{\circ}$C} per hour) cooled down to $650^{\circ}$C in case of undoped FeGa$_3$, RuGa$_3$ and Zn--doped FeGa$_3$, and to $550^{\circ}$C for Mn--doped FeGa$_3$, at which point the molten flux was separated from the crystals by spinning in a centrifuge. The quartz tubes were cracked only after cooling down to room temperature to prevent oxidation of crystals. The remaining flux was removed from crystal faces mechanically using latex sheets.

The microstructures of crystals were examined optically and with a scanning electron microscope (LEO/550) equipped with a Robinson backscatter detector. Chemical composition was determined by means of energy dispersive X--ray spectroscopy (EDXS) studies performed at several points across a crystal surface, based on integrated intensities of the Fe$K$, Mn$K$, Ru$L$ and Ga$L$ X--ray lines.
The average composition for crystals of FeGa$_3$ (Fe$_{0.97(1)}$Ga$_{3.03(1)}$) was taken as a standard to calculate the compositions of Mn-- and Zn--doped crystals. 

\begin{figure}[b] 
\includegraphics[width=0.35\textwidth,angle=-90]{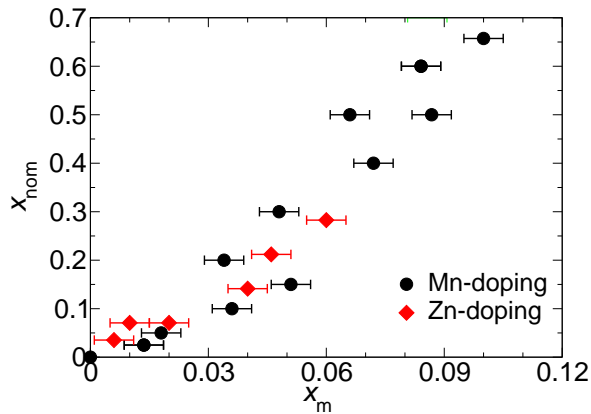}
\caption{\label{fig:Fig1} (Color online) The relation between actual ($x_{\mathrm{m}}$) and nominal $x_{\mathrm{nom}}$ doping levels in Fe$_{1-x}$Mn$_x$Ga$_3$ (black circles) and FeGa$_{3-y}$Zn$_y$ (red diamonds). Error bars indicate one standard deviation of the measured $x_{\mathrm{m}}$ values.}
\end{figure}

The measured concentrations of the dopants ($x_{\mathrm{m}}$) increase gradually with the nominal doping level, and depend to some extent on the Fe:Ga ratio (Fig.~\ref{fig:Fig1}).  
There is no indication for a limit of solubility of Mn or Zn in FeGa$_3$, although we were not able to grow crystals via this approach with $x\apprge 0.12$ and $y\apprge 0.06$.
The compositions discussed here always refer to the measured doping levels.

\begin{figure}[h] 
\includegraphics[width=0.45\textwidth,angle=0]{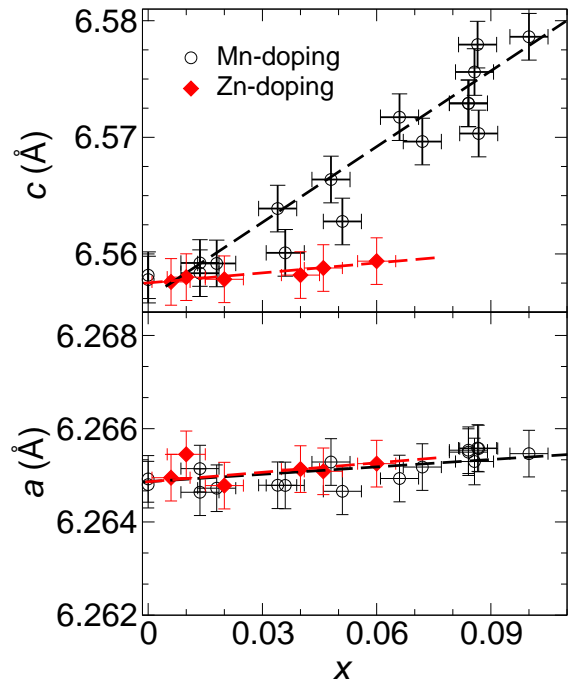}
\caption{\label{fig:Fig2} (Color online)  Lattice parameters $a$ and $c$ as functions of actual doping level ($x$) for Fe$_{1-x}$Mn$_x$Ga$_3$ (black circles) and FeGa$_{3-y}$Zn$_y$ (red diamonds). Dashed lines are guides for the eye.}
\end{figure}

Structural characterization at room temperature was carried out on powdered crystals using a Bruker D-8 Advance diffractometer using Cu~$K\alpha$ radiation. Refinements of the recorded  diffraction patterns were performed using the Jana2006 program.\cite{Jana2006} 
The X-ray powder diffraction measurements confirmed that FeGa$_3$ and RuGa$_3$ form in a tetragonal structure with space group $P$42/$mnm$ (\#136) and lattice parameters that are in agreement with previous reports.\cite{Hauserman, Dasarathy, Hadano}  
Doping of FeGa$_3$ with Mn leads to a monotonic increase in lattice parameter $c$ with increasing $x$, following Vegard's law (Fig.~\ref{fig:Fig2}). For doping with Zn, lattice parameters vary only slightly, in line with recent report on polycrystalline samples.\cite{GrinFeGa3}
X-ray powder diffraction patterns of Fe$_{1-x}$Mn$_x$Ga$_3$ \mbox{($0\textless x \textless 0.12$)} and FeGa$_{3-y}$Zn$_y$ \mbox{($0\textless y\textless 0.06$)} are all consistent with the FeGa$_3$--type structure. No evidence for the presence of an impurity phase or for phase separation in the single crystals was found, apart from trace inclusions of gallium flux.

The magnetization studies in the temperature range between 1.8 and 400~K were carried out in a Quantum Design Magnetic Properties Measurement System (MPMS). At temperatures between 400~K and 900~K, the magnetization was studied using the Vibrating Sample Magnetometer (VSM) option of a Quantum Design Physical Property Measurement System (PPMS). The heat capacity was determined by the relaxation method at temperatures between 0.44 K~and 300~K using the PPMS. Both $ac$ and $dc$ electrical resistivity measurements were performed in the same system with a standard four--probe set--up, where electrical contacts were made using  silver--filled epoxy.

Neutron diffraction was measured on $\sim$6~g powders of FeGa$_3$ and Fe$_{0.95}$Mn$_{0.05}$Ga$_3$ obtained by triturating single crystals. The diffraction measurements were carried out at temperatures ranging from 1.5~K to 300~K using the High Resolution Powder Diffractometer \mbox{BT--1} of the NIST Center for Neutron Research, with the Ge(311) monochromator giving neutrons with wavelength \mbox{$\lambda = 2.079$~\AA.} The diffraction patterns were remeasured in a different cryostat to verify that all the observed diffraction peaks originate with the sample. The Rietveld refinement of the nuclear structure was done with the program Fullprof\cite{Fullprof}.

\subsection{Computational}

The combination of dynamical mean field theory (DMFT) and density functional theory (DFT) in the local density approximation (LDA) was used to compute spectral, magnetic and transport properties of pure and hole doped FeGa$_3$. For the DFT part we employed Wien2k \cite{comput1}.
The DMFT\cite{comput3} impurity problem was solved using a hybridization expansion continuous time quantum Monte Carlo (CTQMC) method.\cite{comput4, comput5} We use a projection-based DFT+DMFT setup with full charge self--consistency.\cite{comput6} 
For FeGa$_3$, we use a Hubbard interaction \mbox{$U=5.0$~eV} and a Hund's rule coupling \mbox{$J=0.7$~eV} for the 3$d$ states of iron. These values were found to be appropriate for iron--based compounds.\cite{FeSiJan, comput7}
For the 4$d$--electron compound RuGa$_3$, we use \mbox{$U=2.5$~eV} and \mbox{$J=0.4$~eV.\cite{jernej}} 
Response functions are obtained within the Kubo formalism.\cite{FeSb2Jan} The doping is simulated through the virtual crystal approximation (VCA).

\section{Results}\label{Results}

\subsection{FeGa$_3$}\label{undoped}

\begin{figure}[b] 
\includegraphics[width=0.38\textwidth,angle=-90]{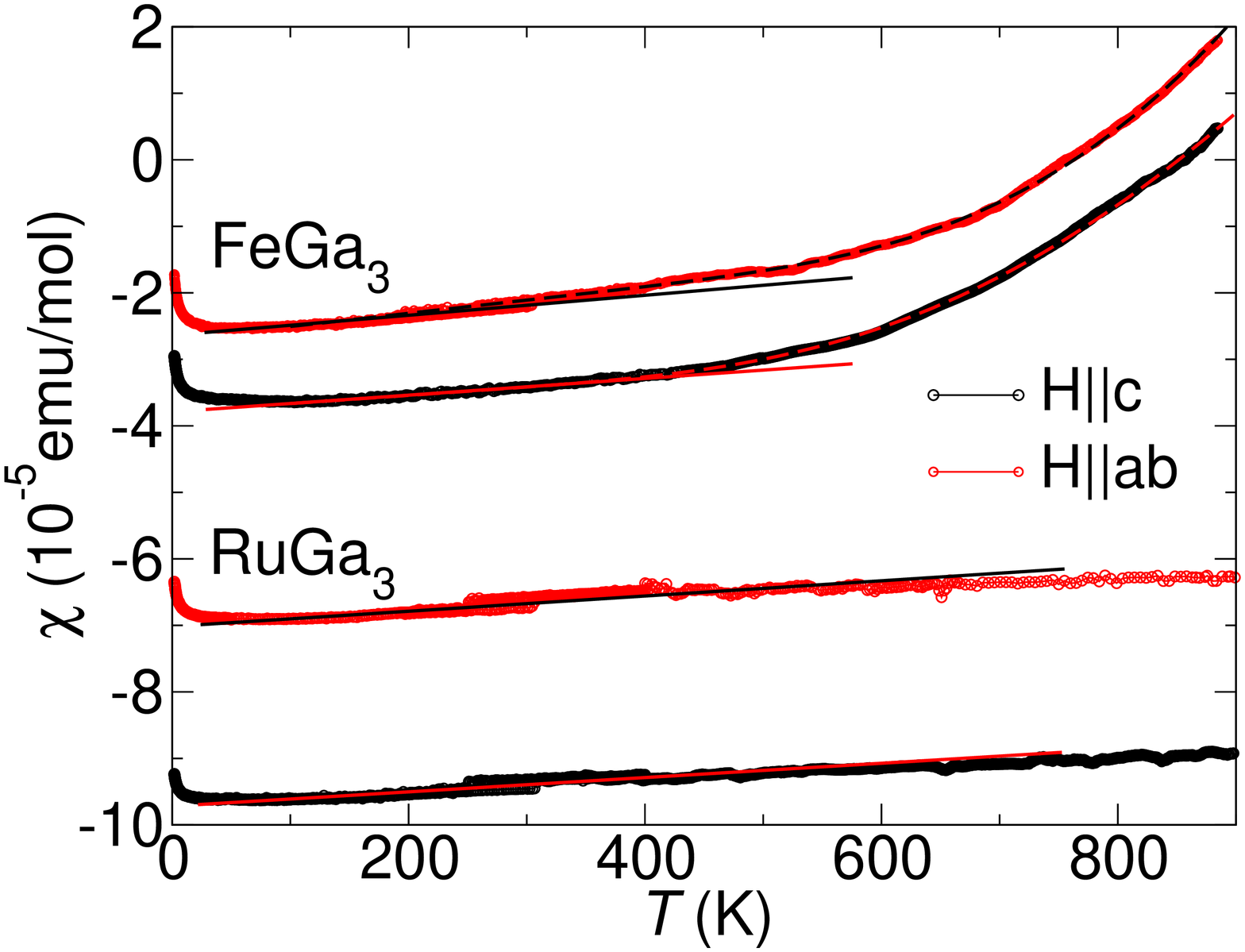}
\caption{\label{fig:Fig3} (Color online)  Magnetic susceptibility of single crystals of FeGa$_3$ and RuGa$_3$ measured in a magnetic field of 10~kOe applied in the tetragonal $ab$~plane (red circles) or along the $c$--axis (black circles). Thin solid lines indicate linear terms $aT$ in $\chi$($T$) for FeGa$_3$ with \mbox{$a=1.3\times 10^{-8}$ emu/(K mol)} and \mbox{$a=1.9\times 10^{-8}$ emu/(K mol)} for $H\| ab$ and $H\| c$, respectively and for RuGa$_3$ with \mbox{$a=0.7\times 10^{-8}$ emu/(K mol)} independent of the field direction. 
Dashed lines represent fits to the activated temperature dependence \mbox{$\chi$($T$)$\propto$ exp(-$\Delta_{\mathrm{S}}$/(k$_{\mathrm{B}}T$))} at \mbox{$T\apprge 500$ K.}}
\end{figure}

Fig.~\ref{fig:Fig3} shows the $dc$ magnetic susceptibility $\chi=M/H$ of FeGa$_3$ and of its isostructural and isoelectronic homolog RuGa$_3$, measured with a magnetic field $H$=10~kOe applied parallel to the $ab$ plane, and along the $c$--axis of the crystal. 
Overall, $\chi$($T$) for FeGa$_3$ is similar to that reported previously for a set of randomly oriented crystals over a more limited temperature range.\cite{Hadano, Tsuji} 
The magnetic anisotropy is mostly reflected in a temperature independent offset between values obtained when the magnetic field is applied perpendicular or parallel to the $c$--axis in the tetragonal crystal lattice, whereas the temperature dependence is almost the same in both directions.
An even larger magnetic anisotropy was found for RuGa$_3$.

For both FeGa$_3$ and RuGa$_3$, $\chi$ is negative at low temperatures and increases nearly linearly with increasing temperature up to $\sim$500~K.
Slight Curie tails in $\chi$($T$) at the lowest temperatures (Fig.~\ref{fig:Fig3}) may be assigned to a small number \mbox{$\sim$0.06\%} of $S=1/2$ paramagnetic impurities. 
After correction for these impurities, an extrapolation of $\chi$($T$) to \mbox{$T=0$~K}  results in \mbox{$\chi_{0(ab)}\approx -26\times 10^{-6}$~emu~mol$^{-1}$} and \mbox{$\chi_{0(c)}\approx -38\times 10^{-6}$~emu~mol$^{-1}$} for FeGa$_3$ and \mbox{$\chi_{0(ab)}\approx -70\times 10^{-6}$~emu~mol$^{-1}$}, \mbox{$\chi_{0(c)}\approx -94\times 10^{-6}$~emu~mol$^{-1}$} for RuGa$_3$. The diamagnetism of the closed-shell ions\cite{increm} gives the values of \mbox{$-37\times 10^{-6}$~emu~mol$^{-1}$} and \mbox{$-47\times 10^{-6}$~emu~mol$^{-1}$} for FeGa$_3$ and RuGa$_3$, respectively, which are comparable with the measured $\chi_0$ values. 
The lattice diamagnetism resulting from the inter--band effect of the magnetic field\cite{inter-band} is likely also responsible for the observed magnetic anisotropies.

The almost linear increase of the magnetic susceptibilities with increasing temperature for both FeGa$_3$ and RuGa$_3$ observed in broad temperature ranges (see Fig.~\ref{fig:Fig3}) can be assigned to a slight shrinking of their band gaps. 
In semiconductors electron--phonon interactions usually lead to a shrinking of the gap with increasing temperature due to an increase in the phonon population.\cite{phe} This, in turn, leads to a small and gradual increase of the magnetic susceptibility.\cite{dia-calc}

The magnetic susceptibility of RuGa$_3$ remains negative and only weakly temperature dependent up to the highest temperatures.
%shows only a weak temperature dependence up to the highest temperatures.
In contrast, for FeGa$_3$ the slope of $\chi$($T$) starts to increase significantly at temperatures above \mbox{$\sim$500~K.} The temperature dependencies of the magnetic susceptibilities can be well described as \mbox{$\chi$($T$)$\propto$ exp(-$\Delta_{\mathrm{S}}$/(k$_{\mathrm{B}}T$)).} Here, $k_{\mathrm{B}}$ denotes the Boltzmann constant. The spin gaps derived from such fits are \mbox{$\Delta_{\mathrm{S-ab}}=0.33$~eV} for \mbox{$H\parallel ab$} and \mbox{$\Delta_{\mathrm{S-c}}=0.41$~eV} for \mbox{$H\parallel c$.}

\begin{figure*}[b] 
\includegraphics[width=0.38\textwidth,angle=-90]{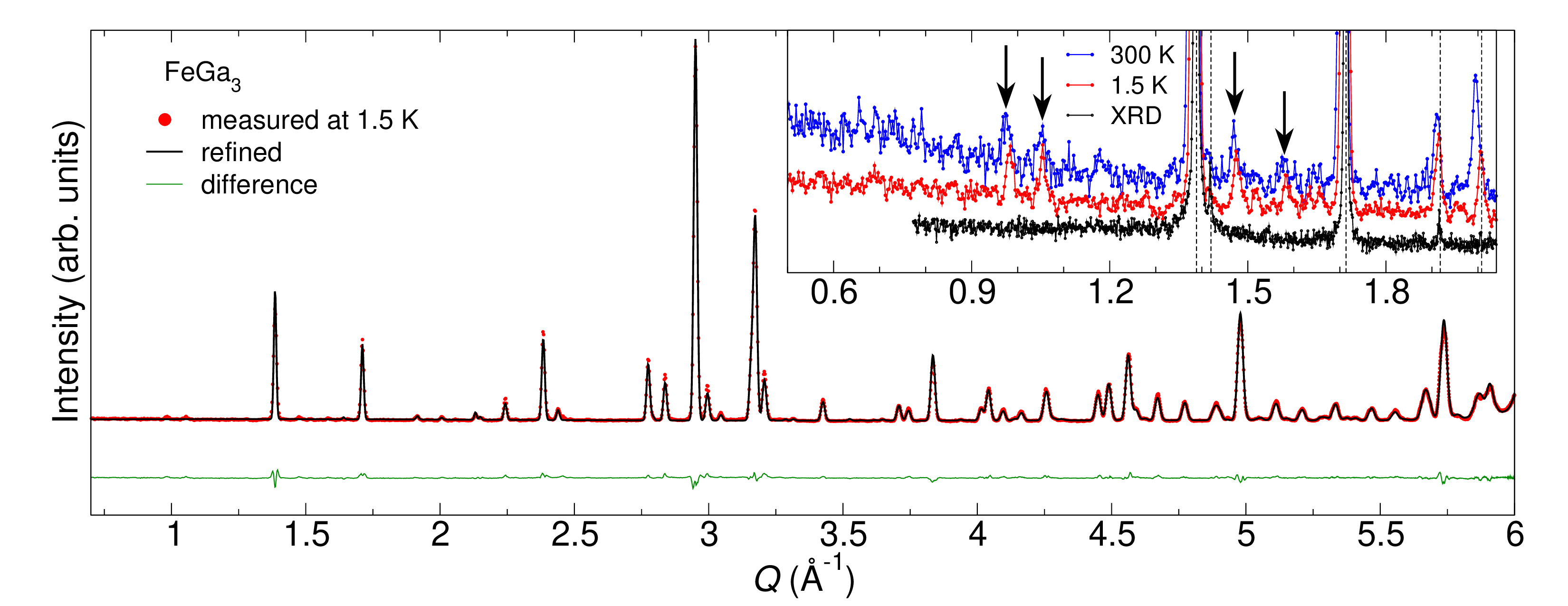}
\caption{\label{fig:Fig4} (Color online)  Results of neutron powder diffraction experiments for FeGa$_3$. The main panel shows the Rietveld refinement (black) of the nuclear structure based on the neutron diffraction pattern (red) collected at 1.5~K. The difference curve is shown in green. The inset shows the low angle range of the neutron diffraction patterns measured at 1.5~K (red) and at room temperature (blue), together with the X--ray powder diffraction data measured at 300~K on the same powder. Dashed vertical lines show the positions of nuclear peaks. The main magnetic peaks are indicated by arrows.}
\end{figure*}

The most direct information about the magnetic state of FeGa$_3$ can be gained from neutron diffraction measurements. Fig.~\ref{fig:Fig4} shows a neutron powder diffraction pattern collected at a temperature of 1.5~K. Rietveld refinement of the nuclear structure confirmed the tetragonal symmetry with space group P4$_2$/mnm. Structural parameters derived from the refinements of the neutron  diffraction patterns measured at temperatures of 1.5~K and 300~K are presented in Table~\ref{table1}. The results obtained at room temperature are in excellent agreement with the structural data attained based on our X--ray powder diffraction patterns (Table~\ref{table1}) and with previous reports.\cite{Dasarathy, Hauserman, Hadano}

In addition to nuclear reflections, there are several clearly discernible extra peaks in the neutron patterns. The strongest additional peaks are at wavevectors \mbox{$Q=0.986$~\r{A}$^{-1}$,} 1.058~\r{A}$^{-1}$, 1.478~\r{A}$^{-1}$ and 1.586~\r{A}$^{-1}$ and are marked in the inset of Fig.~\ref{fig:Fig4}. The absence of diffraction peaks at similar values of $Q$ in room temperature X--ray powder diffraction patterns from the same sample, measured before and after collecting the neutron diffraction data, strongly argues for a magnetic origin of the additional peaks. 
The emergence of the extra peaks only at rather small values of $Q$ provides further support for their magnetic character because the magnetic form factor falls off quite rapidly with diffraction angle.  
The magnetic peaks are almost identical at temperatures of 1.5~K and 300~K and their $Q$--widths are resolution limited, thus indicating long--range magnetic order. This advocates that FeGa$_3$ is an antiferromagnet with the an ordering temperature $T_N$ above 300~K. The presence of distinct magnetic peaks in the collected neutron patterns implies that the size of the staggered moments is well above \mbox{0.1$\mu_B$,} which is the approximate detection limit of the neutron powder diffraction method. Simultaneously, the small size of the magnetic peaks compared to the nuclear peaks makes magnetic moments larger than \mbox{1--1.5 $\mu_{\mathrm{B}}$} per Fe very unlikely.

The magnetic peaks can be indexed only by a propagation vector that is incommensurate with the crystal lattice in at least one direction. The best description of the observed peaks was achieved assuming \mbox{{\bf k} = (0.5, 0.25, 0.44),} indicating that the tetragonal symmetry of the nuclear structure is not shared by the magnetic structure.
A complete solution of this complex magnetic structure requires further experimental studies, in particular neutron diffraction measurements of single crystals.

\begin{center}
\begin{table*} [h]
 \begin{tabular}{|c| c| c c c| c c c| c c c c|}
\hline
\multicolumn{2}{|c|}{Lattice parameters} & \multicolumn{3}{c|}{$T$=1.5 K} & \multicolumn{3}{c|}{$T$=300 K} & \multicolumn{4}{c|}{XRD; $T$=295 K} \\
\hline
\multicolumn{2}{|c|}{$a$ (\r{A})} & \multicolumn{3}{c|}{ 6.2528(2) } & \multicolumn{3}{c|}{ 6.2645(9)}  & \multicolumn{4}{c|}{ 6.2652(1)}\\
\multicolumn{2}{|c|}{$c$ (\r{A})} & \multicolumn{3}{c|}{ 6.5462(8) } & \multicolumn{3}{c|}{ 6.5590(15)} & \multicolumn{4}{c|}{ 6.5586(1)} \\ 
\hline
Atom   & Wyckoff site &  $x$   &  $y$    &  $z$   &  $x$   &  $y$    &  $z$ & $x$   &  $y$    &  $z$ & $B_{\mathrm{iso}}$ \\ 
 \hline
Fe     & 4$f$         &  0.1566(3) & 0.1566(3)   & 0          & 0.1552(8) & 0.1552(8) & 0   & 0.15739(6) & 0.15739(6) & 0   & 0.054(1)  \\
Ga1    & 8$j$         &  0.3448(2) & 0.3448(2)   & 0.2635(2)  & 0.3445(9) & 0.3445(9) & 0.2626(10)   & 0.34462(5) & 0.34462(5) & 0.26193(7)   & 0.045(2)  \\
Ga2    & 4$c$         &  0     & 0.5     & 0                  & 0 & 0.5 & 0 & 0 & 0.5 & 0   & 0.046(1) \\
\hline
 \end{tabular}

 \caption{\label{table1} Structural parameters derived from the Rietveld refinements of the nuclear structure based on our neutron powder diffraction patterns measured for FeGa$_3$ at temperatures of 1.5~K and 300~K. The last columns present results of the Rietveld refinement of the X-ray powder diffraction (XRD) pattern for FeGa$_3$ collected at room temperature. All these refinements resulted in full occupancies of the three atomic sites.} 
 \end{table*}
\end{center}

\begin{figure*}[h] 
\includegraphics[width=0.38\textwidth,angle=-90]{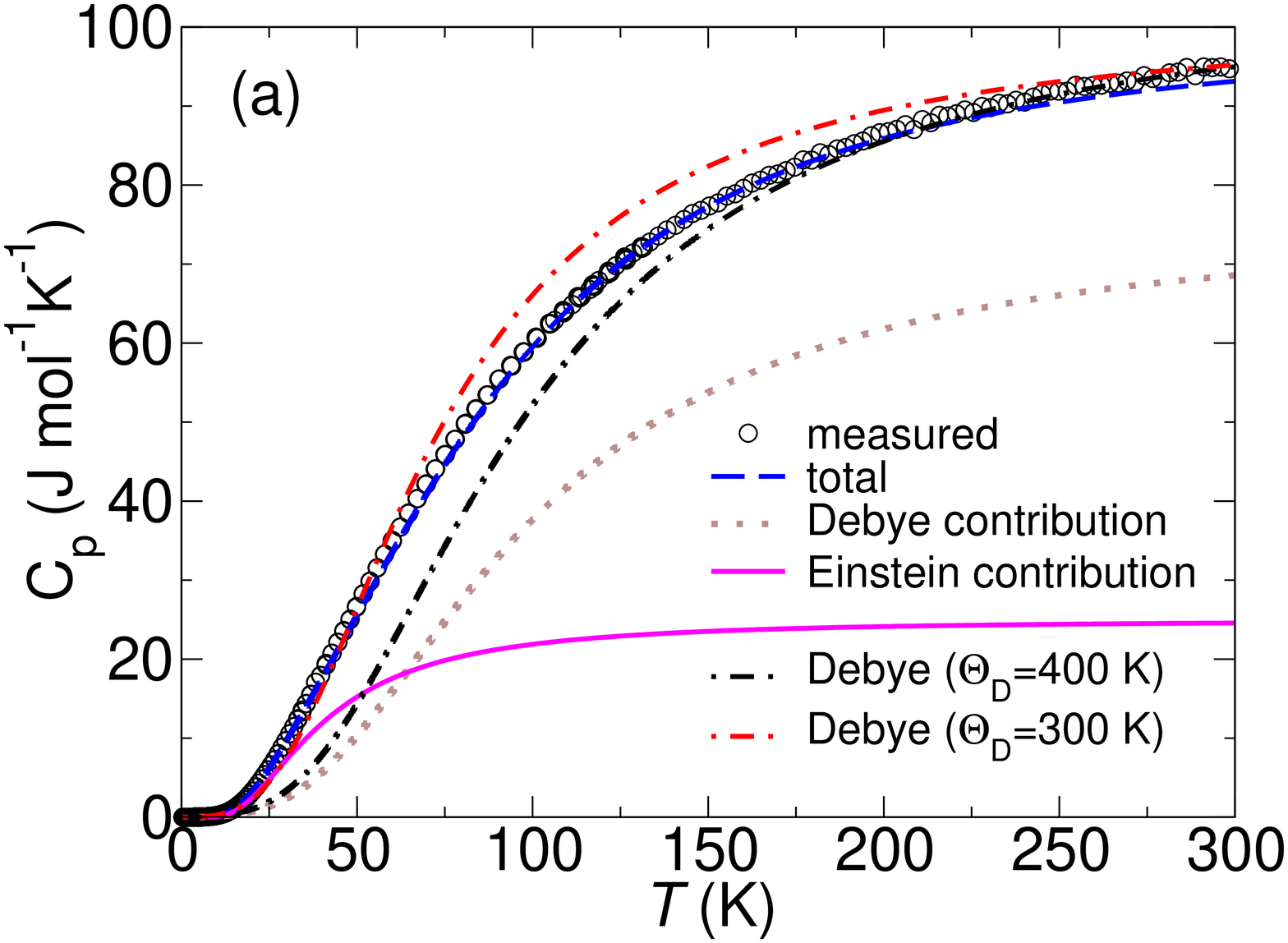}
\includegraphics[width=0.38\textwidth,angle=-90]{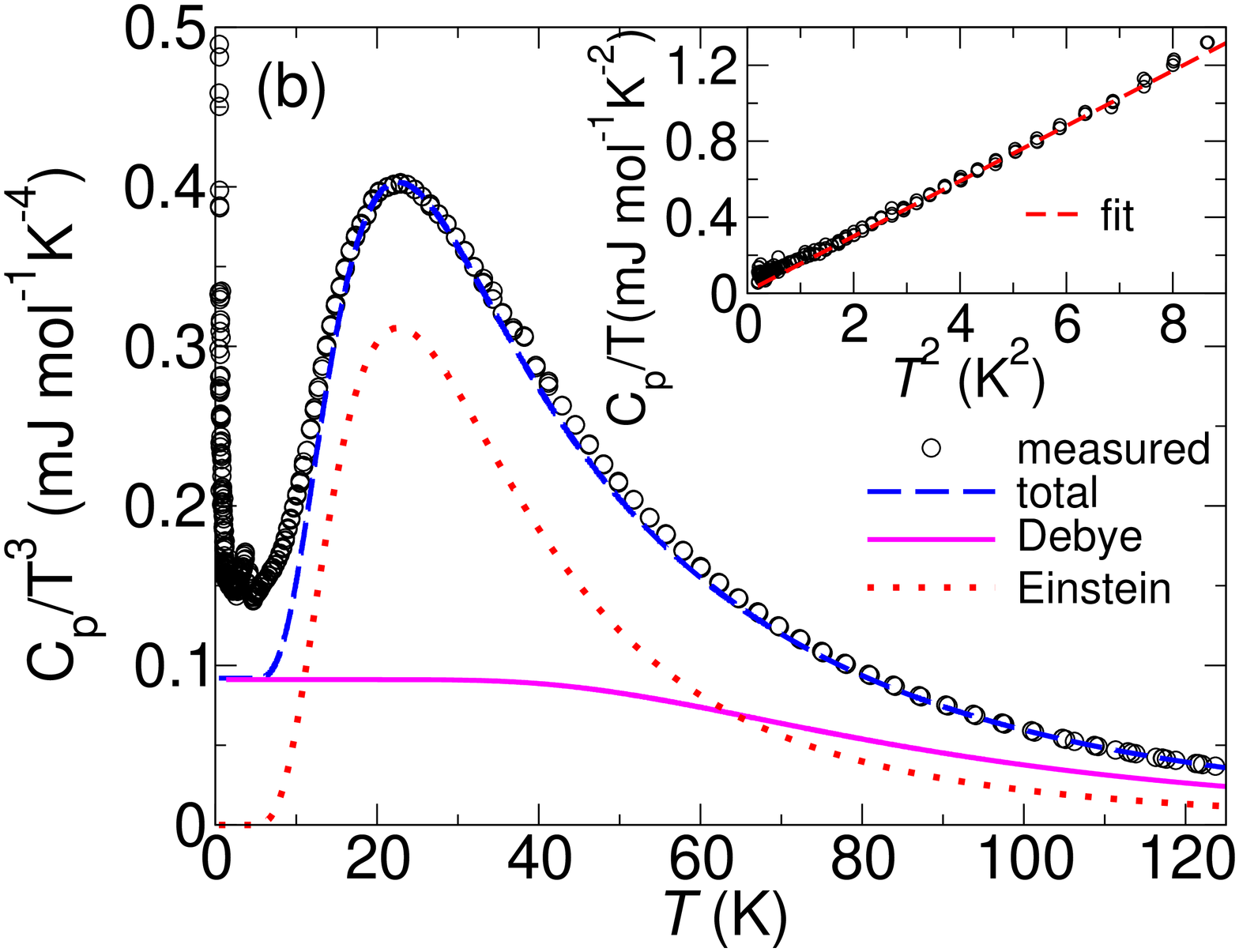}
\caption{\label{fig:Fig5} (Color online)  Specific heat $C_p$ of FeGa$_3$ (black circles) modelled with a series of Einstein--type (solid line) and Debye--type (dotted line) lattice terms (see text). The blue dashed line represents the best fit of the assumed model to the experimental data. Dashed--dotted thin lines in panel (a) represent the ``total'' specific heat calculated based on the Debye model, assuming Debye temperatures of 300~K (red) and 400~K (black). A maximum in $C$/$T^3$($T$) dependence shown in panel (b) is the hallmark of the presence of Einstein modes. The inset in panel (b) shows  $C_p$/T at lowest temperatures plotted versus $T^2$. The red dashed line represents the fit to the experimental data using $C_p$/$T$($T^2$)=$\beta T^2$.} 
\end{figure*}

The overall temperature dependence of the specific heat $C_p$ for FeGa$_3$ is depicted in Fig.~\ref{fig:Fig5}a and is similar to that reported by Hadano $et$  $al.$\cite{Hadano} 
At ambient temperature, the specific heat nearly reaches \mbox{100~J/(mol~K),} which is the result expected from the Dulong--Petit law. 
At $T\textless 3$~K, the experimental data follows the dependence anticipated for the contribution of phonons $C_p$($T$)=$\beta T^3$ (see inset in Fig.~\ref{fig:Fig5}b) with \mbox{$\beta\approx 0.139$~mJ/(mol~K$^4$)}. The Debye model gives \mbox{$\beta=(12/5)Rn\pi^4\theta_{\mathrm{D}}^{-3}$,} where $n$ is the number of atoms per formula unit and $R$ is the gas constant. Thus, we estimate a Debye temperature \mbox{$\theta_{\mathrm{D}}\approx 382$~K.}

For FeGa$_3$, the magnetic contribution to the specific heat is expected to be negligible at temperatures below \mbox{$\sim$300~K} since the magnetic susceptibility is diamagnetic and only weakly temperature dependent in this regime. Consequently, at temperatures above $\sim$3~K, $C_p$($T$) is strongly dominated by the vibrational properties of the Fe and Ga sublattices of FeGa$_3$.    
Attempts to fit the overall temperature dependence of the specific heat as the lattice specific heat \mbox{$C_{\mathrm{lattice}}$($T$)} given by a Debye model, are not satisfactory (Fig.~\ref{fig:Fig5}(a)).
In order to get an acceptable description of the experimental $C_p$($T$), the lattice specific heat of FeGa$_3$ has been modelled by the sum of contributions from Debye-type ($C_{\mathrm{D}}$) and Einstein-type ($C_{\mathrm{E}}$) modes, using the following expressions 
\begin{equation}
C_{\mathrm{D}}(T) = Rn_{\mathrm{D}}\int_0^{\Theta_{\mathrm{D}}/T}\frac{x^4e^x}{(e^x-1)^2}dx
\label{eq:2}
\end{equation} and
\begin{equation}
C_{\mathrm{E}}(T) = Rn_{\mathrm{E}}(\frac{\Theta_{\mathrm{E}}} {T})^2\frac{e^{\frac{\Theta_{\mathrm{E}}}{T}}}{(e^{\frac{\Theta_{\mathrm{E}}}{T}}-1)^2} ,
\label{eq:3}
\end{equation}
where $n_{\mathrm{D}}$ and $n_{\mathrm{E}}$ denote the number of Debye and Einstein modes per formula unit of FeGa$_3$.

The best match with the experimental data is achieved by assuming $n_{\mathrm{D}}=9$ Debye--type modes with $\theta_{\mathrm{D}}=400$~K, one Einstein--type mode with a characteristic temperature $\theta_{\mathrm{E}}=94$~K and two Einstein--type modes with $\theta_{\mathrm{E}}=140$~K, resulting in a total of 12~phonon modes, as expected for FeGa$_3$.  
In Fig.~\ref{fig:Fig5}(b) we plot $C_{p}$/$T^3$ versus $T$. The clear maximum indicates that Einstein modes must be introduced for a reasonable fit.
 
The presence of the Einstein modes in FeGa$_3$ is rather unexpected. Generally, low energy optical modes are associated with specific structure types. For compounds with cage--like crystal structures, for instance, the Einstein term in the specific heat reflects local vibrations of loosely bound atoms inside the cages. Yet, in FeGa$_3$ each atom has several neighbours at distances shorter than the sum of adjacent nominal atomic radii.\cite{Dasarathy} Single crystal X--ray diffraction measurements do not show an enhancement of any displacement parameters in FeGa$_3$\cite{Hauserman} that could otherwise indicate the presence of loosely bound atoms in the crystal lattice. Therefore, the evidence in the specific heat of FeGa$_3$ for local vibrations calls for further investigations, such as thermal expansion measurements.

\begin{figure}
\begin{tabular}{c}
\epsfig{file=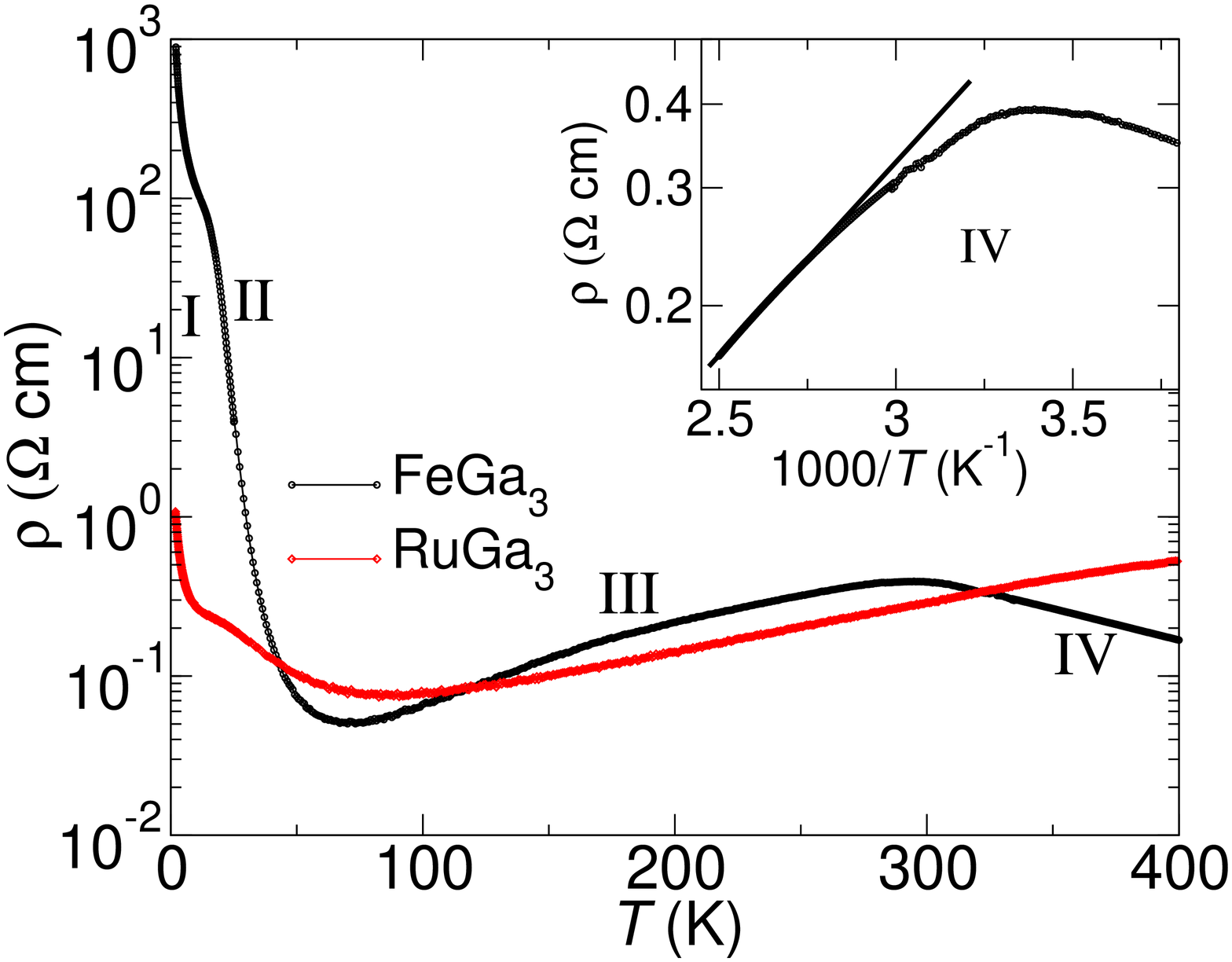,width=0.4\textwidth,angle=-90,clip=} \\
\begin{tabular}{cc}
\epsfig{file=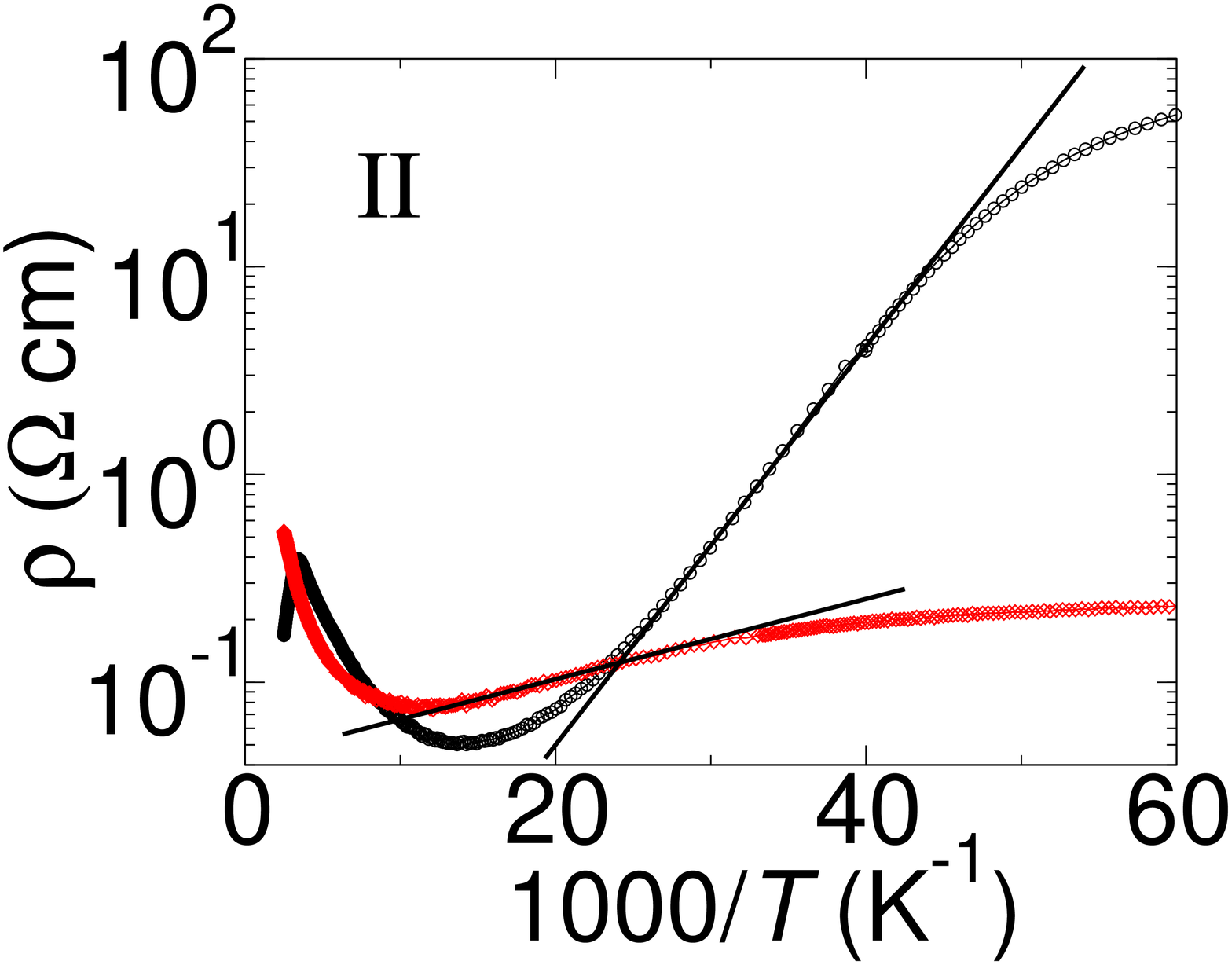,width=0.18\textwidth,angle=-90,clip=} &
\epsfig{file=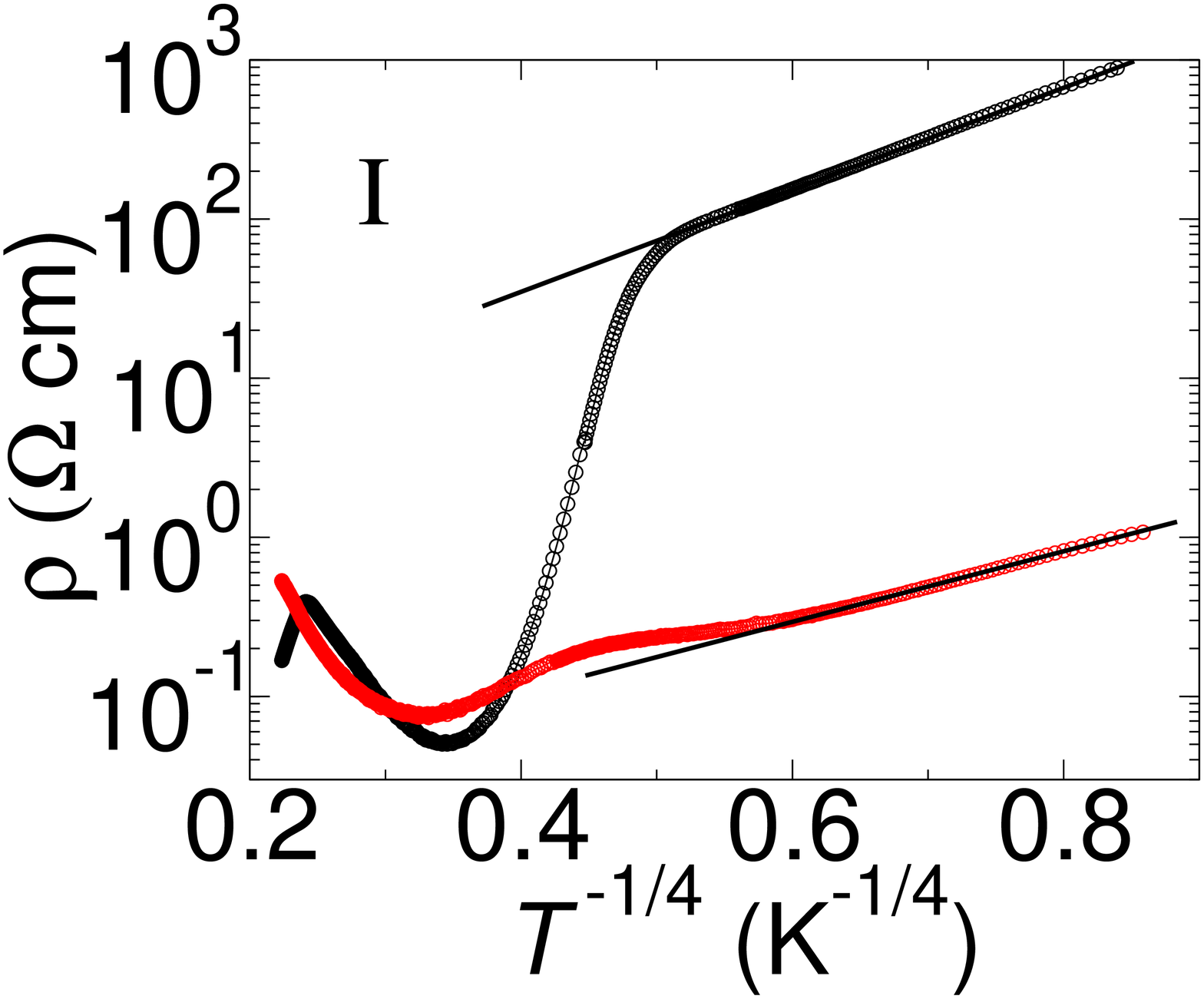,width=0.18\textwidth,angle=-90} \\
\end{tabular}
\end{tabular}
\caption{(Color online)  Electrical resistivity $\rho$ for FeGa$_3$ (black circles) and RuGa$_3$ (red circles). Roman numbers I--IV denote four temperature regimes as described in the text. The inset shows the resistivity for FeGa$_3$ in the temperature range \mbox{260--400~K~(IV)} on a logarithmic scale as a function of inverse temperature. The solid line represents the fit of the experimental data with Arrhenius law \mbox{$\rho$($T$)$=\rho_0$($T$)$ exp$($E_{\mathrm{g}}/$($2k_{\mathrm{B}}T$)),} giving the activation gap \mbox{$E_{\mathrm{g}}=0.4$~eV.} The left small panel displays Arrhenius plots for the range~II with solid lines indicating the best fits of the resistivity data to the Arrhenius expression \mbox{$\rho$($T$)$=\rho_0$($T$)$ exp$($E_{\mathrm{d}}/$($2k_{\mathrm{B}}T$))} at \mbox{24~K$\textless T\textless 44$~K} and \mbox{31~K$\textless T\textless 83$~K,} resulting in activation energies of \mbox{$E_{\mathrm{d}}\approx 40$~meV} and \mbox{$E_{\mathrm{d}}\approx 20$~meV} for FeGa$_3$ and RuGa$_3$, respectively. The right small panel shows the resistivities at the lowest temperatures (I) plotted versus $T^{-1/4}$ to display the 3D~VRH conduction among in--gap states. Solid lines indicate the best linear regression fits.} 
  \label{fig:Fig6} 
\end{figure}

The electrical resistivity $\rho$ was measured on single crystals of FeGa$_3$ and RuGa$_3$ (Fig.~\ref{fig:Fig6}).  
The obtained $\rho$($T$) for FeGa$_3$ is very similar to that found in previous reports based on single crystals,\cite{Hadano, Hauserman} and differs notably from some results for polycrystalline specimens.\cite{Lue, Amagai} In turn, for RuGa$_3$ there are substantial differences between our resistivity data and the $\rho$($T$) curves reported previously for both single crystals\cite{Hauserman} and polycrystalline specimens.\cite{Amagai, Neel} Such a sensitivity of physical properties to structure, morphology and chemical composition is typical for correlated insulators.        

The absolute values of the resistivity are similar for crystals of FeGa$_3$ and RuGa$_3$ grown from Ga flux.
The resistivity of FeGa$_3$ displays four distinct temperature regimes, which we indicate by Roman numbers (see Fig.~\ref{fig:Fig6}):  
At $T\textgreater 350$~K (regime~IV), the resistivity of FeGa$_3$ follows the Arrhenius law $\rho$($T$)$/\rho_0$($T$)$=exp$($E_{\mathrm{g}}/$($2k_{\mathrm{B}}T$)), with an activation energy gap \mbox{$E_{\mathrm{g}}\approx 0.4$~eV} (Fig.~\ref{fig:Fig6}(b)), in line with the size of the band gap derived from a spectroscopic study and electronic structure calculations.\cite{Arita, Hauserman, Picket} 
Below $T\sim 300$~K~(III), the resistivity of FeGa$_3$ decreases with decreasing temperature. For RuGa$_3$ a similar behavior in resistivity is found to extend over an even broader temperature range. 
Since in RuGa$_3$ the orbitals responsible for low energy excitations are of 4$d$ character that are much more delocalized than their 3$d$ counterparts in FeGa$_3$,
the emergence of the metallic--type range~(III) in the resistivity of FeGa$_3$ is likely not associated directly with correlation effects.    
At lower temperatures, the resistivities of both FeGa$_3$ and RuGa$_3$ increase again. The temperature dependencies of the resistivity from 20~K to 45~K for FeGa$_3$ and from 30~K to 80~K for RuGa$_3$ (II, Fig.~\ref{fig:Fig6}(c)) can be again well described by the Arrhenius law \mbox{$\rho$($T$)$/\rho_0$($T$)$= exp$($E_{\mathrm{d}}/$($2k_{\mathrm{B}}T$)),} with activation energies \mbox{$E_{\mathrm{d}}\approx 40$~meV} and \mbox{$E_{\mathrm{d}}\approx 20$~meV} for FeGa$_3$ and RuGa$_3$, respectively. 
These activation gaps are more than one order of magnitude smaller than the band gaps for both compounds.\cite{Arita, Hauserman, Imai}  
%Since the values of the activation gap are more than one order of magnitude smaller than the size of the band gap for both compounds,\cite{Arita, Hauserman, Imai} their conductivities in this temperature range can be assigned to excitations of charge carriers from localized in--gap states to the conduction band. 
Finally, at the lowest temperatures~(I), the resistivities of both FeGa$_3$ and RuGa$_3$ vary as \mbox{$\rho$($T$)$\propto exp$[($T_{\mathrm{M}}$/$T$)$^{1/4}$)]} (Fig.~\ref{fig:Fig6}(d)).  Such a temperature dependence is predicted for the conduction due to variable range hopping (VRH) among localized in--gap states in a 3D system, a dominant conduction mechanism in various classes of lightly doped semiconductors at low temperatures. Although the character of the temperature dependence of the resistivity in this temperature range was similar for several studied crystals of both FeGa$_3$ and RuGa$_3$, considerable variations in values of $T_{\mathrm{M}}$ were observed. 

The temperature dependencies of the resistivities for FeGa$_3$ and RuGa$_3$ observed in regimes I--III can be ascribed to the presence of numerous localized donor--type states in their band gaps.   
Indeed, a recent nuclear spin--lattice relaxation study for FeGa$_3$ unambiguously proved the existence of in--gap states located just below the bottom of the conduction band.\cite{Sascha}
The number of these states estimated from the Hall constant\cite{Hadano} is \mbox{$\sim$1.6$\times$10$^{16}$ cm$^{-3}$}, which is much larger than typical impurity levels observed in intrinsic band semiconductors (e.g. Si, Ge) and even in correlated band insulators (e.g. FeSi, FeSb$_2$).\cite{Sascha}   
Measurements of the Hall effect performed by Hadano $et$~$al$.\cite{Hadano} indicated that in region III mobility of carriers decreases strongly with increasing temperature ($\mu \sim T^{-5/2}$), whereas the number of carriers remains nearly unchanged. Consequently, the emergence of the metallic behaviour of the resistivity can be explained as due to this strong decrease in carrier mobility in the saturation range, where electrons from all the donor levels are thermally excited to the conduction band. 
In turn, a gradual freezing of these donors leads to the activated $\rho$($T$) at temperatures below $\sim$45~K and $\sim$80~K for FeGa$_3$ and RuGa$_3$, respectively.  
Further study is needed to elucidate the origin and character of these numerous in--gap states.

\subsection{Fe$_{1-x}$Mn$_x$Ga$_3$ and FeGa$_{3-y}$Zn$_y$}\label{doped}

\begin{figure*}[t] 
\centering
\begin{tabular}{cc}
\epsfig{file=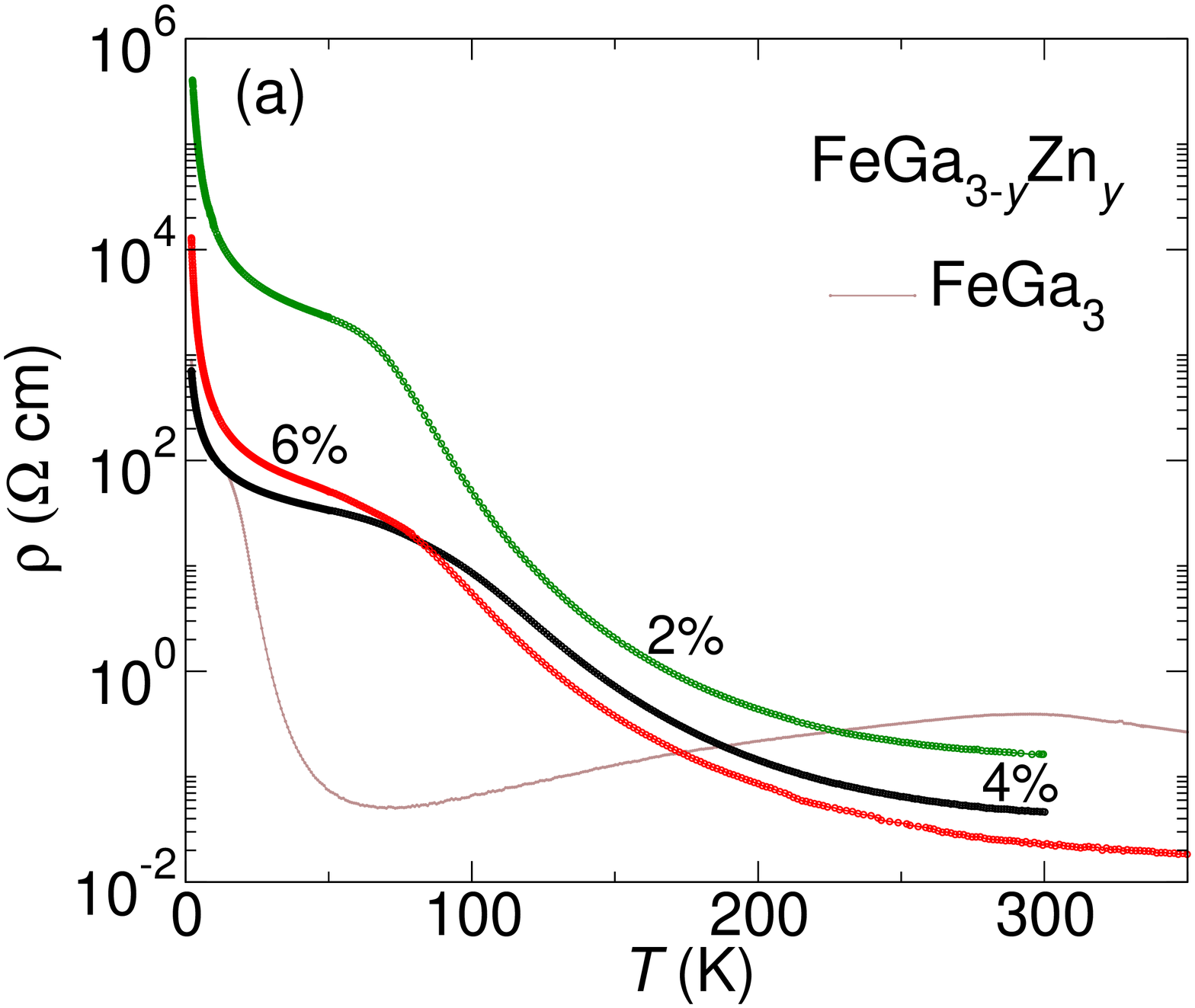,width=0.4\textwidth,angle=-90,clip=} &
\epsfig{file=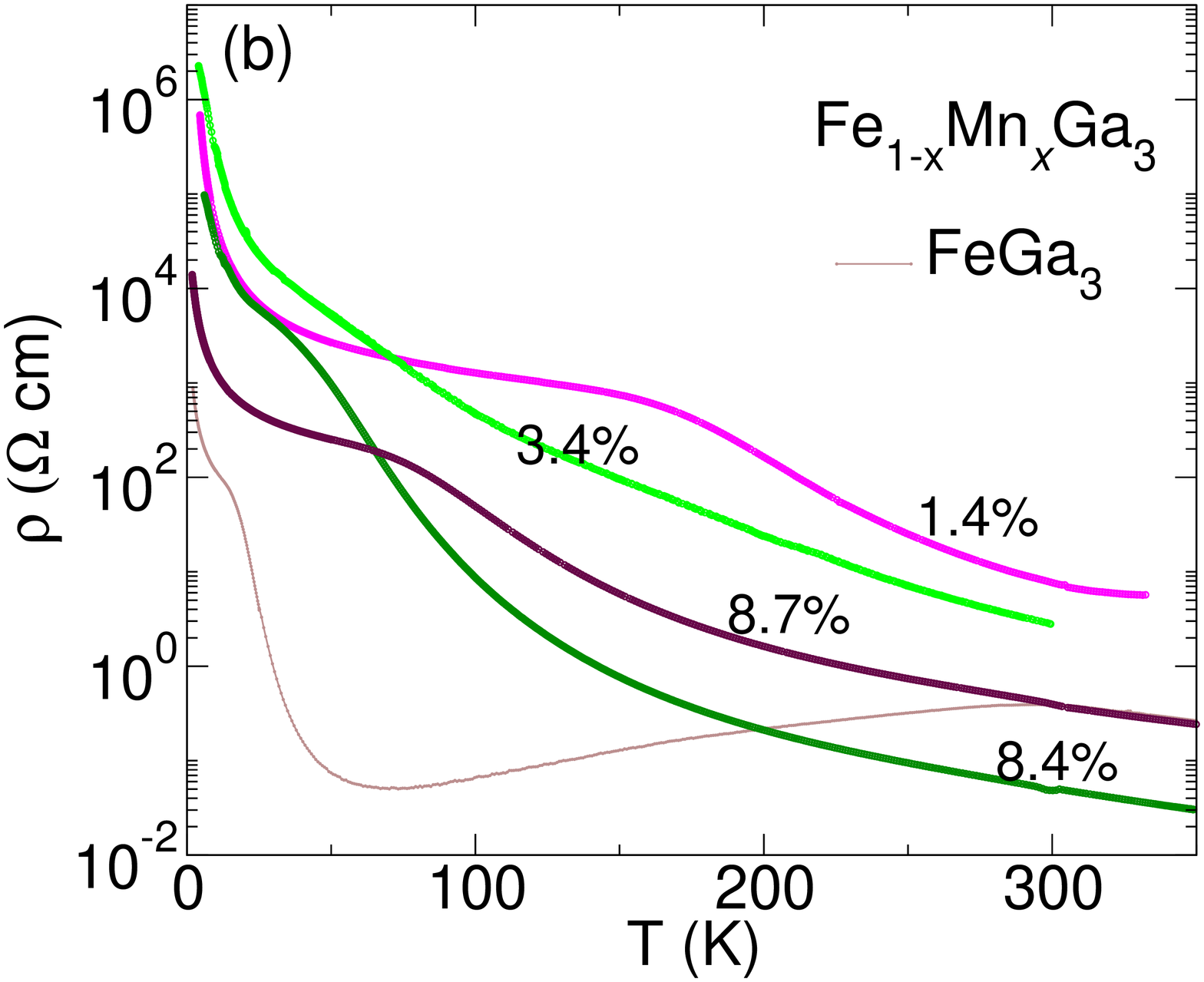,width=0.4\textwidth,angle=-90} \\
\epsfig{file=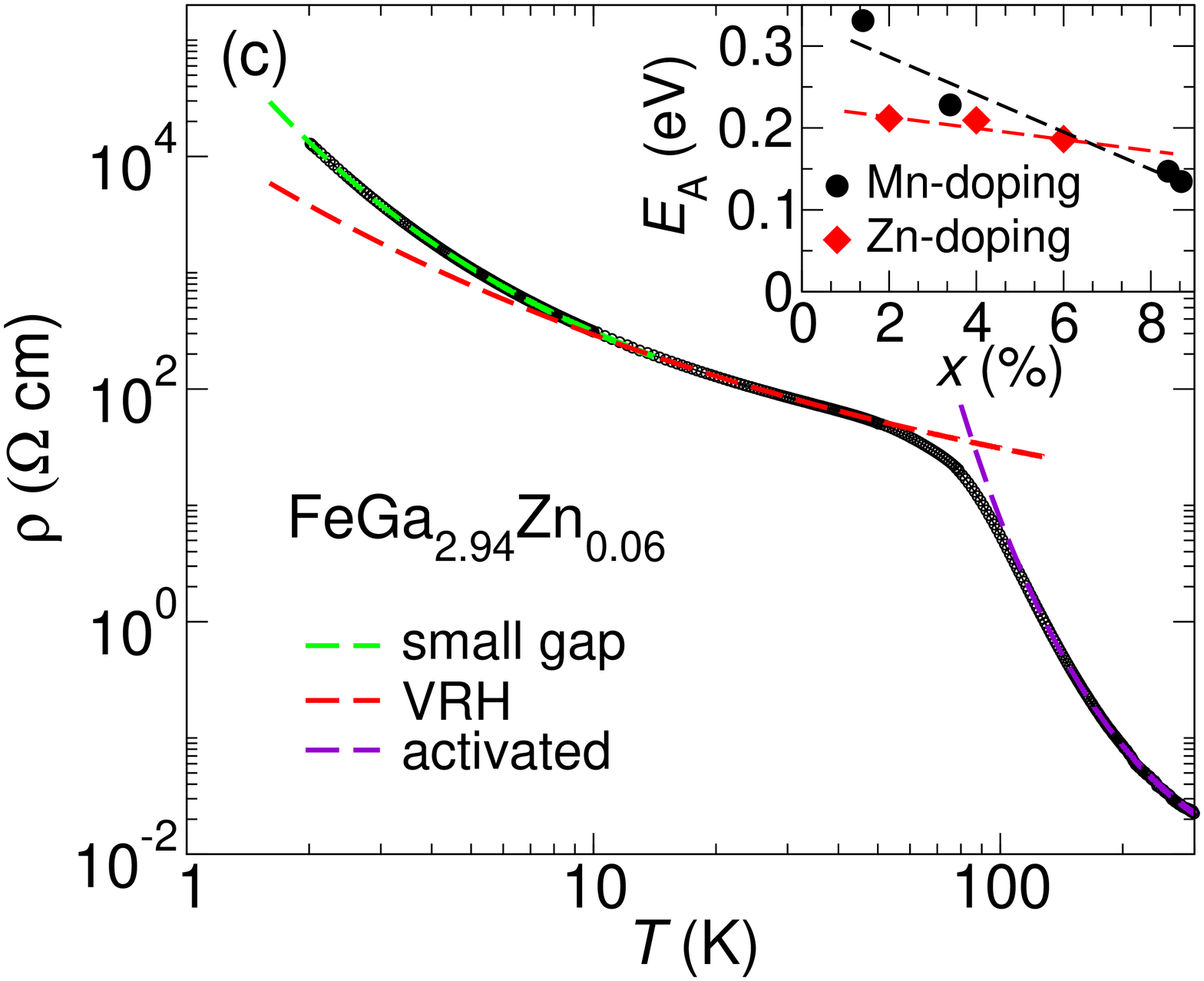,width=0.4\textwidth,angle=-90,clip=} &
\epsfig{file=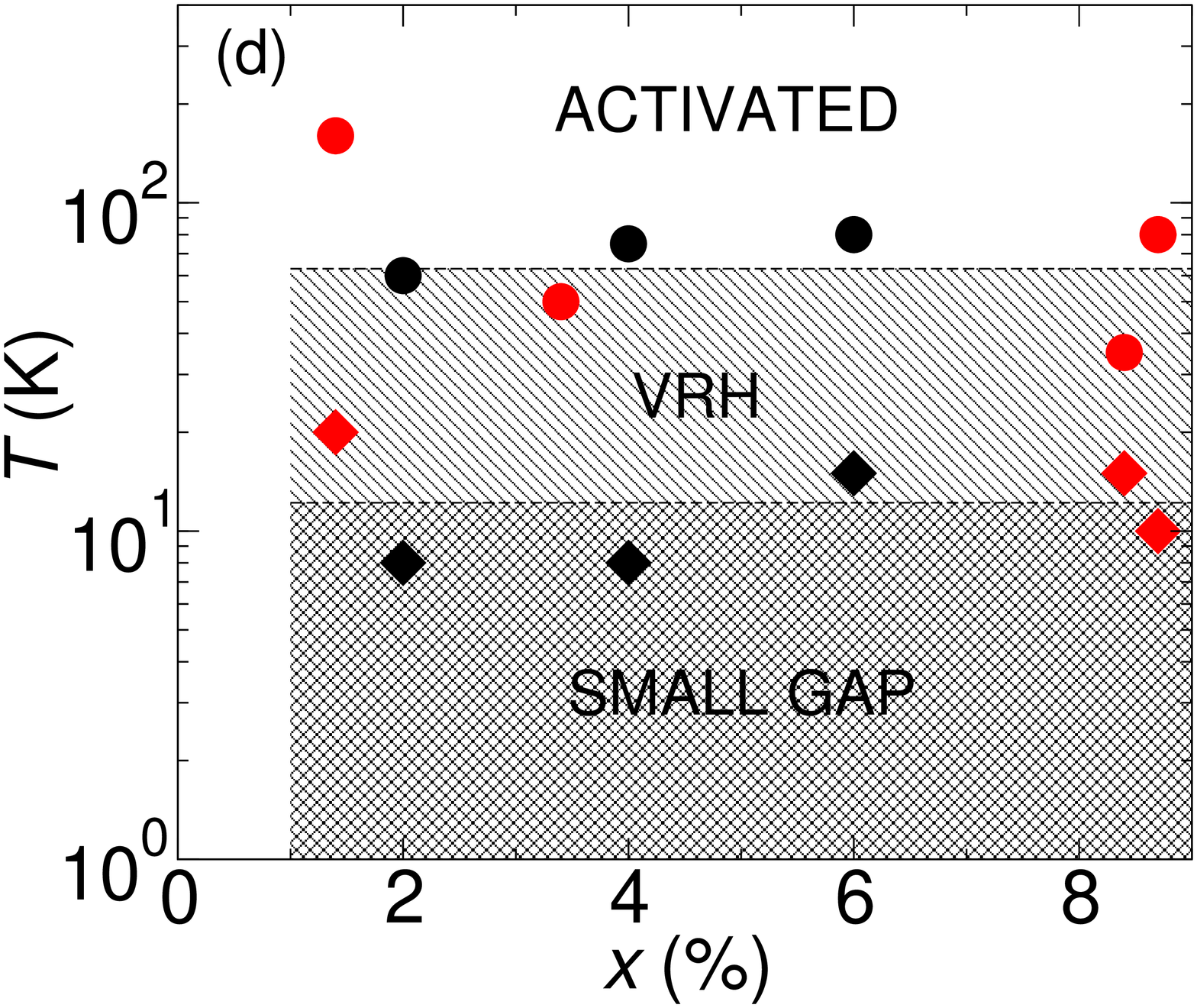,width=0.4\textwidth,angle=-90} \\
\begin{tabular}{cc}
\epsfig{file=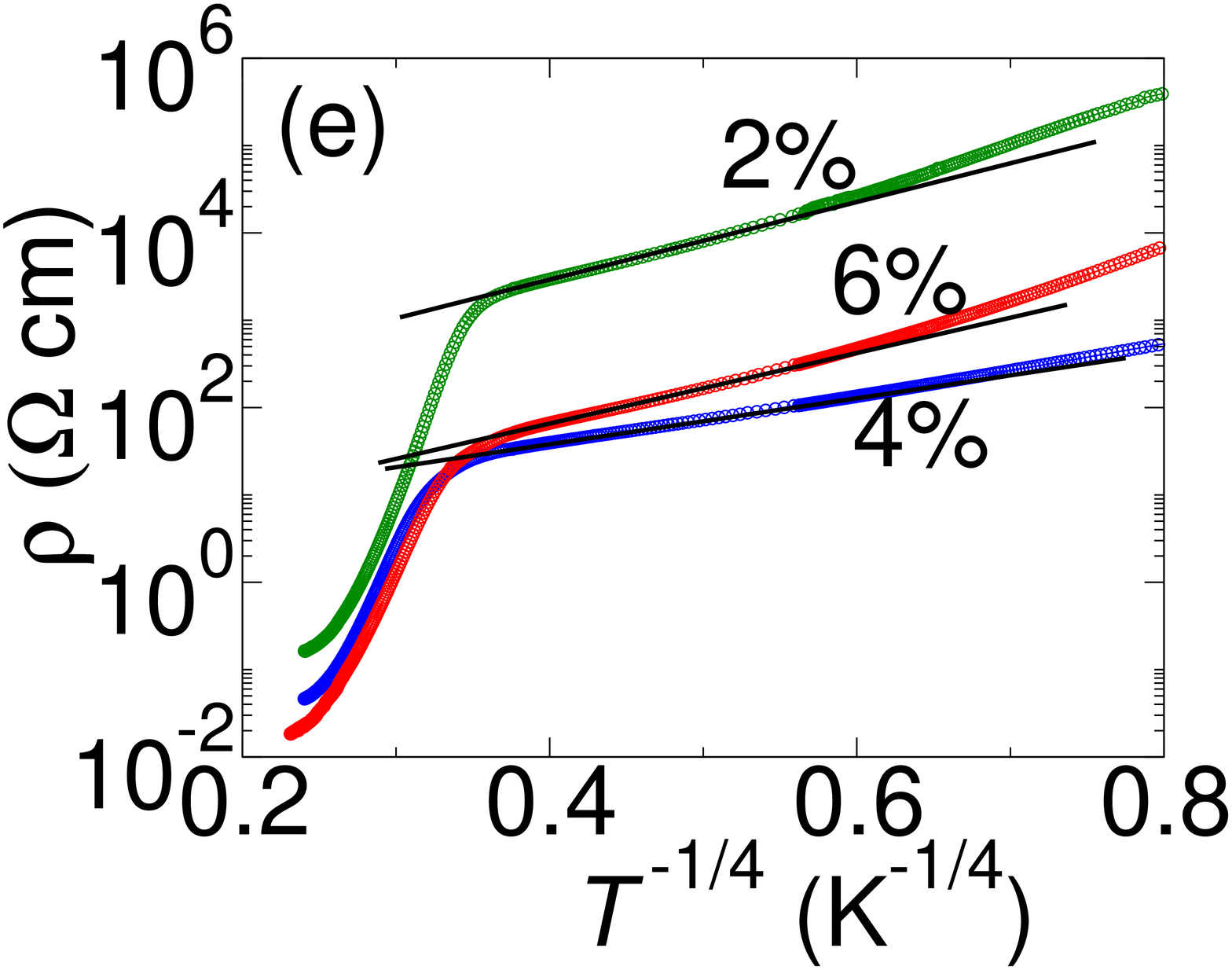,width=0.18\textwidth,angle=-90,clip=} &
\epsfig{file=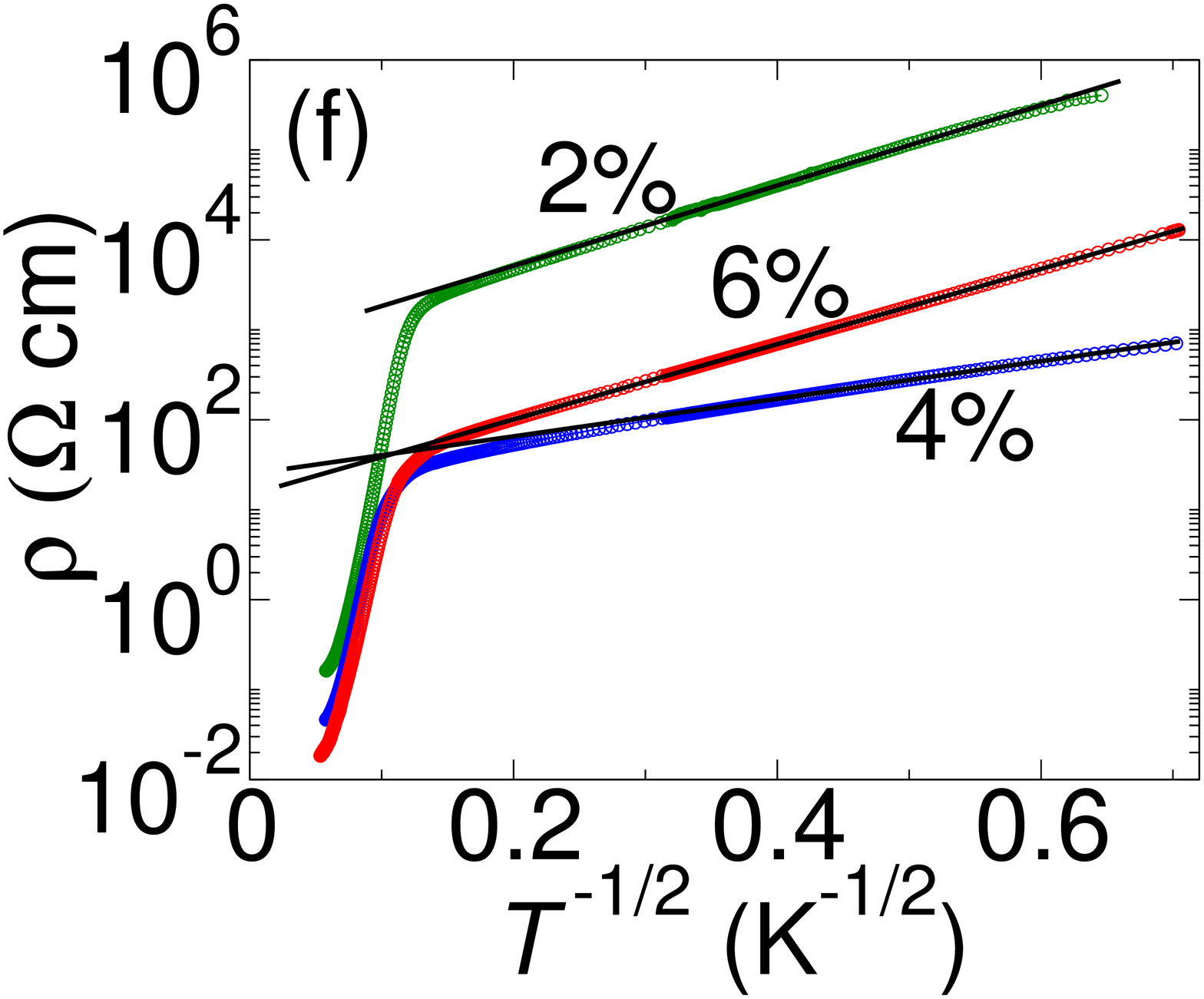,width=0.18\textwidth,angle=-90} \\
\end{tabular} &
\begin{tabular}{cc}
\epsfig{file=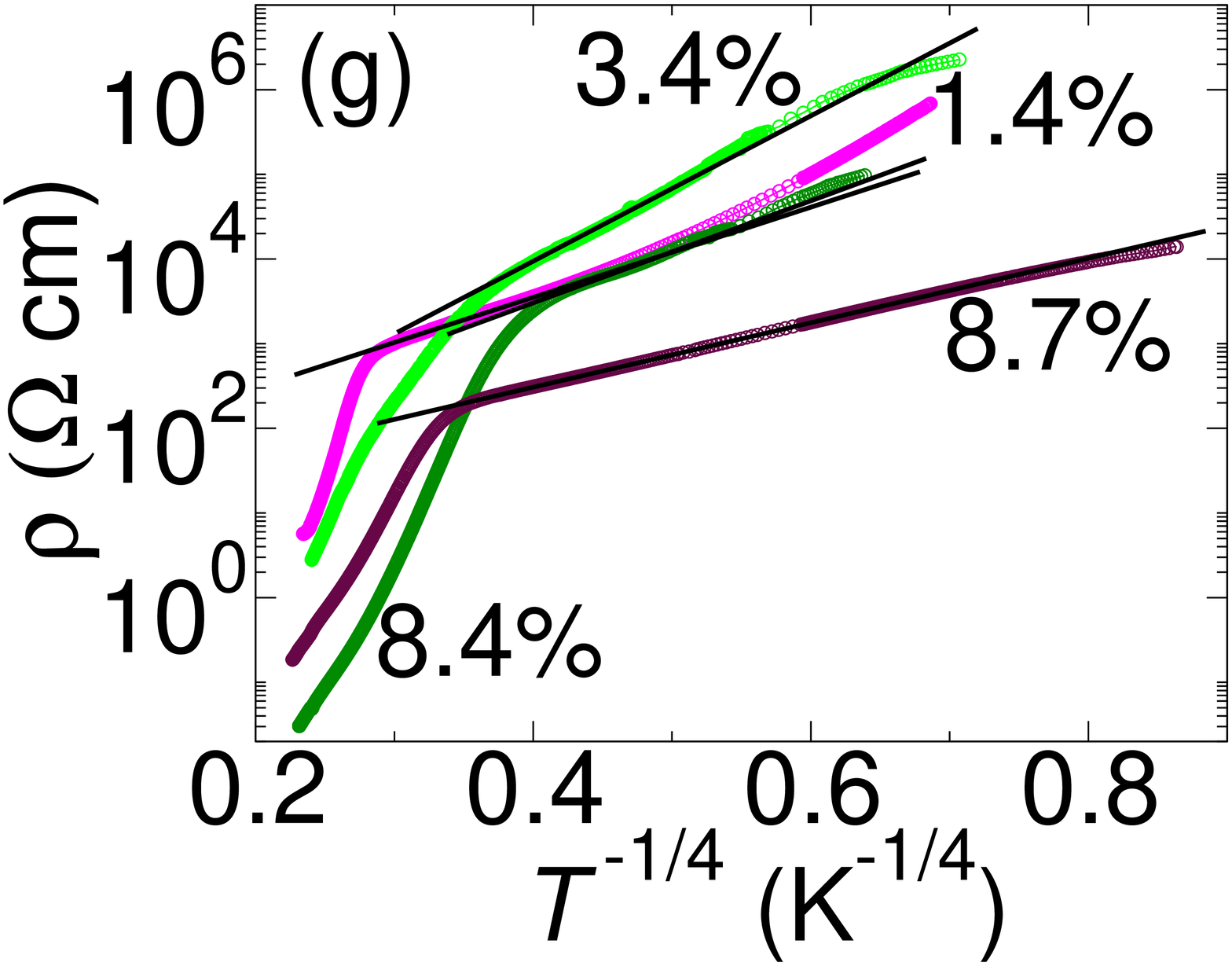,width=0.18\textwidth,angle=-90,clip=} &
\epsfig{file=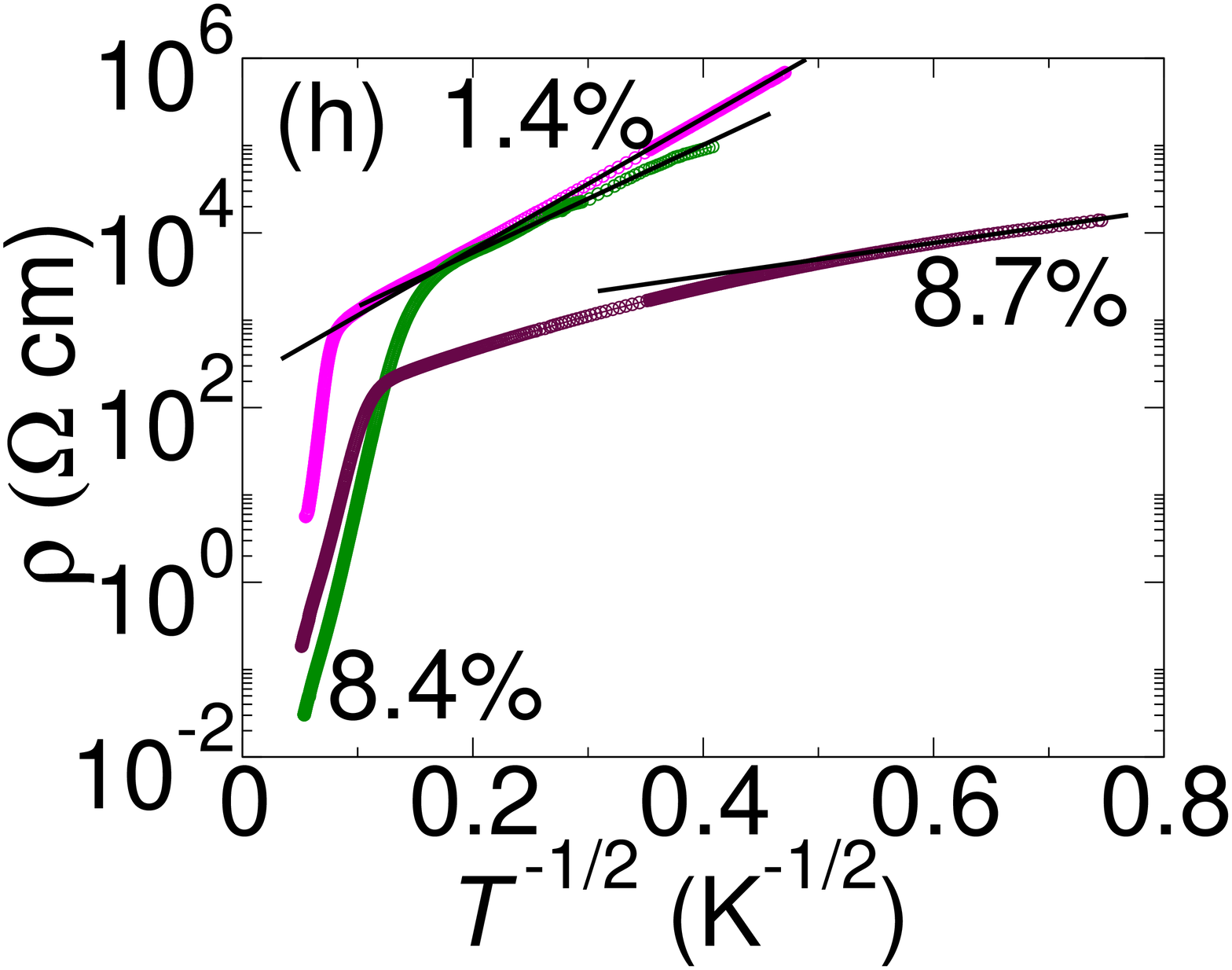,width=0.18\textwidth,angle=-90} \\
\end{tabular} \\
\end{tabular}
\caption{\label{fig:Fig7} (Color online) Electrical resistivity for FeGa$_{3-y}$Zn$_y$ ($y=0.02$; 0.04; 0.06) (a) and Fe$_{1-x}$Mn$_x$Ga$_3$ ($x=0.014$; 0.034; 0.084; 0.087) (b) plotted together with the data for FeGa$_3$ (brown solid lines). 
Panel (c) presents the resistivity of FeGa$_{2.94}$Zn$_{0.06}$ (black circles) together with fits using Arrhenius law \mbox{$\rho$($T$)$\propto exp$($E_{\mathrm{d}}/$($2k_{\mathrm{B}}T$))} for \mbox{$T\apprge 100$~K} (dashed violet line), 3D VRH conduction \mbox{$\rho$($T$)$\propto exp$[($T_{\mathrm{M}}/T$)$^{1/4}$]} at \mbox{10~K$\apprge T\apprge 50$~K} (dashed blue line) and based on the formula \mbox{$\rho$($T$)$\propto exp$[($T_M$/$T$)$^{1/2}$]} (dashed red line) describing 3D VRH conduction of the Efros--Shklovskii type\cite{ESkl} associated with the opening of a small gap within the doping induced density of states at \mbox{$T\apprge 10$~K.} The temperature ranges in which the resistivities of FeGa$_{3-y}$Zn$_y$ and Fe$_{1-x}$Mn$_x$Ga$_3$ follow the described temperature dependencies are indicated in panel (d). The points denote crossover temperatures estimated as the midpoints between the corresponding temperature ranges.      
The resistivities of FeGa$_{3-y}$Zn$_y$ and Fe$_{1-x}$Mn$_x$Ga$_3$ for different $x$ and $y$ are plotted versus $T^{-1/4}$ in panels (e) and (g), respectively, and as functions of $T^{-1/2}$ in panels (f) and (h), respectively. Solid lines in panels (e, f, g, h) indicate the best regression fits. The inset in panel (c) shows values of the average activation gap obtained from the fits to the Arrhenious expression at highest temperatures as a function of the actual doping level $x$. }
\end{figure*}

Doping FeGa$_3$ with both Mn and Zn results in drastic changes in $\rho$($T$), compared to undoped FeGa$_3$.   
Fig.~\ref{fig:Fig7} presents the electrical resistivity of crystals doped with Zn (panel (a)) and with Mn (panel (b)). 
Already for the lowest doping levels, the metallic range in $\rho$($T$) that was observed for undoped crystals of FeGa$_3$ from $\sim$60~K to $\sim$300~K gives way to an activated temperature dependence followed by broad maxima at temperatures of 30--150~K. Although qualitatively similar behaviors of the resistivities were observed for different crystals, the details of the shape and position of the maxima in $\rho$($T$) were found to be sample dependent, even for crystals within a given batch, where EDX microanalysis indicated the same chemical composition and a homogeneous distribution of the dopant atoms across the crystals.   

Since the shape of the $\rho$($T$) for FeGa$_3$ was assigned to the presence of impurity donor--type states inside the band gap,\cite{Hadano, Sascha} the drastic change in $\rho$($T$) upon doping likely results from a compensation mechanism. Our results suggest that the number of holes introduced by doping almost immediately exceeds the number of impurity donor states present in the nominally undoped FeGa$_3$. Consequently, the presence of localized acceptor--type states in the band gap may determine $\rho$($T$) for both Fe$_{1-x}$Mn$_x$Ga$_3$ and FeGa$_{3-y}$Zn$_y$, at least at temperatures up to $\sim$400~K.  
This is in line with $p$--type conduction found by means of thermopower measurements for polycrystals of FeGa$_{2.97}$Zn$_{0.03}$ and FeGa$_{2.94}$Zn$_{0.06}$ at temperatures until 400~K and 500~K, respectively.\cite{GrinFeGa3}

We notice that there are three distinct temperature ranges in the resistivities of both series of compounds Fe$_{1-x}$Mn$_x$Ga$_3$ and FeGa$_{3-y}$Zn$_y$, as indicated in Fig.~\ref{fig:Fig7}(c). At high temperatures, the resistivity is activated. Application of the Arrhenius law to fit the resistivity data of the doped crystals leads to values of the activation gap $E_{\mathrm{A}}$ presented in the inset of Fig.~\ref{fig:Fig7}(c). For both dopants, the average energy for these excitations decreases slightly with increasing doping level.

At lower temperatures, the resistivity is dominated by VRH among the localized in--gap states. 
The temperature dependence of the resistivity can be well described as  
\begin{equation}
\rho(T)\propto exp[(T_{\mathrm{M}}/T)^{p})]
\label{eq:VRH}
\end{equation}   
The value of the exponent $p$ in Eq.~\ref{eq:VRH} distinguishes different conduction mechanisms. For temperatures down to $\sim$10--15~K, $p\approx 0.25$ (see Fig.~\ref{fig:Fig7}(e,g)) as expected for the Mott's type VRH in a 3D~semiconductor and also seen in nominally undoped FeGa$_3$. At the lowest temperatures, the resistivity of all the doped crystals becomes larger than predicted by Mott's expression (see Fig.~\ref{fig:Fig7}(c)). Such an increase in the resistivity hints at the opening of a small gap within the doping induced density of states at the Fermi level. 
Indeed, a good description of the experimental data for FeGa$_{3-y}$Zn$_y$ at temperatures below $\sim$10~K was achieved assuming $p=0.5$ (see Fig.~\ref{fig:Fig7}(f), as predicted for the 3D~VRH conduction of the Efros--Shklovskii type.\cite{ESkl} For Fe$_{1-x}$Mn$_x$Ga$_3$, the values of the exponent $p$ for the resistivity at lowest temperatures are somewhat ambiguous. Attempts to describe the experimental data assuming $p=0.5$ are shown in Fig.~\ref{fig:Fig7}(h).  
The temperature ranges on which the electrical resistivities of FeGa$_{3-y}$Zn$_y$ and Fe$_{1-x}$Mn$_x$Ga$_3$ follow the three distinct temperature dependences are indicated in Fig.~\ref{fig:Fig7}(d).

\begin{figure*}
\begin{tabular}{cc}
\epsfig{file=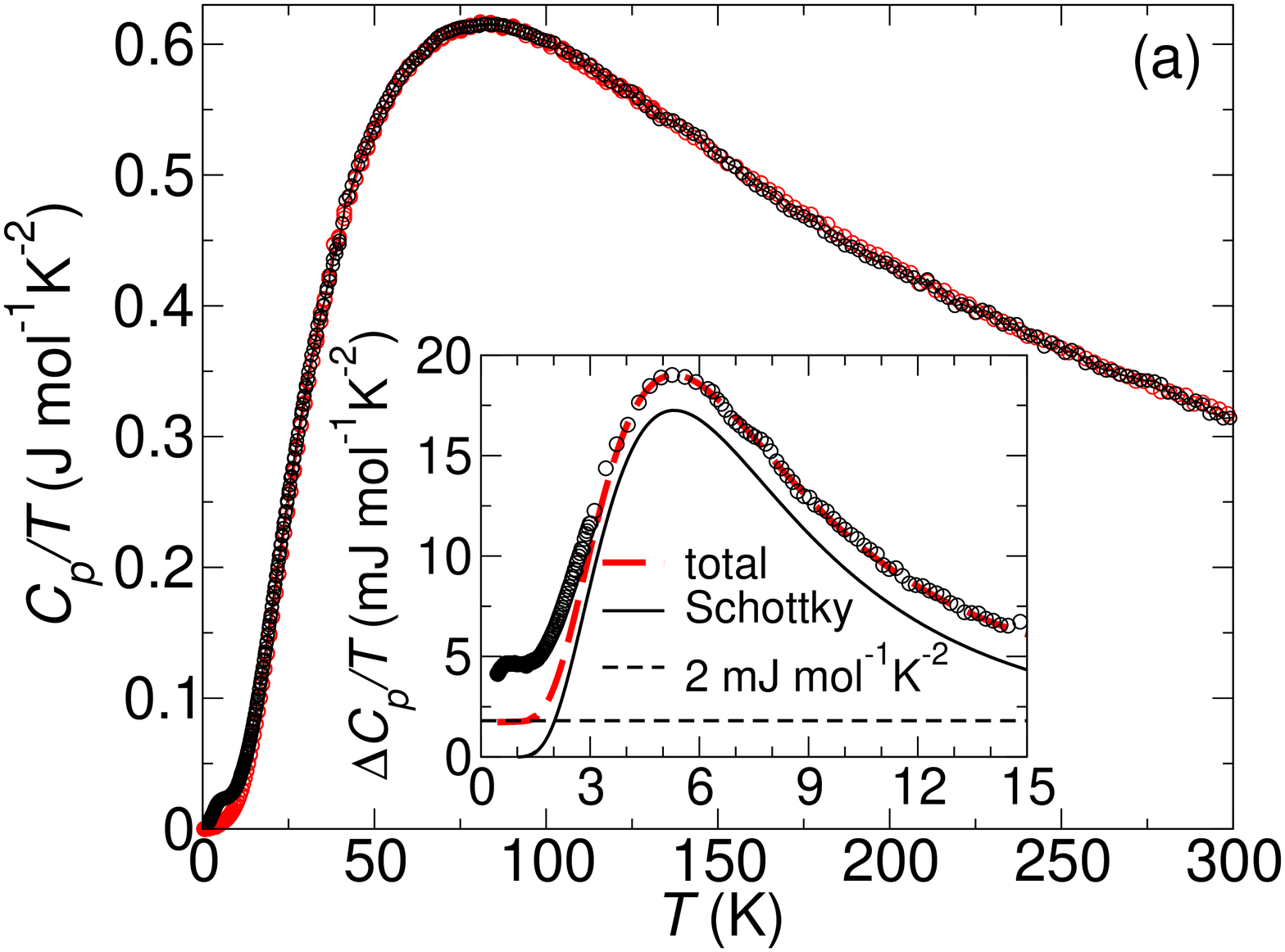,width=0.38\textwidth,angle=-90,clip=} &
\epsfig{file=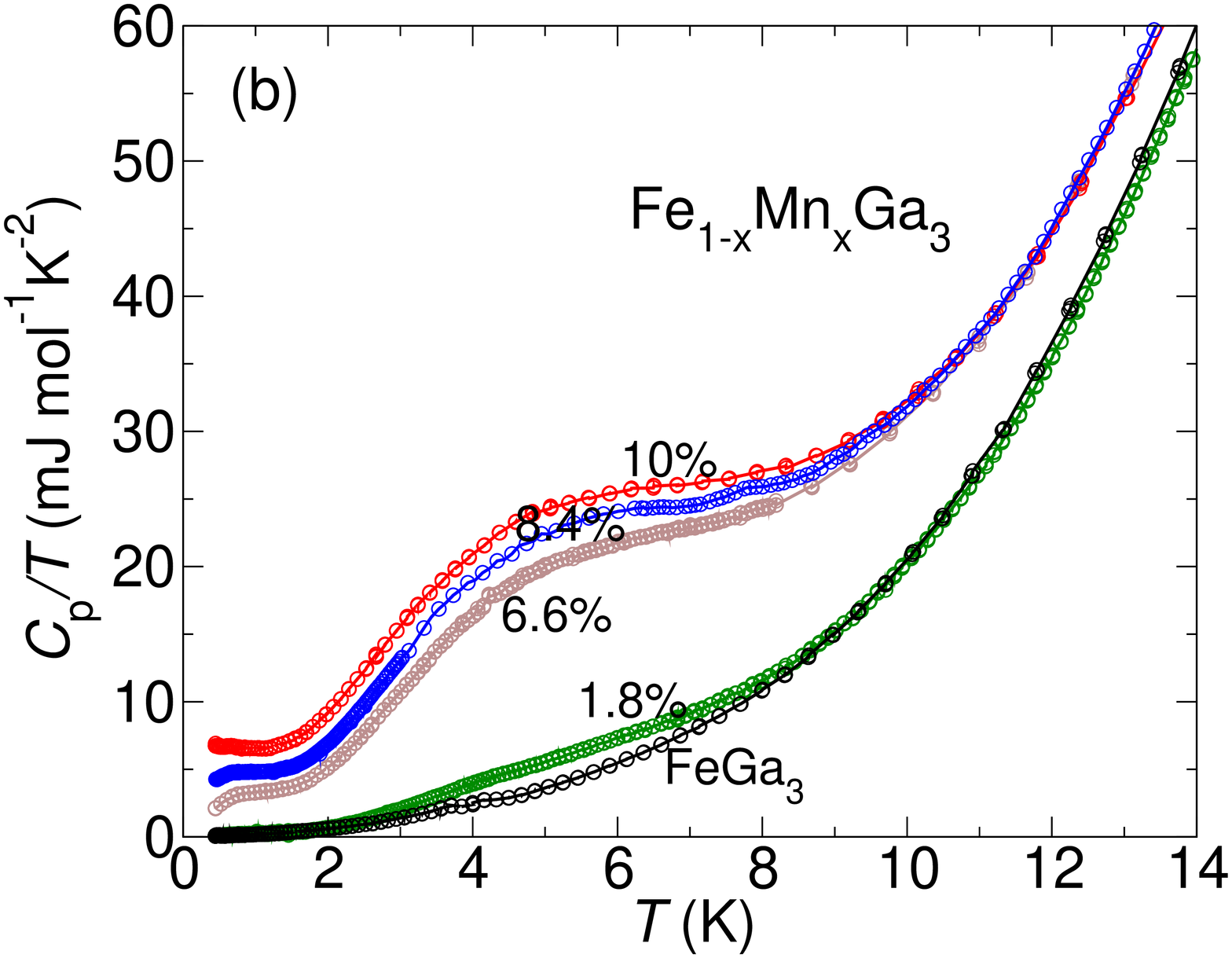,width=0.38\textwidth,angle=-90} \\
\epsfig{file=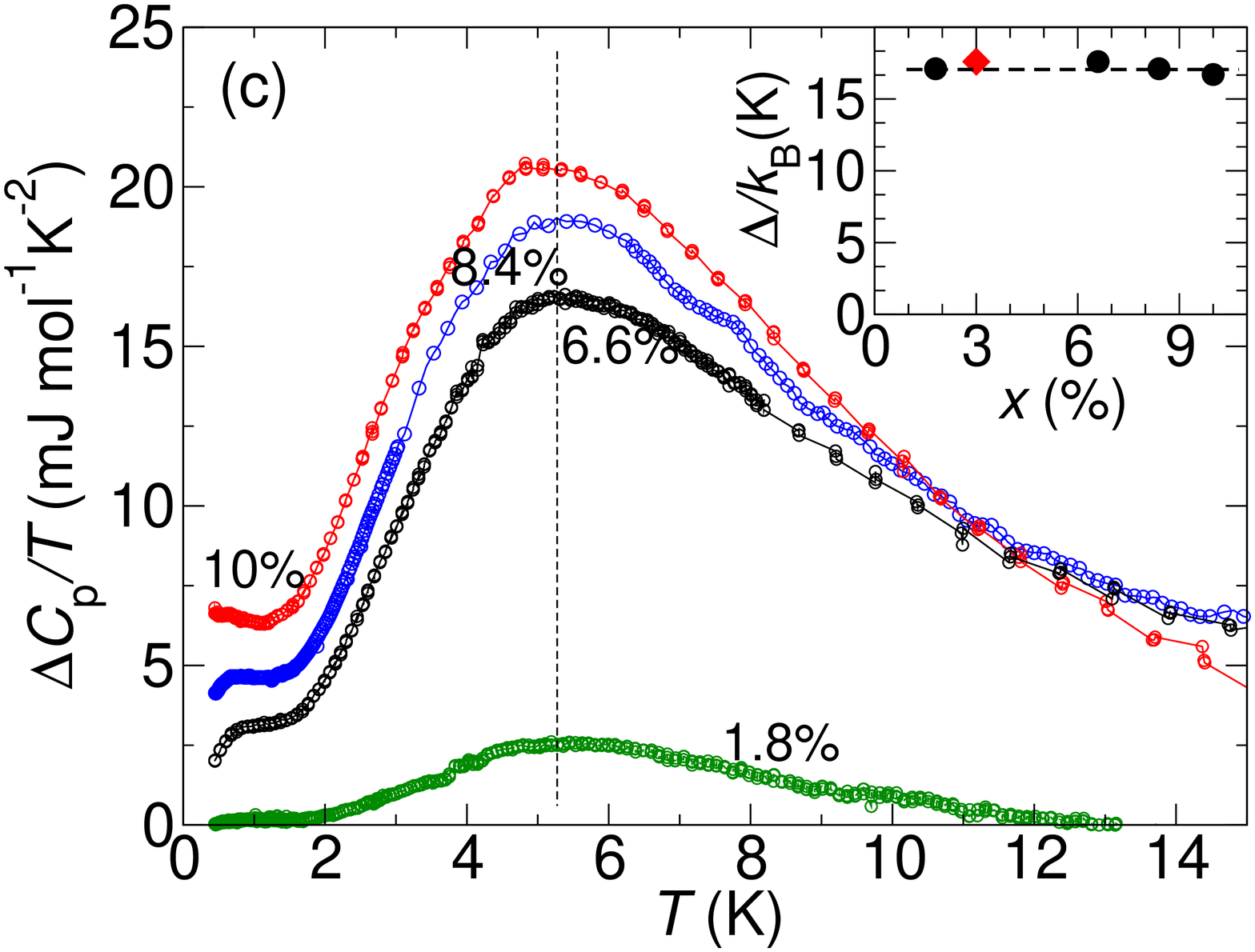,width=0.38\textwidth,angle=-90,clip=} &
\epsfig{file=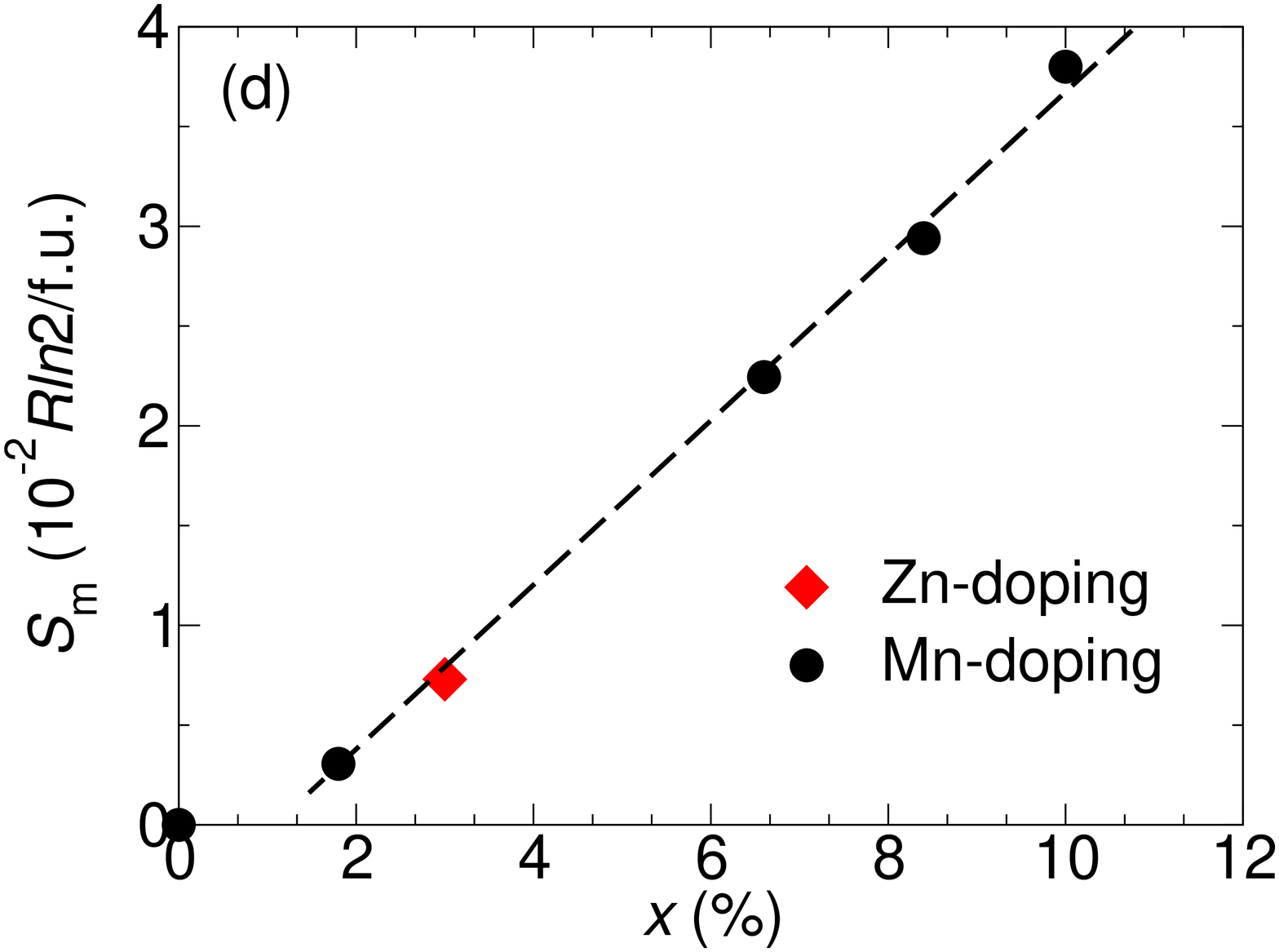,width=0.38\textwidth,angle=-90} \\
\end{tabular}
\caption{(Color online)  (a) Specific heat $C_p$/$T$ of Fe$_{0.92}$Mn$_{0.08}$Ga$_3$ (black circles) together with the data of FeGa$_3$ (red squares). The inset displays the low temperature magnetic specific heat, with a fit of data based on the two--level Schottky model as described in the text. (b) Low temperature $C_p$/$T$($T$) for the series of Mn--doped FeGa$_3$ together with the data for undoped FeGa$_3$ (black empty circles). (c) $\Delta C_p$/$T$($T$) for the series of Fe$_{1-x}$Mn$_x$Ga$_3$ ($x=0$; 1.8\%; 6.6\%; 8.4\%; 10\%) estimated as described in the text. The thin dashed vertical line indicates the position of the maxima in $\Delta C_p$/$T$($T$). The inset shows the values of the energy gap $\Delta$ derived from the fits based on the two--level Schottky model. (d) Values of the magnetic entropy at a temperature of 14~K for various doping levels of Mn (black circles) and Zn (red square). The dashed line corresponds to the entropy of $0.4Rln2$ per dopant. The dashed lines are guides for the eye.}
  \label{fig:Fig8} 
\end{figure*}

Fig.~\ref{fig:Fig8}(a) compares $C_p$/$T$ as a function of temperature for Fe$_{0.92}$Mn$_{0.08}$Ga$_3$ to the data for undoped FeGa$_3$.  
The specific heat does not give any indication for a phase transition in either compound. 
Overall, the specific heat for the doped crystal is very similar to that of FeGa$_3$, indicating that their phonon densities of states are very much alike.  
However, there is an additional contribution to $C_p$, that is present in the doped sample for $T\textless 20$~K.
In order to isolate it, we subtracted from the total specific heat of Fe$_{0.92}$Mn$_{0.08}$Ga$_3$ the corresponding data for undoped FeGa$_3$, taking advantage of their similar lattice specific heats and the negligible electronic contribution for insulating FeGa$_3$.
The resulting \mbox{$\Delta C_p$($T$)$=C_p-C_p$(FeGa$_3$)} is presented in the inset of Fig.~\ref{fig:Fig8}(a) in a standard \mbox{$\Delta C_p$/$T$($T$)} representation. It shows a broad feature with a maximum at about 5~K that resembles a two--level Schottky anomaly. A reasonable description of the experimental data is achieved assuming that the anomaly in \mbox{$\Delta C_p$/$T$} originates from thermally activated transitions between two discrete levels having the same degeneracy, separated by an energy gap \mbox{$\Delta/k_{\mathrm{B}} \approx 16$~K} (see the inset of Fig.~\ref{fig:Fig8}(a)). An additional small constant term of \mbox{2 mJ mol$^{-1}$K$^{-2}$} has also been included to improve the quality of the fit. Although this small contribution to the $C_p$ superimposed on the Schottky anomaly mimics a Fermi Liquid--like specific heat, the insulating temperature dependence of the resistivity overrules the presence of a finite density od states at the Fermi level in Fe$_{0.92}$Mn$_{0.08}$Ga$_3$. Therefore, we suppose that this additional specific heat overlapping with the gap anomaly has a more complex temperature dependence and originates from low--energy magnetic excitations and/or localized in--gap states.

Similar Schottky--like anomalies in the specific heat are visible for all the compounds in the Fe$_{1-x}$Mn$_x$Ga$_3$ series (see Fig.~\ref{fig:Fig8}(b)) as well as for FeGa$_{3-y}$Zn$_{y}$ (not shown). 
They occur at the same temperature for both dopants, regardless of the doping level $x$. 
In order to estimate the entropy associated with this excess specific heat in the doped crystals, we calculated \mbox{$\Delta C_p$($T$)} for each crystal by subtracting from the total $C_p$($T$) the lattice contribution approximated by the specific heat of FeGa$_3$. We extrapolated \mbox{$\Delta C_p$/$T$} to $T=0$ and then integrated \mbox{$\Delta C_p$($T$)/$T$} up to \mbox{$T^{*}=15$~K,} at which temperature the specific heat of the doped crystals becomes very similar to that of FeGa$_3$:
\begin{equation}
S_{\mathrm{m}} = \varint_{0}^{T^{*}} \frac{\Delta C_p}{T} dT.
\label{eq:5}
\end{equation}  
The values of the entropy $S_{\mathrm{m}}$ recovered at 15~K increase nearly linearly with increasing doping level, starting from \mbox{$x\approx 1.5$\%,} and follow the same straight line for both Mn-- and Zn--doped FeGa$_3$, as indicated in the inset of Fig.~\ref{fig:Fig8}(c). The slope of this line corresponds to an entropy of about 0.4$Rln2$ per Mn/Zn atom introduced by the doping. 

Although this entropy is strongly dominated by transitions between two states, it contains also additional contributions originating presumably from low--energy magnetic excitations and localized in--gap states in the vicinity of the Fermi level.  
The insulating behavior of the resistivity with a VRH conduction at low temperatures in Mn-- and Zn--doped FeGa$_3$ excludes the presence of delocalized holes in the whole investigated doping ranges.
The broad features in \mbox{$\Delta C_p/T$} with maxima at the temperature of 0.7~K grow with doping in both FeGa$_{3-y}$Zn$_y$ and Fe$_{1-x}$Mn$_x$Ga$_3$, which may be suggestive of an approaching quantum critical point.

\begin{figure*}
\begin{tabular}{cc}
\epsfig{file=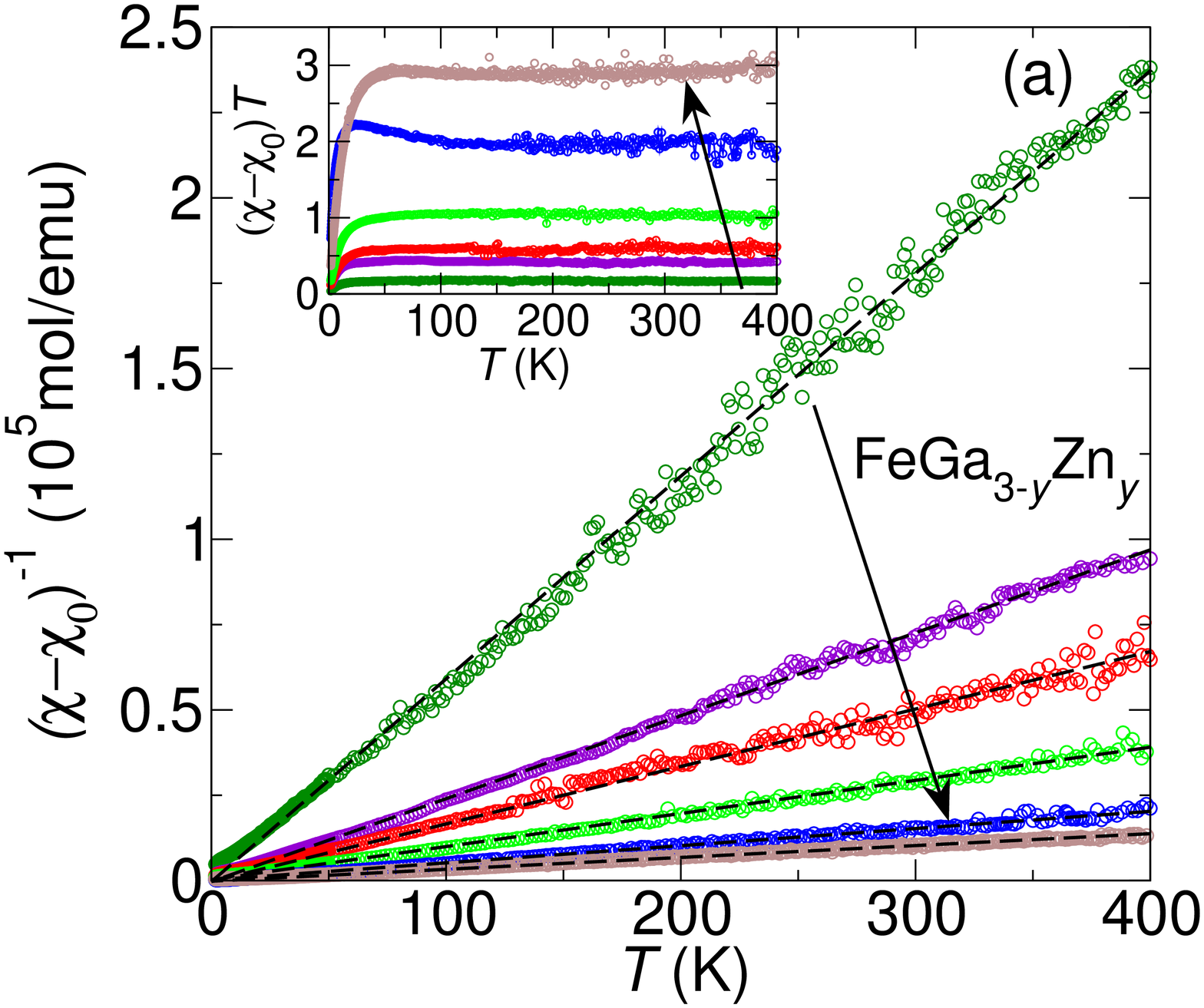,width=0.385\textwidth,angle=-90,clip=} &
\epsfig{file=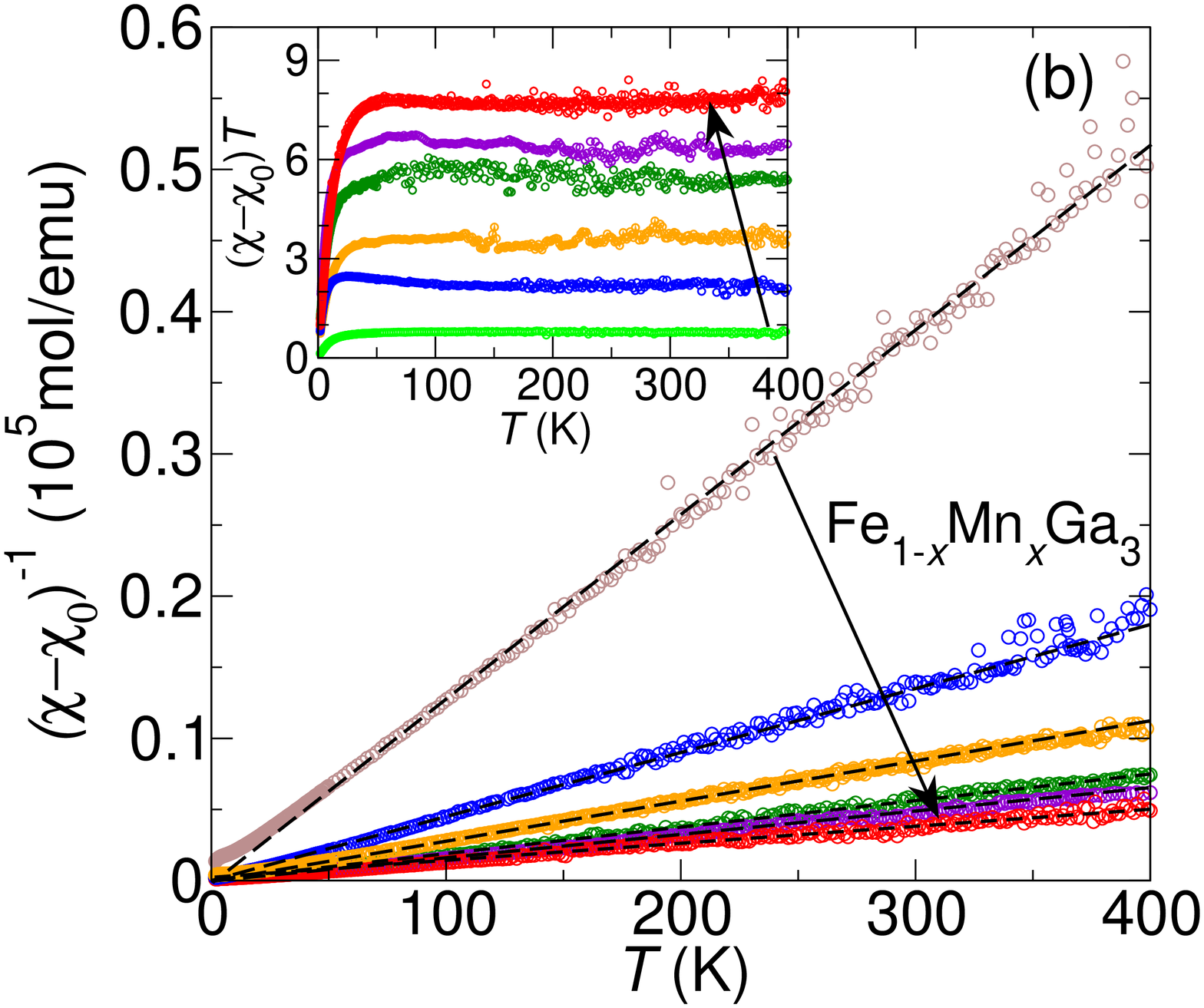,width=0.385\textwidth,angle=-90} \\
\epsfig{file=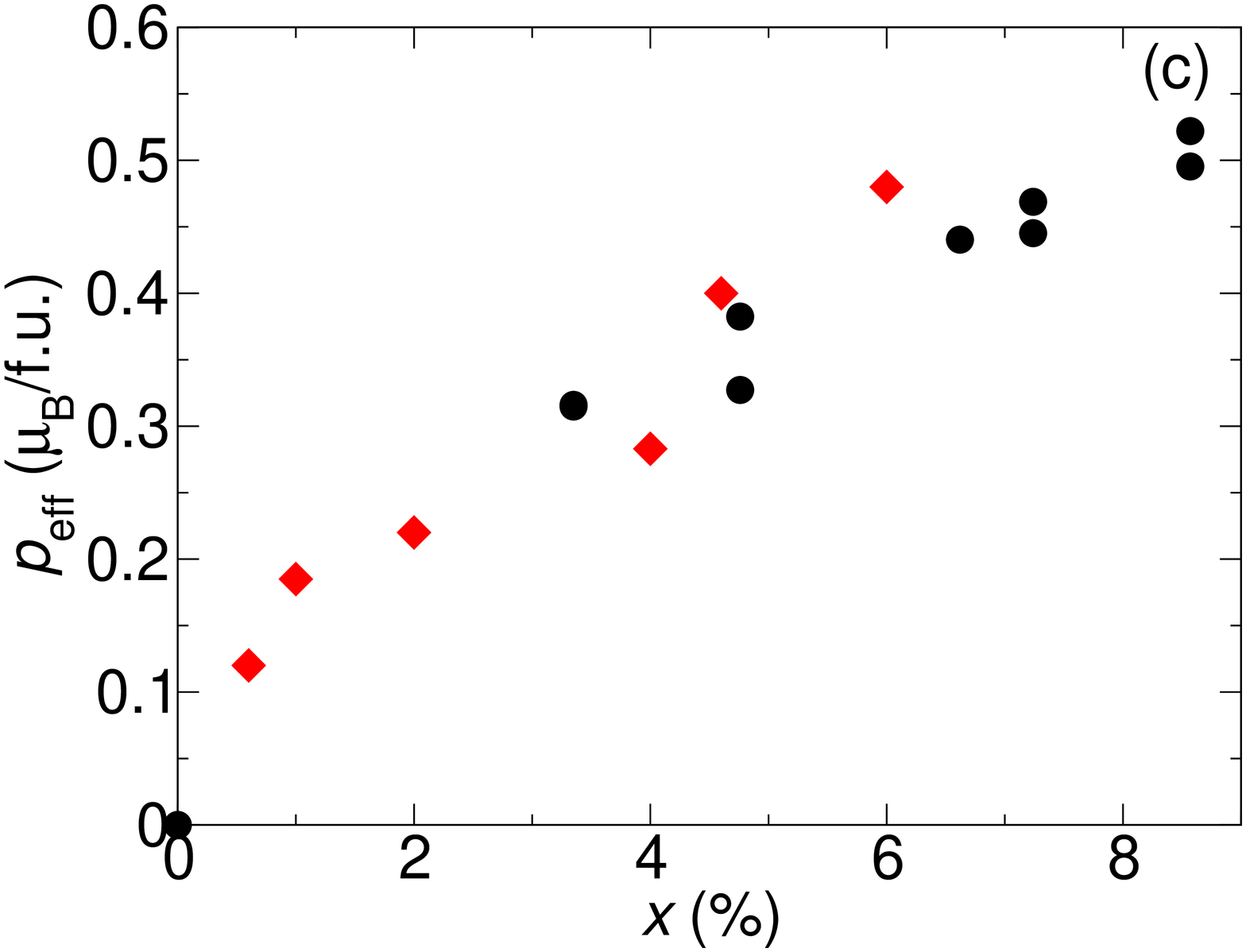,width=0.35\textwidth,angle=-90,clip=} &
\epsfig{file=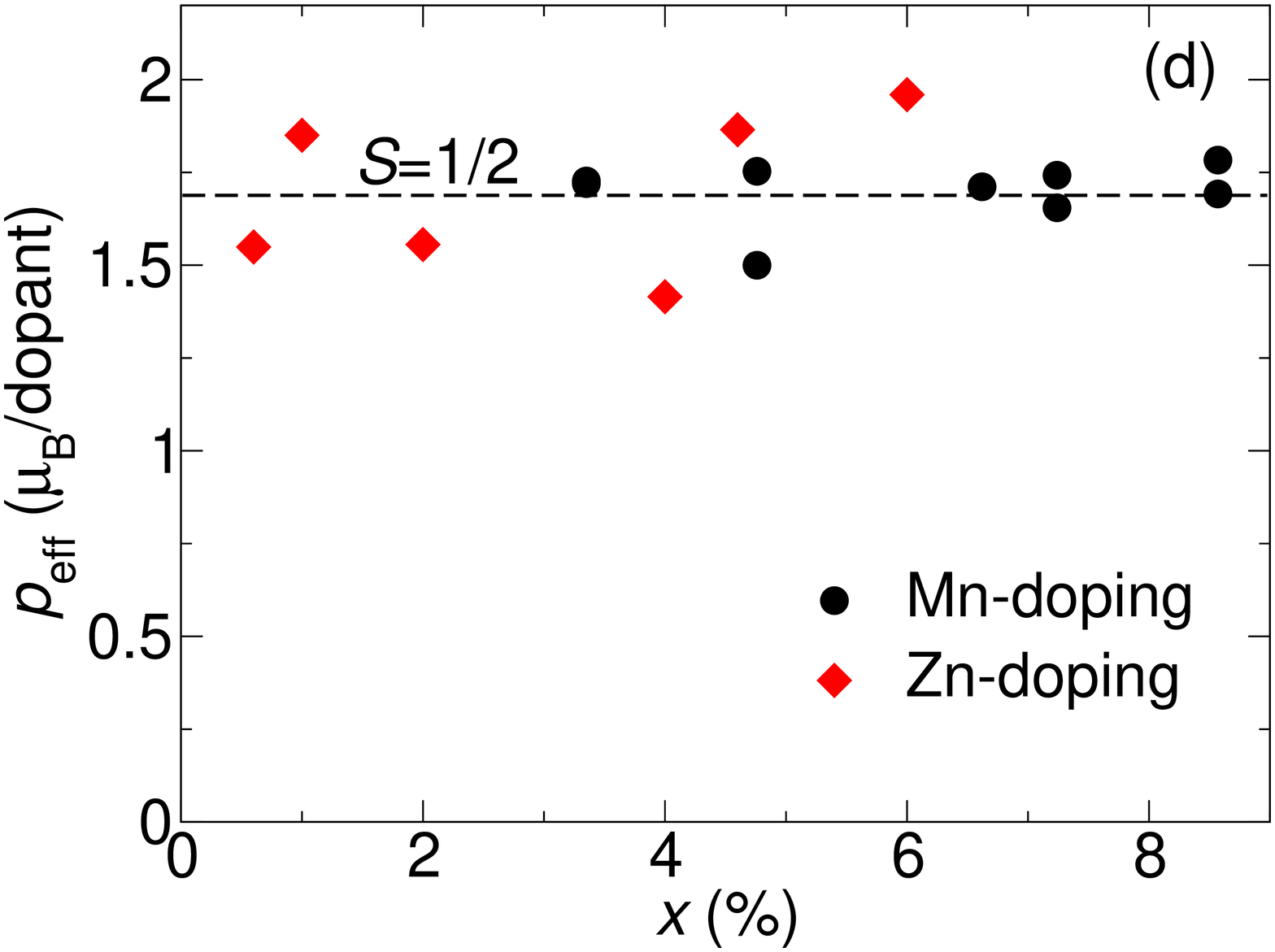,width=0.35\textwidth,angle=-90} \\
\end{tabular}
\caption{(Color online) Inverse magnetic susceptibilities for FeGa$_{3-y}$Zn$_y$ ($y=0.006$; 0.01; 0.02; 0.04; 0.046; 0.06) (a) and Fe$_{1-x}$Mn$_x$Ga$_3$ ($x=0.032$; 0.043; 0.044; 0.061; 0.065; 0.066; 0.085; 0.087) (b) after subtracting a constant term $\chi_0$ as obtained from fits using a modified Curie law (see details in text). The insets in panels (a,b) show the values of the product $\chi T$ as functions of temperature for the field of $H=70$~kOe obtained after subtracting the temperature--independent terms $\chi_0$. Arrows indicate the direction of the increase in measured doping level. Panel (c) depicts the values of the effective magnetic moment per formula unit obtained from the fits as a function of the doping level $x$ for FeGa$_{3-y}$Zn$_y$ (red diamonds) and Fe$_{1-x}$Mn$_x$Ga$_3$ (black circles). The values of the effective moment recalculated per dopant are shown in panel (d), together with a dashed line indicating an effective moment of 1.73~$\mu_{\mathrm{B}}$, as expected for $S=1/2$.}
  \label{fig:Fig9} 
\end{figure*}

Doping of FeGa$_3$ with Mn and Zn has also very similar effects on the overall magnetic properties. For both dopants, the magnetic susceptibility in the temperature range of \mbox{$\sim$100--400~K} can be described by a modified Curie law: 
\begin{equation}
\chi(T)=\chi_0 + \frac{C}{T},
\label{eq:4}
\end{equation}  
where $\chi_0$ denotes the temperature independent part of the magnetic susceptibility and $C$ is the Curie constant defined as \mbox{$C=N_{\mathrm{m}}p_{\mathrm{eff}}^2$/($3k_{\mathrm{B}}$)} with $N_{\mathrm{m}}$ being the number of magnetic atoms per formula unit and $p_{\mathrm{eff}}$ denoting their effective magnetic moment. The least--squares fits to the experimental data (see Fig.~\ref{fig:Fig9}(a,b)) yield values of $\chi_0$ ranging from \mbox{$-5\times 10^{-5}$emu/mol} to \mbox{$-2\times 10^{-5}$emu/mol,} which are similar to the diamagnetic signals observed for crystals of undoped FeGa$_3$. 
Values of the effective magnetic moment per formula unit $p_{\mathrm{eff}}$ derived from the Curie constant are plotted as a function of the measured doping level in Fig.~\ref{fig:Fig9}(c). 

For both Fe$_{1-x}$Mn$_x$Ga$_3$ and FeGa$_{3-y}$Zn$_y$, the effective moment increases gradually with increasing doping level. 
The effective moment calculated per dopant is close to the theoretical value of \mbox{1.73 $\mu_{\mathrm{B}}$/dopant} expected for spin $S=1/2$ (see Fig.~\ref{fig:Fig9}(d)). 
Inclusion of a paramagnetic Weiss temperature $\theta$ did not lead to any significant improvement of the fits. Thus, the magnetic susceptibility measurements indicate that doping with both Zn and Mn introduces moments of $S=1/2$ per dopant atom, which are freely fluctuating at temperatures above $\sim$100~K. 
The magnetic anisotropy in this temperature range is reflected only by smaller absolute values of $\chi_0$ for $H$ applied in the $ab$ plane than in case of \mbox{$H\parallel c$,} imitating the magnetic anisotropy found for the undoped FeGa$_3$.   

At lower temperatures, the magnetic susceptibilities of the doped crystals deviate from the Curie--Weiss law in a manner that may indicate an increasing role for antiferromagnetic interactions. 
To follow the variation of the effective moments with temperature, we plot the temperature dependence of the product \mbox{($\chi -\chi_0$)$T$} measured in a magnetic field of 70~kOe (insets, Fig.~\ref{fig:Fig9}(a,b)). For Fe$_{1-x}$Mn$_x$Ga$_3$ and FeGa$_{3-y}$Zn$_y$, at temperatures above $\sim$80~K the product \mbox{($\chi -\chi_0$)$T$} is nearly temperature independent, defining the Curie law regime. At lower temperatures, \mbox{$(\chi -\chi_0)T$} decreases strongly with decreasing temperature, implying a decrease in the effective fluctuating moment.

Below $\sim$30~K, the magnetic susceptibilities of both Fe$_{1-x}$Mn$_x$Ga$_3$ and FeGa$_{3-y}$Zn$_y$ develop substantial magnetic anisotropies and become field dependent in magnetic fields $H$ applied along the $c$--axis (see Fig.~S2 in the Supplemental Material).  Fig.~\ref{fig:Fig10}(a) presents the low temperature magnetic susceptibility data for Fe$_{0.92}$Mn$_{0.08}$Ga$_3$.  
The $\chi$($T$) measured in weak magnetic fields applied along the $c$--axis continues to increase with decreasing temperature. Its temperature dependence at \mbox{$T\apprle 5$~K} can be approximated by \mbox{$\chi(T)\sim T^{-\alpha}$} with \mbox{$\alpha\approx 0.5$,} as indicated in the inset of Fig.~\ref{fig:Fig10}(a).  Application of \mbox{$H\apprge 1$~kOe} leads to a gradual suppression of the upturn in $\chi$($T$) at temperatures below $\sim$10~K. 
In contrast, the magnetic susceptibility for \mbox{$H\parallel$[110]} is almost field independent from 1~kOe to 70~kOe even at lowest temperatures. It shows a broad and very weak maximum at \mbox{$T\approx 5$~K.}

To check for the presence of a spontaneous magnetization, we measured the magnetization as a function of magnetic field at constant temperatures (Fig.~\ref{fig:Fig10}(b)). 
In the inset of Fig.~\ref{fig:Fig10}(b), we re--plotted the isothermal $M$($H$) curves obtained in fields applied along the $c$--axis as $M^2$ against $H$/$M$.
The absence of a positive $M^2$ intercept in this Arrott plot indicates that there is no spontaneous magnetization in Fe$_{0.92}$Mn$_{0.08}$Ga$_3$.
Importantly, the $M$($H$) curves at $T=1.8$~K do not saturate in magnetic fields up to 70~kOe, as expected for a system with noninteracting moments. They increase nearly linearly with increasing field strength in the high field range, independent of the field direction, and the maximum measured moment for Fe$_{0.92}$Mn$_{0.08}$Ga$_3$ in the field of 70~kOe is only \mbox{$\sim$0.028 $\mu_{\mathrm{B}}$/f.u.} and \mbox{$\sim$0.023 $\mu_{\mathrm{B}}$/f.u.} for \mbox{$H\parallel$[001]} and \mbox{$H\parallel$[110]}, respectively. Implications of these field dependencies of the magnetization will be discussed in Section \ref{Discussion}.

Similar results were obtained for crystals of Fe$_{1-x}$Mn$_x$Ga$_3$ and FeGa$_{3-y}$Zn$_y$ regardless of the dopant type and level, with nonsaturating $M$($H$) at $T=1.8$~K, $\chi$($T$) showing broad maxima at \mbox{$T\lesssim 5$~K} for $H\|$[110] and \mbox{$\chi$($T$)$\sim T^{-\alpha}$} at \mbox{$T\apprle 5$~K} in case of an applied field along the $c$--axis with $\alpha$ between 0.2 and 0.6. 
Although the exact values of the exponent $\alpha$ may be affected by slight crystal misalignment and the presence of a small amount of paramagnetic impurities, they are always below 1. 
Such weak power laws in $\chi$($T$) at \mbox{$T\apprle 5$~K} may hint at the emergence of critical fluctuations associated with a proximity to a quantum critical point in the presence of disorder introduced by doping.\cite{VojtaT} 
Further, no hysteresis was observed in field--cooled (FC) and zero--field--cooled (ZFC) $\chi$($T$) data for Fe$_{1-x}$Mn$_x$Ga$_3$ and FeGa$_{3-y}$Zn$_y$, that could indicate a spin glass behavior at low temperatures.  
The values of the magnetic susceptibility, measured at a temperature of 2~K, increase systematically with increasing doping level for both investigated directions of magnetic field, following the same straight lines for crystals of Mn and Zn--doped FeGa$_3$ (Fig.~\ref{fig:Fig11}).
These results need to be contrasted with previous studies on polycrystals of Fe$_{1-x}$Mn$_x$Ga$_3$ which revealed that already in samples with $x=0.05$ there are either two magnetic phase transitions at temperatures of $\sim$10~K and $\sim$30~K according to Ref. \onlinecite{NeelPhD} or there is an onset of ferromagnetism with $T_{\mathrm{C}}\approx 160$~K superimposed on a Curie--Weiss--type paramagnetic behavior with the effective magnetic moment of \mbox{4.9 $\mu_{\mathrm{B}}$/Mn}.\cite{Kotur}

\begin{figure*}[h] 
\centering
\includegraphics[width=0.38\textwidth,angle=-90]{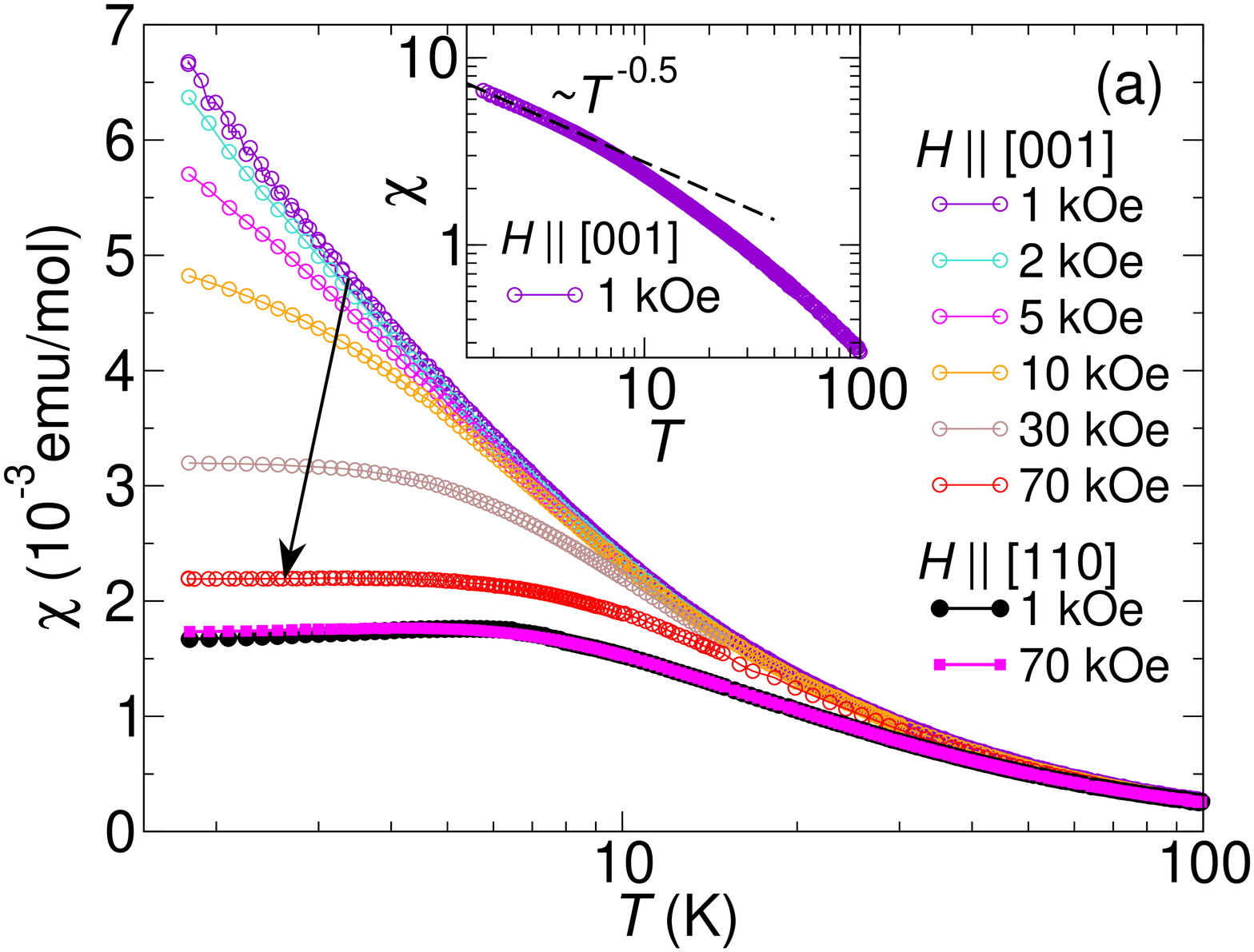}
\includegraphics[width=0.38\textwidth,angle=-90]{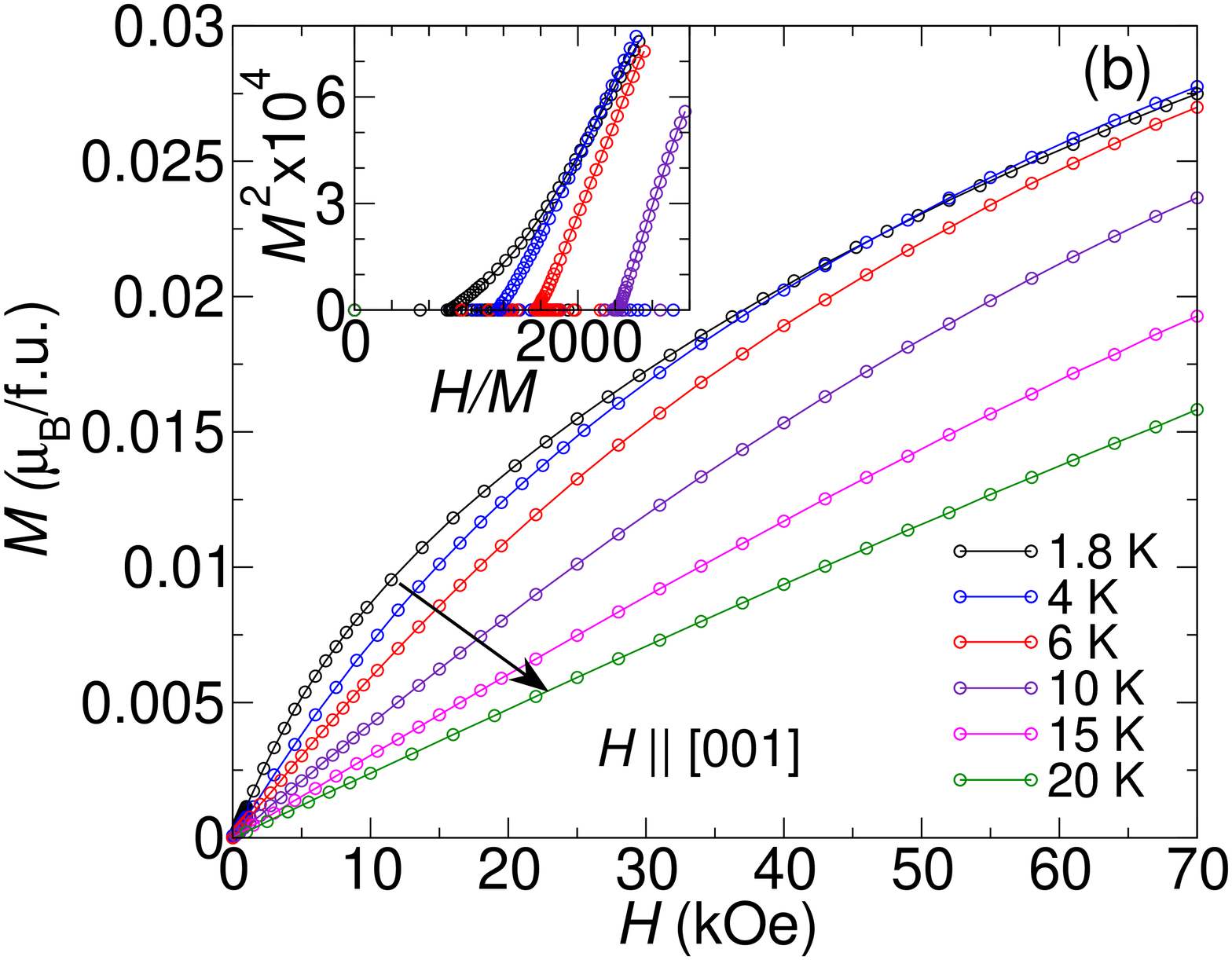}
\caption{\label{fig:Fig10} (Color online) (a) Magnetic susceptibility of Fe$_{0.92}$Mn$_{0.08}$Ga$_3$ measured in various magnetic fields applied either along the $c$--axis (empty circles) or along the [110] direction (full squares). Arrows point in the direction of the increasing magnetic field applied along the $c$--axis. For $H\|$[110], $\chi$($T$) is only weakly field--dependent. The inset shows  $\chi$($T$) measured in $H=1$~kOe applied along the $c$--axis in a log--log plot. The dashed line represents the fit to the data at \mbox{$T\apprle 5$~K} using \mbox{$\chi$($T$)$\sim T^{-\alpha}$.} (b) Magnetization measured for Fe$_{0.92}$Mn$_{0.08}$Ga$_3$ in a magnetic field applied along the $c$--axis at indicated temperatures. The inset shows the $M$($H$) curves plotted in a standard Arrott representation.  
}
\end{figure*} 

\begin{figure}[h] 
\centering
\includegraphics[width=0.38\textwidth,angle=-90]{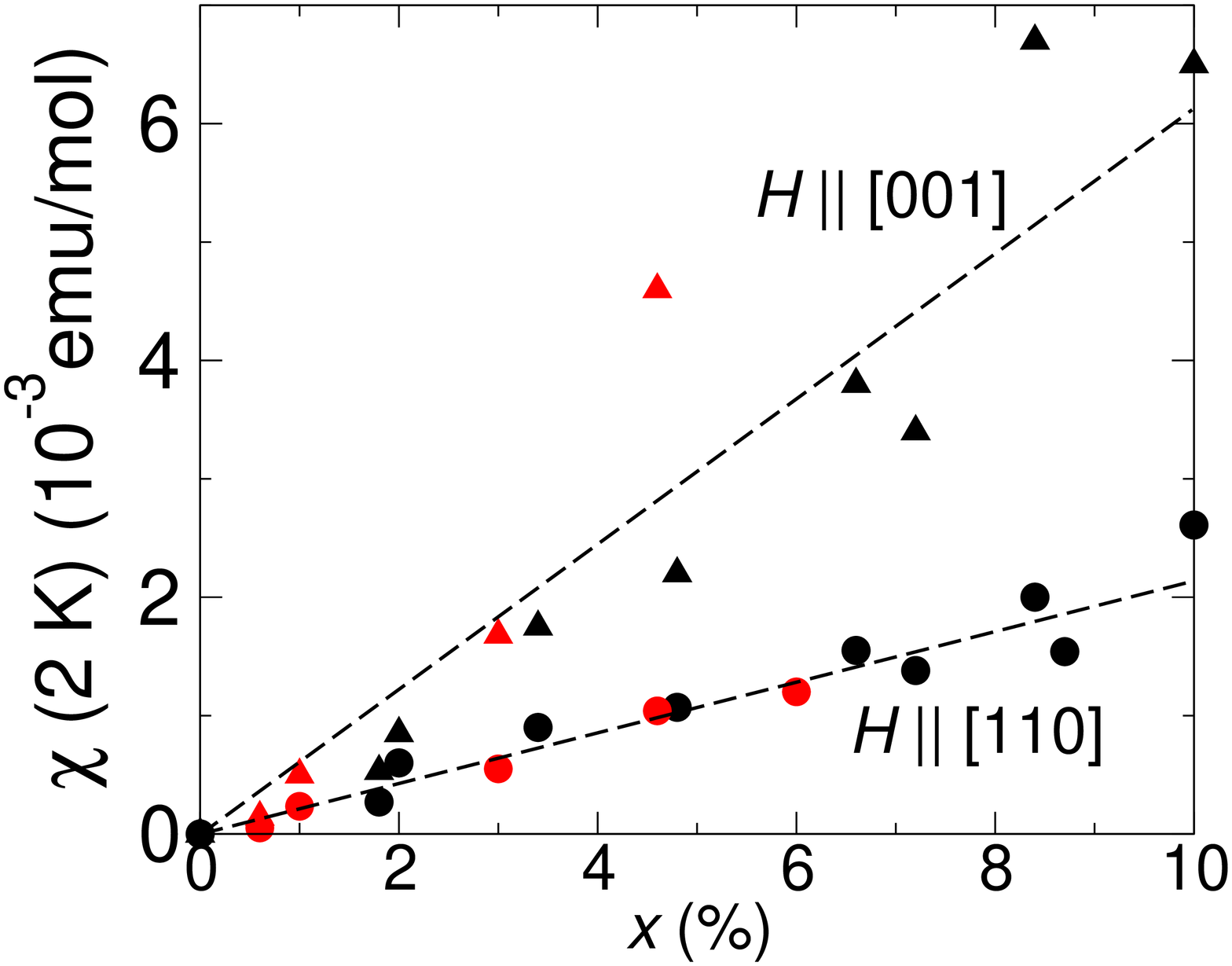}
\caption{\label{fig:Fig11} (Color online) Values of the magnetic susceptibility for Fe$_{1-x}$Mn$_x$Ga$_3$ (black) and FeGa$_{3-y}$Zn$_y$ (red) measured at 2~K in a magnetic field 
applied either along the [110] direction (circles) or along the $c$--axis for $H$=1~kOe (triangles). Dashed lines are guides for the eye.
}
\end{figure}

\begin{figure}[h] 
\centering
\includegraphics[width=0.38\textwidth,angle=-90]{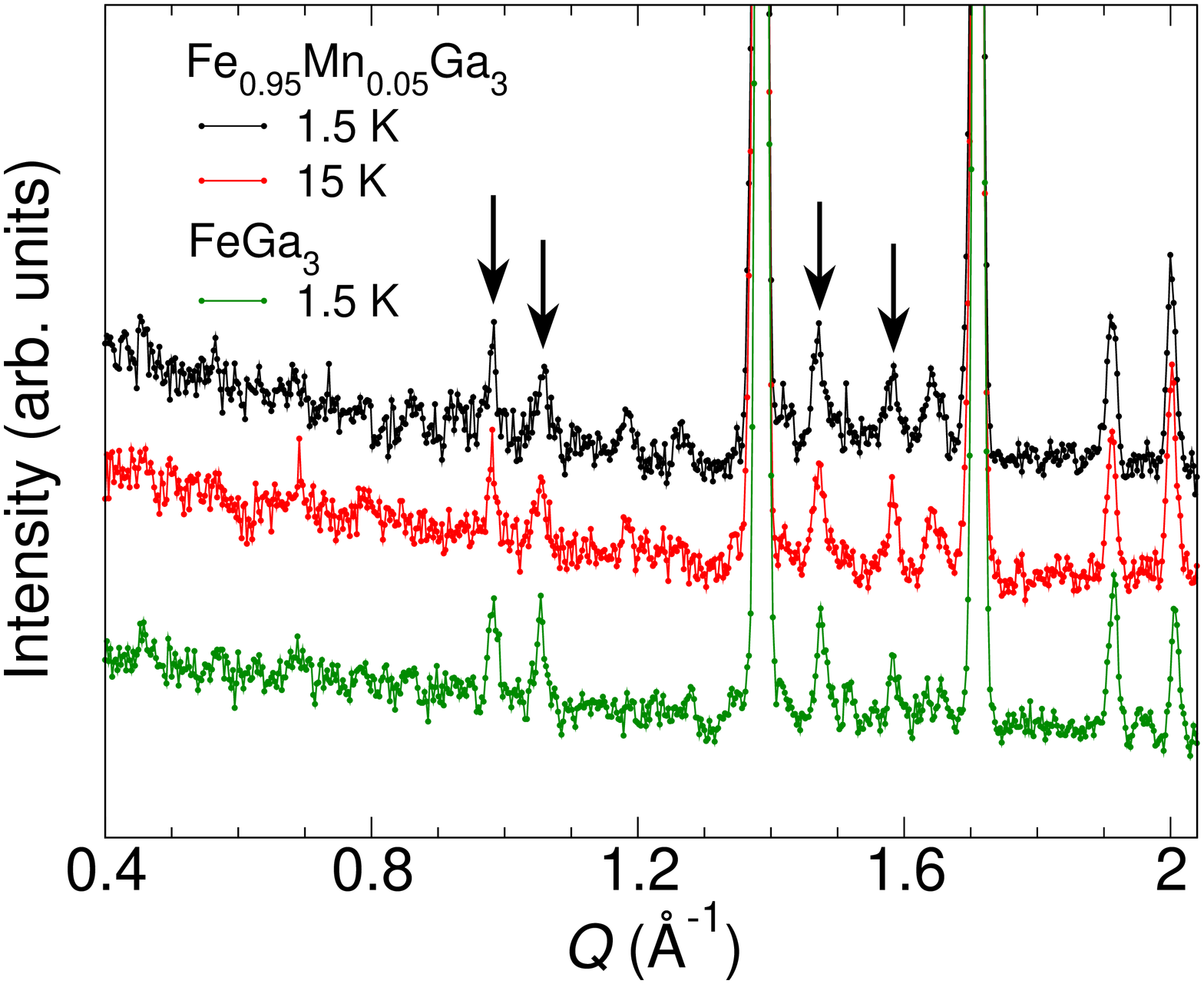}
\caption{\label{fig:Fig12} (Color online) Neutron powder diffraction patterns for Fe$_{0.95}$Mn$_{0.05}$Ga$_3$, obtained at temperatures of 1.5~K (black) and 15~K (red), together with the data for undoped FeGa$_3$ collected at 1.5~K.  The main magnetic peaks are indicated by arrows.
}
\end{figure}

Finally, neutron powder diffraction experiments have been performed to inspect the magnetic and structural properties of hole--doped FeGa$_3$ at low temperatures. The diffraction patterns for Fe$_{0.95}$Mn$_{0.05}$Ga$_3$ measured at temperatures of 1.5~K and 15~K are basically identical (Fig.~\ref{fig:Fig12}). There are no differences in the sizes and positions of either magnetic or nuclear peaks, that could indicate a change in the long--range magnetic order or a structural transition in the doped FeGa$_3$ at low temperatures. 
Furthermore, the magnetic peaks observed in the neutron powder diffraction patterns for Fe$_{0.95}$Mn$_{0.05}$Ga$_3$ persist at temperatures up to 300~K and, most importantly, are similar to those found for undoped FeGa$_3$ (Fig.~\ref{fig:Fig12}), indicating that their magnetic structures are closely related.

\begin{figure}[h] 
\centering
\includegraphics[width=0.38\textwidth,angle=-90]{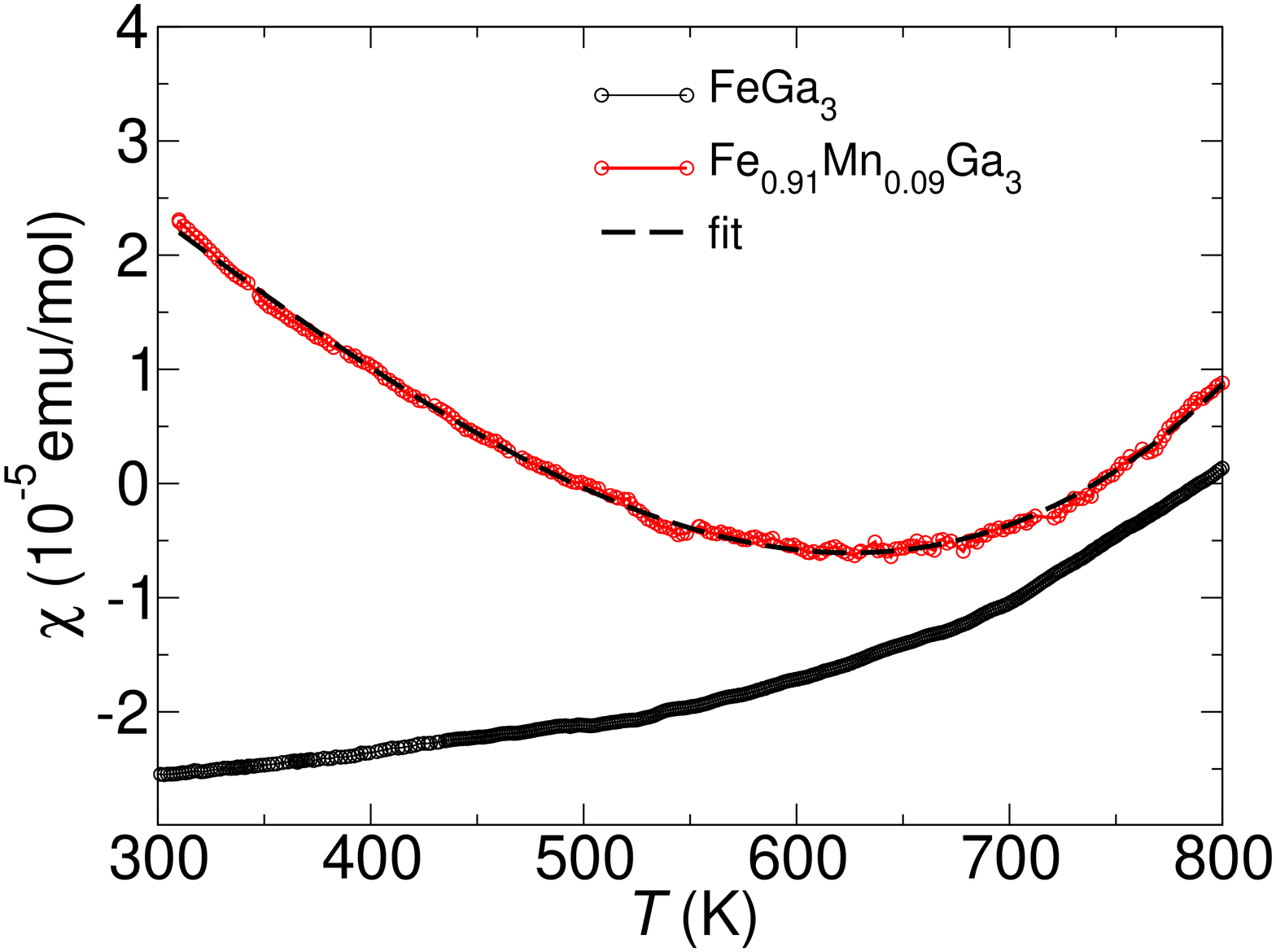}
\caption{\label{fig:Fig13} (Color online) Magnetic susceptibility of Fe$_{0.91}$Mn$_{0.09}$Ga$_3$ (red circles) measured at temperatures above $\sim$300~K in a magnetic field of 10~kOe on a set of randomly oriented crystals, together with the data for FeGa$_3$ (black circles) calculated as explained in the text. Dashed line represents the fit to a model described in the text.
}
\end{figure}

A magnetization study carried out at high temperatures shows that for hole--doped FeGa$_3$, the fluctuating moment behavior observed at $T\apprge$100~K gives way to a notable increase in $\chi$($T$) at temperatures above $\sim$600~K.
Fig.~\ref{fig:Fig13} presents the temperature dependence of the magnetic susceptibility for Fe$_{0.91}$Mn$_{0.09}$Ga$_3$, measured on a set of randomly oriented single crystals in a magnetic field of 10~kOe, together with the inferred polycrystalline magnetic susceptibility for undoped FeGa$_3$ estimated as \mbox{$\chi(T)=\frac{1}{3}(2\chi_{ab}(T)+\chi_c(T$)),} where $\chi_{ab}$($T$) and $\chi_c$($T$) denote the magnetic susceptibilities measured in magnetic fields applied in $ab$ plane and along the $c$--axis, respectively.
The $\chi$($T$) for Fe$_{0.91}$Mn$_{0.09}$Ga$_3$ can be well described by an activation--type behavior, extended by an additional term $C/T$ that is introduced to account for the presence of fluctuating moments (Fig.~\ref{fig:Fig13}).  
The value of the Curie constant $C$ derived from this fit is very similar to that obtained from the analysis of the magnetic susceptibility at temperatures between 100~K and 300~K, and \mbox{$\chi_0= -5.4\times 10^{-5}$ emu mol$^{-1}$.} Most importantly, the obtained spin gap $\Delta_{\mathrm{S}}$=0.31~eV is only slightly smaller than that estimated for the parent compound FeGa$_3$, implying that the doping with Mn in Fe$_{1-x}$Mn$_x$Ga$_3$ falls short of closing the spin gap.

\section{Discussion}\label{Discussion}

There is an experimental consensus that FeGa$_3$ is an insulator with a narrow gap of 0.4~eV.\cite{Arita, Hadano, Hauserman} 
The band gap can be well reproduced in first--principles electronic structure calculations based on DFT in its standard LDA and GGA implementations.\cite{Imai, Picket, Singh} These calculations, however, result in a nonmagnetic state, whereas our neutron diffraction measurements revealed a complex type of magnetic order in FeGa$_3$, even at room temperature.
The existence of staggered moments ultimately questions the LDA description of FeGa$_3$ as a nonmagnetic band insulator and calls for more sophisticated computational techniques. 

Both the LDA calculations and the ARPES study show a narrow Fe 3$d$--derived state located near the top of the valence band,\cite{Arita, Hauserman, Picket} indicating the relevance of electronic correlation effects. 
Following this observation, Z. P. Yin and W. E. Pickett\cite{Picket} applied a static mean--field like LSDA+U method and found a magnetic insulating ground state for FeGa$_3$, independent of which double counting scheme was used, with an antiferromagnetic order being lowest in energy.\cite{Picket} Further, the size of the band gap obtained assuming moderate values of the on--site Coulomb interaction for the Fe 3$d$ states, \mbox{$U\sim 2$~eV,} coincides with experimental results.\cite{Picket} However, a non--magnetic state is stabilized for \mbox{$U\apprle 1.5$ eV} and even for larger values of $U$ if screening effects are included in the LSDA+U formalism via a Yukawa ansatz.\cite{pressure} 
These findings indicate that the 3$d$ electrons in FeGa$_3$ are not close to the well understood limits of being either localized or itinerant. Consequently, neither LDA nor L(S)DA+U methods are able to provide a satisfying description of its electronic band structure. Therefore we employed DMFT to interpolate between the strongly localized and delocalized limits pertinent to our analysis of FeGa$_3$. 
We applied the combination of the DFT in its LDA approximation with DMFT to investigate the charge and spin states of the iron atoms as well as temperature dependent many--body renormalizations in FeGa$_3$. 
In order to separate effects originating from correlated 3$d$ states of iron, we performed calculations also for the isostructural and isoelectronic compound RuGa$_3$.

\begin{figure*}[b] 
\centering
\includegraphics[width=0.32\textwidth,angle=-90]{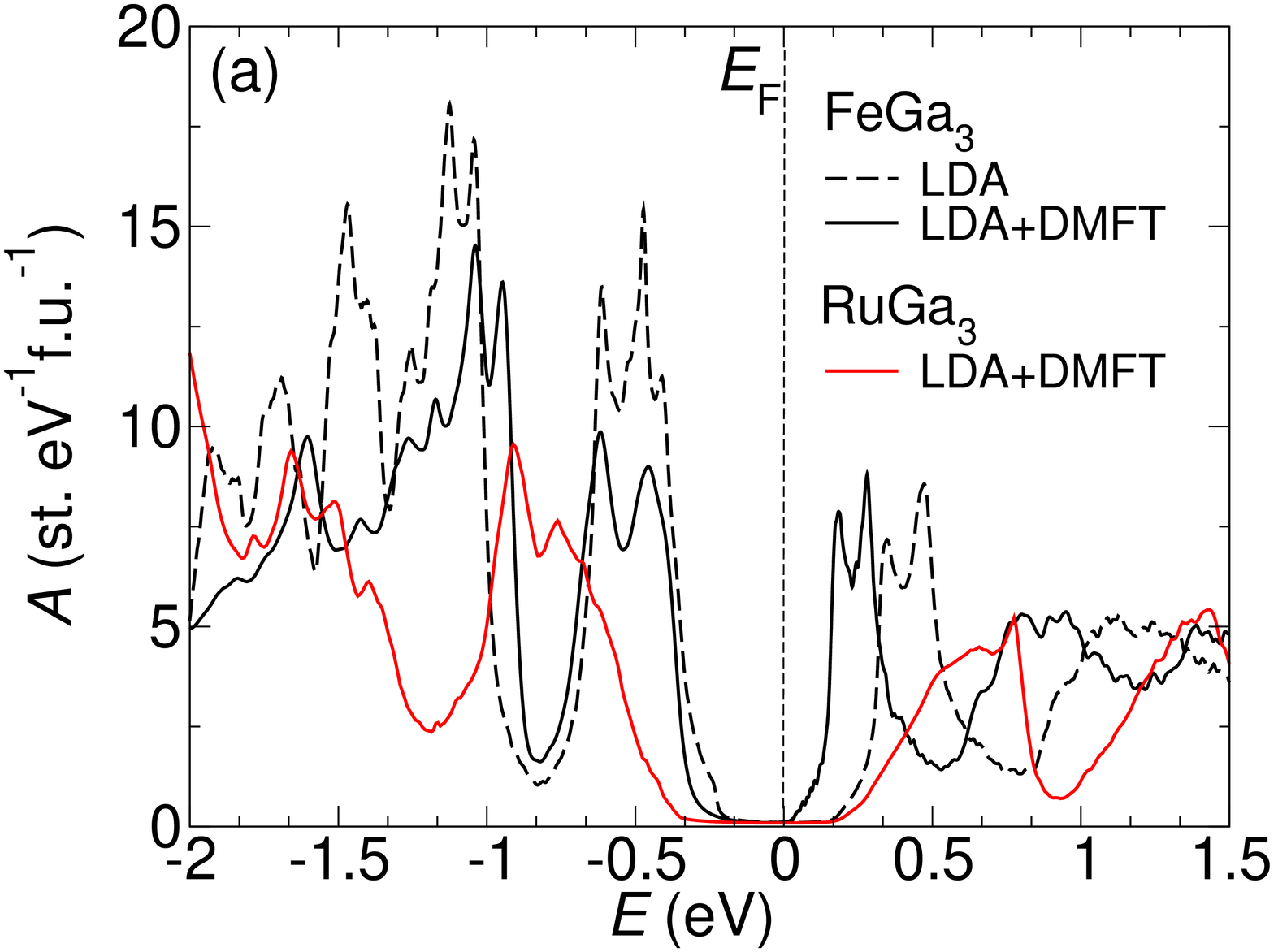}
\includegraphics[width=0.32\textwidth,angle=-90]{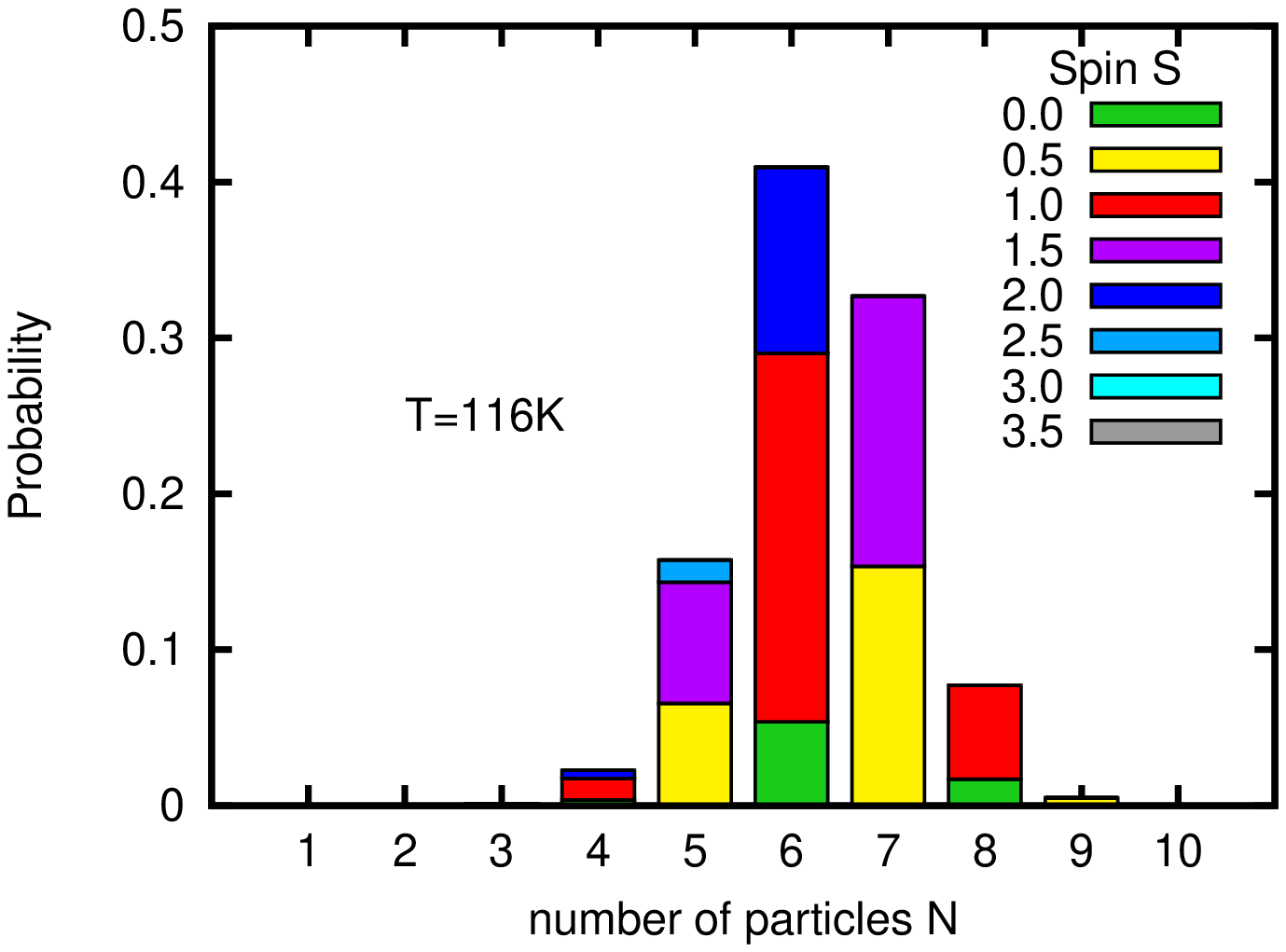}
\caption{\label{fig:Fig14} (Color online) Left panel shows LDA+DMFT spectral functions calculated at $T=116$~K (solid lines) for FeGa$_3$ (black) and for RuGa$_3$ (red) compared to the LDA results (dashed line). Right panel presents probability of atomic states of the DMFT impurity at $T=116$~K, decomposed in number of particles $N$ ($x$-axis) and spin--state $S$, as indicated.}
\end{figure*}

The left panel of Fig.~\ref{fig:Fig14} shows the LDA+DMFT spectral functions for both FeGa$_3$ and RuGa$_3$, together with the band--theory (LDA) results.
Overall, our calculations reproduce insulating states for FeGa$_3$ and RuGa$_3$ with a narrow gap produced by a strong hybridization between $d$ states of Fe or Ru and $p$ states of Ga. The size of the band gap obtained from band--theory for FeGa$_3$ of 0.35~eV is close to results of the combined PES and inverse PES study,\cite{Arita} and also previous works.\cite{Hauserman,Arita,Singh} 
Further, the shape of the DOSs calculated within the LDA approximation for both FeGa$_3$ and RuGa$_3$ is in a very good agreement with antecedent reports.\cite{Hauserman, Arita, Singh, Takagiwa, Picket}

\begin{figure}[h] 
\centering
\includegraphics[width=0.35\textwidth,angle=-90]{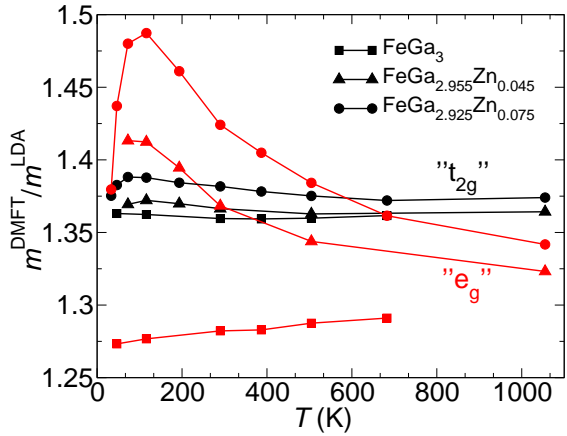}
\caption{\label{fig:Fig15} (Color online) LDA+DMFT effective masses for doped and undoped FeGa$_3$ as obtained from the energy derivative of the self-energy. Orbital components that account for the majority of spectral weight for the valence (conduction) states are denoted ``$t_{\mathrm{2g}}$'' in black (``$e_{\mathrm{g}}$'' in red).} 
\end{figure}

Inclusion of many--body effects via LDA+DMFT leads to a slight decrease in the size of the charge gap in FeGa$_3$, as compared to the LDA results (Fig.~\ref{fig:Fig14}(a)).
To further inspect the strength of dynamical correlations, we calculate the effective masses from the energy--derivative of the DMFT self--energy at the Fermi level.
The formal evaluation based on the first linear regime in the DMFT self--energy shows that the effective masses are rather low and nearly temperature independent, \mbox{$m^{\hbox{\tiny{DMFT}}}/m^{\hbox{\tiny{LDA}}}\approx 1.3$--1.4,} (see Fig.~\ref{fig:Fig15}). 
This indicates a large degree of delocalization of the states at the bottom of the valence band and only weak dynamical correlations.

To address the charge and spin states of Fe in FeGa$_3$, we analyse the probability distribution of the many--body wave--function with respect to the eigenstates of the effective iron atom, decomposed into the number of particles $N$ and the spin state $S$. The histogram of the CTQMC (the right panel of Fig.~\ref{fig:Fig14}) shows the largest probability for the spin state $S=1$, with a variance \mbox{$\delta S=\langle(S-\langle S\rangle)^2\rangle\approx 0.34$} that reflects strong fluctuations at short time--scales. The dominance of the $S=1$ configuration implies that FeGa$_3$ is not a nonmagnetic insulator with Fe$^{2+}$ in a low spin--state, as previously suggested.\cite{Hadano, Tsuji, Imai, Hauserman, Amagai, pressure, Singh, Takabatake, Neel, Bittar}  
Further, the number of Fe~3$d$ electrons \mbox{$N_{\mathrm{d}}\approx 6.3$} has a large variance \mbox{$\delta N=\langle(N-\langle N\rangle)^2\rangle\approx 0.88$}, implying a strong mixed valent character.
Although the variances in $S$ and $N$ are large, they do not evolve with temperature. Thus, the charge and spin states of Fe are temperature independent in FeGa$_3$. 

The experimental magnetic susceptibility of FeGa$_3$ is only weakly temperature dependent below $\sim$500~K and increases strongly at higher temperatures, indicative of a gap in the spin excitation spectrum (Fig.~\ref{fig:Fig3}). Early attempts to explain this behaviour were based on thermal excitations of electrons across the band gap.\cite{Tsuji} 
These models, however, require unreasonably large values of the density of states near the band gap edges to account for the size of the experimental $\chi$($T$).\cite{Tsuji} 
This, together with the absence of similar spin excitations in the magnetic susceptibility of RuGa$_3$ - an insulator with a slightly larger band gap of $\sim$0.5~eV (Fig.~\ref{fig:Fig14}) - makes these density of states models implausible.
In turn, the lack of temperature dependence of the spin state of Fe in FeGa$_3$, together with magnetic order revealed by our neutron diffraction measurements, overrules the so--called localized moment model proposed by Tsujii $et$ $al.$ \cite{Tsuji}, in which thermally induced transitions from the nonmagnetic ground state to the first excited state with $S$=1 determine the shape of $\chi$($T$) for FeGa$_3$ at high temperatures.  
 
\begin{figure*}[b] 
\centering
\includegraphics[width=0.34\textwidth,angle=-90]{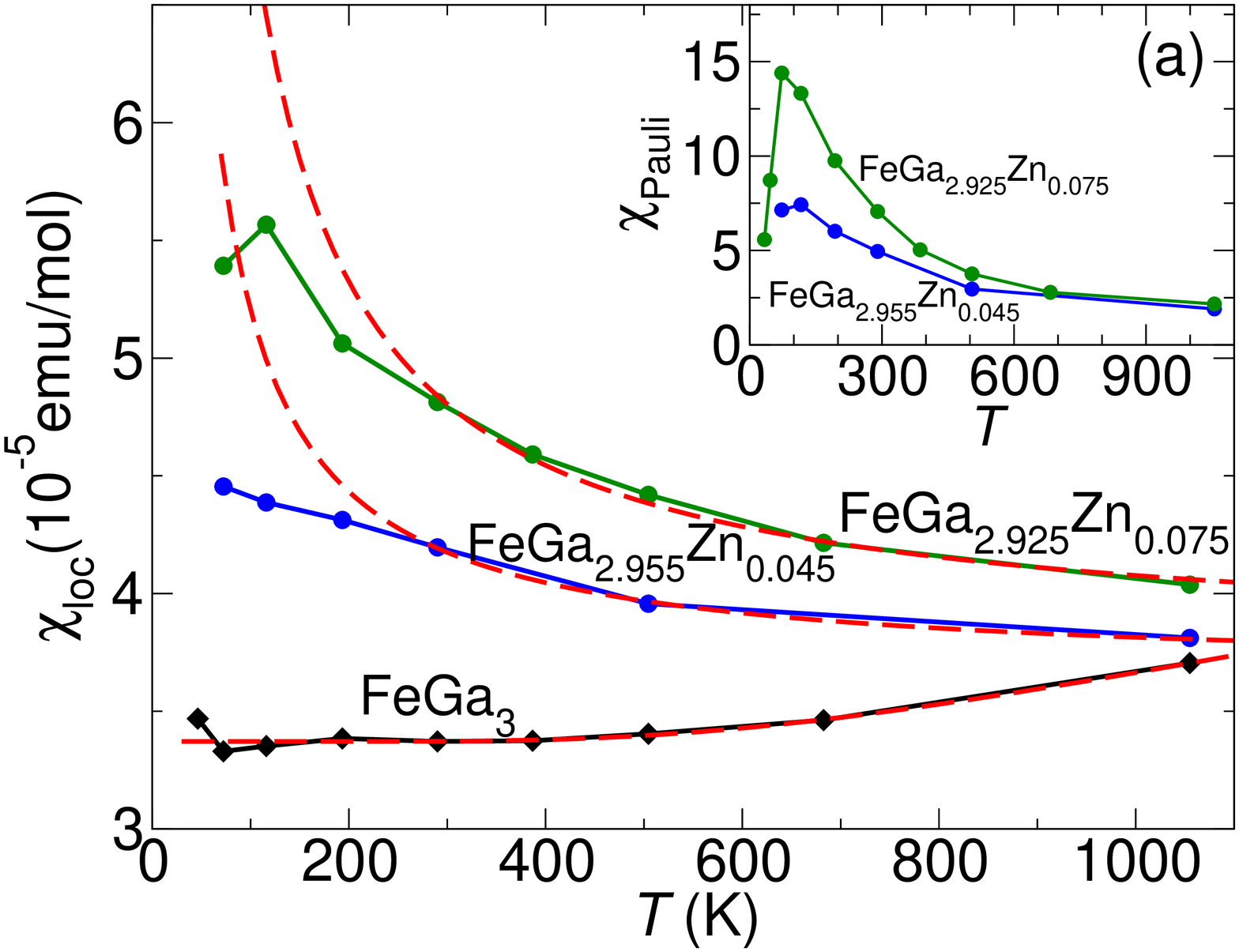}
\includegraphics[width=0.34\textwidth,angle=-90]{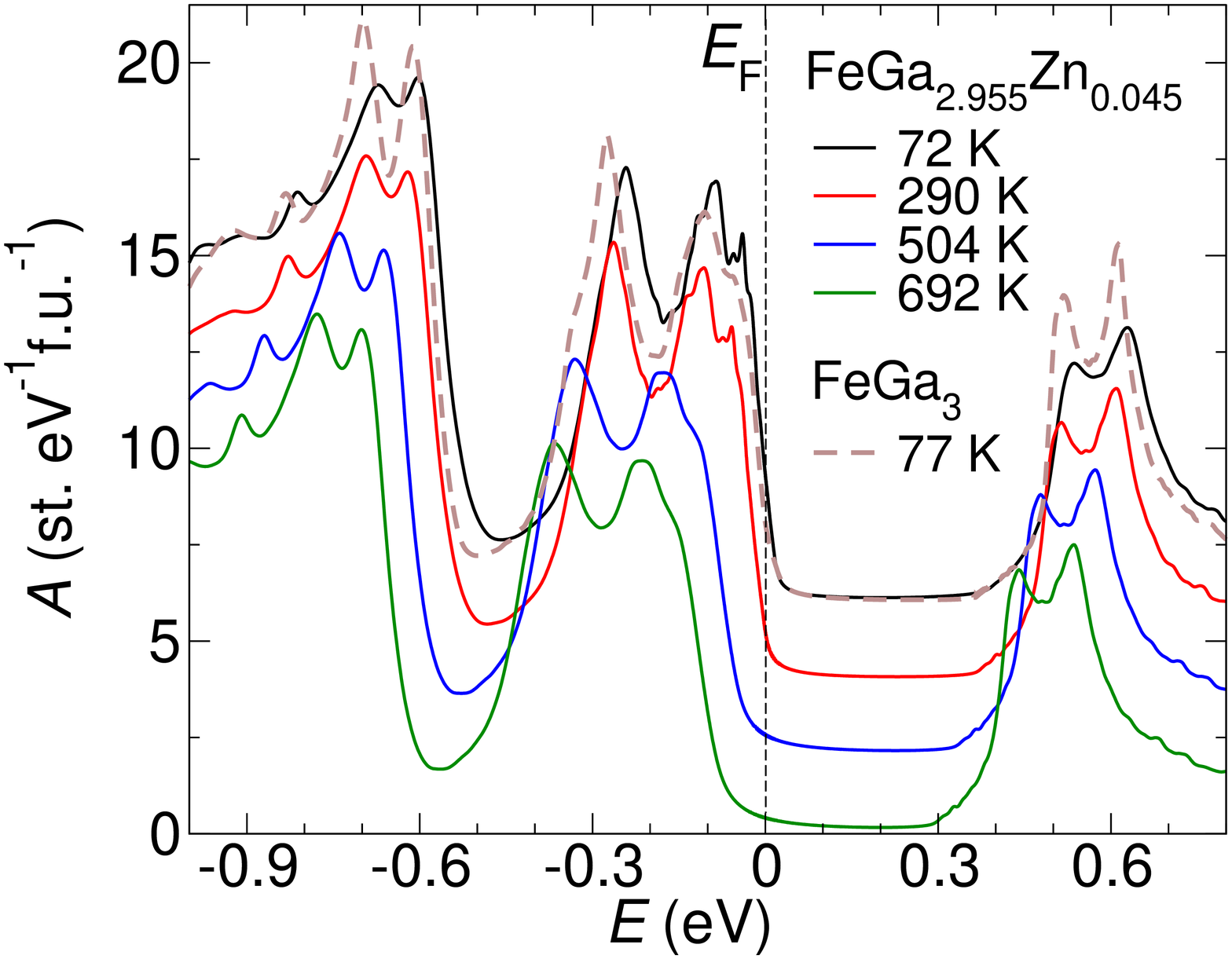}
\caption{\label{fig:Fig16} (Color online) (a) LDA+DMFT local spin susceptibility as functions of temperature (solid lines) for FeGa$_{2.955}$Zn$_{0.045}$ (blue), FeGa$_{2.925}$Zn$_{0.075}$ (green) and FeGa$_3$ (black) together with fits (dashed lines) to either the modified Curie law or to the activation law $\chi$($T$)$=\chi_0+exp$(-$\Delta_{\mathrm{S}}$/($k_{\mathrm{B}}T$)) with a spin gap $\Delta_{\mathrm{S}}=0.21$~eV. 
The inset displays Pauli susceptibilities computed from the spectral function for the two simulated Zn--doping levels in the same units. (b) LDA+DMFT spectral functions (solid lines) for FeGa$_{2.925}$Zn$_{0.075}$ at different temperatures compared to those for FeGa$_3$ at 77~K (dashed line). An offset (multiples of 2~st.~eV$^{-1}$f.u.$^{-1}$) was added to each curve so that all the presented curves can be easily viewed and compared. The $A$($E$) curve for FeGa$_3$ was shifted on the energy scale by \mbox{$-0.365$ eV} to facilitate visual comparison of its shape with the calculated $A$($E$) for FeGa$_{2.955}$Zn$_{0.045}$.}
\end{figure*}

To get insight into the energy scales for spin excitations in FeGa$_3$ based on our LDA+DMFT study, we calculated local spin susceptibilities for both FeGa$_3$ and RuGa$_3$, \mbox{$\chi_{\mathrm{loc}}(\omega=0)\sim\varint d\tau \langle S_z(\tau)S_z(0)\rangle$}. For FeGa$_3$, the $\chi_{\mathrm{loc}}$($T$) can be approximated by the sum of a small constant term of \mbox{$3.4\times 10^{-5}$emu/mol} and a contribution with an activation--type temperature dependence due to a gap of 0.21~eV (Fig.~\ref{fig:Fig16}(a)).   
This spin gap results from quenching of spin fluctuations at finite timescales due to the formation of a coherent insulating state at low temperatures.\cite{FeSiJan} 
It is smaller than the spin gap of 0.33--0.41~eV that was estimated from the activation expression applied to the experimental $\chi$($T$). 
The larger size of the experimental spin gap as compared to that obtained from the local spin susceptibility suggests the relevance of non--local exchange interactions that are not treated in our single--site DMFT setup. We suppose that these intersite exchange interactions are also responsible for the antiferromagnetic order revealed by our neutron diffraction measurements since the LDA+DMFT calculations do not give any hint for this antiferromagnetism in FeGa$_3$.
Importantly, the temperature dependence of the local susceptibility, derived from the LDA+DMFT study, is more than one order of magnitude smaller than the experimental values for the uniform susceptibility (Fig.~\ref{fig:Fig3}). Hence, the temperature--induced fluctuating moment of the underlying spin state is presumably not local,\cite{Singh} and a considerable momentum dependence in $\chi$ is expected. A similar disparity in size between the measured uniform magnetic susceptibility and the calculated $\chi_{\mathrm{loc}}$($T$) was recently reported for FeSi, an archetypal correlated band insulator.\cite{FeSiJan} Neutron experiments for FeSi found a significant magnetic scattering at "ferromagnetic reciprocal lattice vectors,\cite{FeSineutrons} confirming the non--local charater of the temperature--induced magnetic moment. 
In contrast, for RuGa$_3$ the local susceptibility is very small \mbox{($\sim$1.1$5\times 10^{-5}$emu/mol),} as expected for its more delocalized 4$d$ electrons, and it increases only very weakly with increasing temperature, in--line with the experimental results (see~Fig.~\ref{fig:Fig3}).

Although our neutron powder diffraction study for FeGa$_3$ revealed a complex magnetic ordering with an onset above room temperature, there is no signature of a magnetic phase transition in $\chi$($T$) measured up to 900~K.  
For numerous low--dimensional spin systems, however, features in $\chi$($T$) associated with magnetic ordering are not pronounced, especially in case of magnetic transitions at high temperatures. The shape of the temperature dependence of the magnetic susceptibility of such systems is determined mostly by the strongest exchange interaction. Consequently, signatures of magnetic transitions in $\chi$($T$) can be only barely detected even for compounds with staggered moments as large as 2--3 $\mu_{\mathrm{B}}$ per magnetic ion.\cite{PNASJack, RuSb2} 
For FeGa$_3$ the expected ordered moments are much smaller, the ordering temperature is above 300~K and the temperature dependence of the magnetic susceptibility is determined by the presence of the large gap in the spin excitation spectrum. Further, LSDA+U calculations suggest that there is a very strong antiferromagnetic coupling \mbox{($J\approx 3000$~K)} between Fe atoms within structural dimers, that leads to the formation of spin--singlets.\cite{Picket} 
Nevertheless, we can not exclude that the N{\'e}el temperature for FeGa$_3$ is higher than 900~K and therefore no signature of the magnetic ordering was found in $\chi$($T$) in the investigated temperature range.

Given the rather small staggered moments and an incommensurate ordering wavevector, we considered the possibility that the additional peaks in neutron diffraction patterns originate from extrinsic sources, such as the artifacts of the measurement process or a small amount of an impurity phase. The former can be excluded because neutron diffraction measurements performed in two different cryostats gave the same results. Furthermore, the diffraction patterns collected for samples of FeGa$_3$ and Fe$_{0.95}$Mn$_{0.05}$Ga$_3$ show very similar peaks at small values of the wavevector $Q$.  
Below we discuss the possibility of secondary phases as extrinsic sources of these additional diffraction peaks.

Both the neutron and X--ray powder diffraction patterns collected on the same sample of FeGa$_3$  did not show any unexpected peaks or extra intensities that could be assigned to the nuclear structure of an impurity phase. We found only a few small reflections that can be unambiguously assigned to the presence of $\sim$2~vol.\% of $\alpha$--Ga, which is nonmagnetic. 
Moreover, magnetization measurements performed on several random crystals gave only a weakly temperature dependent diamagnetic signal, further ruling out the presence of magnetic impurity phases, including iron oxides. Such measurements are much more sensitive to magnetic impurities than the neutron powder diffraction experiments.
Microstructures of several crystals were also studied using a scanning electron microscope with resolutions down to $\sim$10~nm. Examples of high resolution images of as grown crystalline surfaces taken in the backscattered electrons mode are presented in Fig.~S1 in the Supplemental Material. They show no trace of a secondary phase or compositional inhomogeneities, apart from small and single inclusions of gallium flux. Since backscattered electrons are sensitive to local variations in the atomic number, the presence of an impurity phase would have been visible as either brighter or lighter fields in the surface images. 
Finally, the sample of FeGa$_3$ was prepared only from crystals grown in batches with nominal ratio \mbox{Fe:Ga = 1:12.5.} As expected based on the Fe--Ga phase diagram,\cite{Dasarathy} the only binary phase found in these batches was FeGa$_3$. 
The next binary phase, Fe$_3$Ga$_4$, was observed only in growths with nominally less than 85~at.\% Ga. The presence of this compound in the sample investigated by neutron diffraction was ruled out based on X--ray powder diffraction, where the detection limit is of the order of \mbox{1~vol.\%.} 
Furthermore, magnetic peaks expected for Fe$_3$Ga$_4$\cite{Fe3Ga4neutrons} do not match the extra peaks observed in the neutron patterns of FeGa$_3$. 
Taken together, our results argue strongly against extrinsic sources of the additional peaks in the neutron powder diffraction patterns of FeGa$_3$ and Fe$_{0.95}$Mn$_{0.05}$Ga$_3$. 

We note that room temperature $^{57}$Fe M$\ddot{o}$ssbauer spectra for FeGa$_3$ did not show a distinct magnetic splitting, thus ruling out the presence of a sizable hyperfine--field on Fe site. For an incommensurate system with small ordered moments, however, the presence of a finite internal magnetic field may manifest itself only in a broadening of spectral lines due to a distribution of hyperfine fields arising from the incommensurately modulated magnetic structure.\cite{Moessbauer19} This effect superposed on a quadrupole splitting \mbox{$\Delta E$=0.33~mm/s} in FeGa$_3$,\cite{Tsuji, Moessbauer} may be too weak to be detected on top of broadening originating from life--time effects and experimental resolution.

The antiferromagnetic order in FeGa$_3$ with the ordering wavevector incommensurate with the nuclear structure and only small staggered moments suggests a close similarity of FeGa$_3$ to Slater insulators. 
In case of Slater insulators, however, an opening of a band gap is induced by the formation of a spin--density--wave, whereas our computational study shows that the magnetic order is not needed to cause the insulating behavior in FeGa$_3$. Moreover, electronic structure calculations based on local density methods give a similar gap of 0.4 eV for both a nonmagnetic and a magnetically ordered state in FeGa$_3$,\cite{Picket} thus making a spin--density--wave--type order resulting from a Fermi surface instability in FeGa$_3$ highly unlikely. Therefore we conclude that the modulated magnetic structure in FeGa$_3$ results rather from competing exchange interactions.

It is of interest to compare FeGa$_3$ with the prototypal correlated band insulator, FeSi.
Both compounds are band insulators with hybridization gaps of 0.4~eV for FeGa$_3$ and 0.05~eV for FeSi.\cite{Arita,FeSi1,FeSiJan}  
Electronic structure calculations based on LDA+DMFT indicate a strong mixed valence of Fe with a similar average occupancy for the Fe 3$d$ shell of 6.3 and 6.2 in FeGa$_3$ and FeSi\cite{FeSiJan}, respectively. Furthermore, the spin state with $S=1$ dominates in both FeGa$_3$ and FeSi, with strong fluctuations on short time scales. 
Even the remarkably uniform distribution of states is very similar in the two compounds (see Fig.~\ref{fig:Fig14} and Fig.~S5 of Ref.~\onlinecite{FeSiJan}). For both systems, the spin excitations are gapped, and the temperature--induced fluctuating moment is, to a large extent, not local. The large degree of delocalization and strong momentum dependence is typical for correlated band insulators and differentiates them from $f$-electron based Kondo insulators such as Ce$_3$Bi$_4$Pt$_3$, SmB$_6$, YbB$_{12}$, or CeNiSn.\cite{FeCoSi}
Also the ratio between the size of the band gap $E_g$ and the width of the bands around the gap \mbox{$E_g$/$W \gg 1$} for FeGa$_3$ and FeSi, whereas for Kondo insulators \mbox{$E_g$/$W \ll 1$.\cite{FeCoSi}} 
Further, values of the $U$/$W$ ratio estimated for both FeGa$_3$ and FeSi are of the order of 1, much smaller than those found in strongly correlated electron systems.\cite{FeSiJan} The latter require strong electron--electron interactions to open a band gap at the Fermi level.
%Both FeGa$_3$ and FeSi are in the correlated band insulator regime.
The effective mass renormalization resulting from many--body effects in FeGa$_3$ is rather low, \mbox{$m^{\hbox{\tiny{DMFT}}}/m^{\hbox{\tiny{LDA}}}\approx 1.3$--1.4,} but it is comparable to that in FeSi \mbox{$m^{\hbox{\tiny{DMFT}}}/m^{\hbox{\tiny{LDA}}}\approx 1.5$.\cite{FeSi1, FeSiXCo,FeSiJan}} 
For FeGa$_3$, however, the effective masses are nearly independent of temperature (Fig.~\ref{fig:Fig15}). This has to be contrasted with FeSi, for which the same local interaction vertex gives rise to a temperature dependence in the self-energy that is pertinent for experimental observables.\cite{FeSiJan} Consequently, calculations based on DMFT indicate that FeSi metallizes with increasing temperature through correlation induced incoherence.\cite{FeSiJan} The primary difference between FeGa$_3$ and FeSi lies in the size of the band gap, which is about eight times larger for FeGa$_3$ than for FeSi. 
Therefore, for FeSi all the relevant energy scales are of similar magnitude resulting in low-energy properties being dominated by correlation effects,\cite{FeSiJan} whereas for FeGa$_3$ the effects of local physics are rather small on a relative energy scale. 
Furthermore, in FeGa$_3$ non--local exchange interactions are presumably responsible for the antiferromagnetic order revealed by our neutron diffraction study. In contrast, for FeSi the influence of non--local exchange is reflected mostly in an increase in size of the band gap.\cite{NewFeSi}  
Neutron diffraction measurements did not give any sign of a spin ordering in FeSi even at lowest temperatures.\cite{FeSiNEUTRONS}

We were successful in incorporating sizable amounts of Zn and Mn into FeGa$_3$. This p--type doping on both Fe and Ga sites did not lead either to a notable change in the long--range magnetic order or to a metal--insulator transition. Remarkably, the effect of doping is very similar for both Mn and Zn. 
Electrical resistivity measurements show that the doping creates localized states in the semiconducting gap. 
Our magnetization study indicates that both a substitution of Zn onto the Ga site and a replacement of Fe by Mn lead to formation of an essentially noninteracting magnetic moment of $S=1/2$ that fluctuates freely at temperatures above $\sim$80~K (Fig.~\ref{fig:Fig9}).
At lower temperatures, the effective moment decreases strongly with decreasing temperature, regardless of the dopant type and level. 

Similar effects of doping were observed for simple semiconductors such as silicon and germanium.
The thermodynamic properties of these systems in the insulating regime can be well described by the phenomenological model introduced by Bhatt and Lee.\cite{BL1} 
The Bhatt--Lee model\cite{BL2} considers local moments of $S=1/2$ associated with dopant atoms, which lead to the Curie--like behavior in $\chi$($T$) at high temperatures and a weaker temperature dependence of the magnetic susceptibility at lower temperatures due to antiferromagnetic exchange interactions between these randomly distributed spins.  
Bhatt and Lee iteratively divided the system into a hierarchy of antiferromagnetically coupled spin--pairs and showed that at low temperatures $\chi$($T$)$\propto T^{-\alpha}$ and $C_p$($T$)$\propto T^{1-\alpha}$ with $\alpha\textless 1$.\cite{BL2}
For Fe$_{1-x}$Mn$_x$Ga$_3$ and FeGa$_{3-y}$Zn$_y$, however, at temperatures below $\sim$30~K the magnetic susceptibilities develop notable magnetic anisotropies. Additionally, the magnetic susceptibilities measured in $H\parallel ab$ are nearly field independent and saturate at \mbox{$T\apprle 6$ K.} Furthermore, estimates of the magnetic specific heat from the experimental $\chi$($T$) based on the Bhatt--Lee model (Eqs 2 and 3 of Ref.~\onlinecite{BL2}) give values that are at least about one order of magnitude smaller than the experimentally obtained magnetic specific heat even at the lowest temperatures (not shown). 
Finally, there is an important difference between FeGa$_3$ and the simple semiconductors regarding doping levels required to drive the system through an insulator--to--metal transition:
For Si doped with phosphorus (boron), the critical concentration of dopant atoms \mbox{$N_{\mathrm{c}}=3.52\times 10^{18}$ cm$^{-3}$} (\mbox{$4.06\times 10^{18}$ cm$^{-3}$)} corresponds to about 0.007~at.\% (0.009~at.\%), whereas even the introduction of about 2.5~at.\% of acceptors in Fe$_{0.9}$Mn$_{0.1}$Ga$_3$ does not lead to a metallic state.  
Also for the archetypal correlated band insulators, FeSi and FeSb$_2$, similar or smaller concentrations of dopants were found to induce a metallic conduction in FeSi$_{1-x}$Al$_x$, Fe$_{1-x}$Co$_x$Si, Fe$_{1-x}$Mn$_x$Si, Fe$_{1-x}$Co$_x$Sb$_2$, FeSb$_{2-x}$Sn$_x$, Fe$_{1-x}$Ni$_x$Sb$_2$ and FeSb$_{2-x}$Te$_x$.\cite{Al1,Co1,FeSigen, FeCoSi, FeCoSb2, FeSb2Sn, FeSiAl, FeNiSb, FeSbTe}

The development of a broad Schottky--type anomaly in specific heat of both Fe$_{1-x}$Mn$_x$Ga$_3$ and FeGa$_{3-y}$Zn$_y$ indicates the formation of a gap (Fig.~\ref{fig:Fig8}). 
The magnitude of this gap \mbox{$\Delta/k_{\mathrm{B}}\approx 16$~K} is very similar for Mn-- and Zn--doped FeGa$_3$ and does not change appreciably with increasing doping level. 
Since the gap--anomalies in specific heat are accompanied by the broad features in magnetic susceptibility, we considered their common origin as due to thermally induced changes in spin state of dopant atoms from $S=0$ to $S=1$/2. Attempts to describe the $\chi$($T$) based on this model, however, do not lead to satisfactory fits (not shown). Instead, the magnetic susceptibilities revealed notable magnetic anisotropies in this temperature range, very similar for both Fe$_{1-x}$Mn$_x$Ga$_3$ and FeGa$_{3-y}$Zn$_y$. 
We propose that holes introduced to FeGa$_3$ by doping with Mn and Zn localize at low temperatures into magnetic ``droplets'' on the scale of the lattice spacing due to strong exchange interactions between spins of these carriers and spins of Fe in the antiferromagnetic lattice. 
Such localized charge carriers accompanied by reorientations of surrounding lattice spins are usually referred to as small spin polarons. Thus, we interpret the gap found in specific heat measurements as the binding energy of these states forming at low temperatures.

The linear increase of the excess entropy associated with the Schottky anomalies with increasing doping levels observed for Mn-- and Zn--doped FeGa$_3$ (Fig.~\ref{fig:Fig8}(c)) is consistent with a growing number of doping--induced polarons. Also, the magnetic susceptibilities at the lowest temperatures increase nearly linearly with the dopant levels for both Fe$_{1-x}$Mn$_x$Ga$_3$ and FeGa$_{3-y}$Zn$_y$ (Fig.~\ref{fig:Fig11}), in--line with the scenario of magnetic polarons. 
Further, the field dependent magnetization $M$($H$) curves measured at 1.85~K do not saturate up to the highest field of 70~kOe, and the maximum measured moments correspond to only $\sim$0.3~$\mu_{\mathrm{B}}$ per dopant atom (Fig.~\ref{fig:Fig10}(b)). Since the doping with Mn and Zn falls short of closing the spin gap of 0.4~eV and essentially does not influence the antiferromagnetic ordering as revealed by our neutron diffraction measurements, the shape of the $M$($H$) curves reflects the magnetization associated with the dopant atoms. The lack of saturation and only small values of the field--induced magnetic moment per dopant atom that are independent of the doping level, suggest the presence of strong exchange interactions between spins of carriers introduced by the doping and spins of Fe atoms in the antiferromagnetic material. 
This exchange coupling is presumably responsible for the formation of the spin polarons. 
Recent muon spin rotation study gave an evidence for the presence of anisotropic spin polarons in FeGa$_3$ at temperatures below $\sim$10~K.\cite{Petrovic}     
However, we caution that spectroscopic measurements on Mn-- and Zn--doped FeGa$_3$ are needed to place our proposal on firmer footing.

Our experimental study for Mn-- and Zn--doped FeGa$_3$ is in strong contradiction with predictions based on electronic structure calculations within DFT, which indicated the formation of an itinerant ferromagnetic state with half--metallic properties in hole--doped FeGa$_3$ over an extended composition range, independent of the presence of pre--formed Fe moments in the undoped semiconducting phase.\cite{Singh} To test this theoretical prediction and to further study the influence of hole doping on magnetic properties and many--body renormalizations, we employed DMFT.
We simulated doping of FeGa$_3$ with holes on both Fe and Ga sites in LDA+DMFT calculations using the VCA approximation.

Our calculations indicate that hole doping on either Fe or Ga sites has very similar effects on the electronic structure and magnetic properties of FeGa$_3$. 
Fig.~\ref{fig:Fig16}(a) shows $\chi_{\mathrm{loc}}$($T$) calculated for a few selected doping levels. 
At high temperatures, the local susceptibilities can be well described using a modified Curie law. With decreasing temperature a notable deviation from \mbox{$\chi_{\mathrm{loc}}$($T$)$\sim T^{-1}$} develops. The values of $\chi_{\mathrm{loc}}$($T$) become smaller than expected from the Curie law, capturing the behavior observed in the experiments.
For both Fe$_{1-x}$Mn$_x$Ga$_3$ and FeGa$_{3-y}$Zn$_y$, however, modified Curie fits to $\chi_{\mathrm{loc}}$($T$) at temperatures above $\sim$300~K result in small values of the effective moment per dopant, much smaller than the moment of \mbox{1.73 $\mu_{\mathrm{B}}$} expected for $S=1/2$. For example, for FeGa$_{2.925}$Zn$_{0.075}$, we obtained only \mbox{$p_{\mathrm{eff}}=0.55\pm 0.02$~$\mu_{\mathrm{B}}$} per Zn. 

As seen in Fig.~\ref{fig:Fig15}, the deviation of $\chi_{\mathrm{loc}}$($T$) from the modified Curie law is accompanied by a notable increase in the effective mass for the orbital character that accounts for the majority of spectral weight at the Fermi level, near the top of the valence band. Although the overall values of $m^{\hbox{\tiny{DMFT}}}/m^{\hbox{\tiny{LDA}}}$ for the simulated doping levels are not very big, 
they are comparable with those found for FeSi.\cite{FeSiJan}
Most importantly, the temperature induced changes are as large as 15\% an increase in the shown range for FeGa$_{2.925}$Zn$_{0.075}$. Conceptually, growing effective masses can be expected to increase the validity of a local picture of fluctuating spin moments. Therefore the experimentally observed decrease of the effective moment when lowering temperature points towards a predominantly itinerant mechanism for the spin response at all temperatures.

Moreover, the changes in $\chi_{\mathrm{loc}}$ coincide with the temperature dependence of the spectral weight at the Fermi level displayed in Fig.~\ref{fig:Fig16}(b).  
The Pauli susceptibility (see the inset in Fig.~\ref{fig:Fig16}(a)) calculated from
\begin{equation}
\chi_{\mathrm{Pauli}}(T) = \mu_{\mathrm{B}}^2 N_{\mathrm{A}} \varint dE (-\frac{\partial f(E)}{\partial E}) A(E)   
\label{eq:7}
\end{equation}
provides an even better match to the experimental uniform susceptibility than the LDA+DMFT local spin susceptibility. 
Here, $N_{\mathrm{A}}$ is Avogadros number, $\mu_{\mathrm{B}}$ is the Bohr magneton, $f(E)$ denotes the Fermi function, $A$(E) is the spectral function.  
These calculations imply that the physics of local moments is not predominant in the Mn-- and Zn--doped FeGa$_3$.

To further address the strength of magnetic correlations in the hole--doped FeGa$_3$ based on DMFT, we calculate also the Sommerfeld--Wilson ratio:
\begin{equation}
R_{\mathrm{SW}} = \frac{4\pi^2k_{\mathrm{B}}^2}{3(g\mu_{\mathrm{B}})^2} \frac{\chi_{\mathrm{p}}}{\gamma}.   
\label{eq:SW}
\end{equation}
Here, $g=2$ is the electron Lande factor, $\gamma$ is the Sommerfeld coefficient obtained based on the calculated spectrum at the Fermi level and $\chi_{\mathrm{p}}$ is approximated by the local spin susceptibility $\chi_{\mathrm{loc}}$.
A tentative $T\rightarrow 0$ extrapolation gives values of the $R_{\mathrm{SW}}$ in the range of 1--3 for both FeGa$_{2.955}$Zn$_{0.045}$ and FeGa$_{2.925}$Zn$_{0.075}$, thus indicating the absence of strong electronic spin--spin interactions which could lead to Stoner--type magnetic instability, in--line with our experimental results.

Our experimental and computational study puts forward the complex magnetic properties of the Fe$_{1-x}$Mn$_x$Ga$_3$ and FeGa$_{3-y}$Zn$_y$ solid solutions.
Even though the inclusion of dopant atoms basically does not influence the long range antiferromagnetic order and falls short of closing the spin (and charge) gap, it triggers an intriguing magnetic behavior.  
Both the experimental $\chi$($T$) and the calculated $\chi_{\mathrm{loc}}$ exhibit a Curie--like behavior at high temperatures and become smaller than expected based on the Curie law at lower temperatures. 
However, the experimental magnetic susceptibility suggests the formation of local moments of $S=1/2$ associated with the dopant atoms that fluctuate freely at temperatures above $\sim$100~K (Fig.~\ref{fig:Fig9}), whereas the LDA+DMFT calculations point to the itinerant character of the spin response, that however mimic a Curie--like fluctuating moment bahavior at high temperatures.
These findings, together with the presence of an incommensurate magnetic order with rather small staggered moments as well as the formation of anisotropic spin polaronic--like states at low temperatures,  
indicate that Fe$_{1-x}$Mn$_x$Ga$_3$ and FeGa$_{3-y}$Zn$_y$ combines features of both itinerant and localized magnetism. 

We notice that our LDA+DMFT calculations as well as a previous computational study based on DFT\cite{Singh} indicate that $p$--type doping shifts the Fermi level to the valence band, whereas transport measurements for all Fe$_{1-x}$Mn$_x$Ga$_3$ and FeGa$_{3-y}$Zn$_y$ revealed only insulating behavior with electrical conductivities determined by the presence of localized in--gap states even for the highest doping levels. 
It is known, however, that computational methods based on the local density and the VCA approximation can not address this type of conduction in doped semiconductors.\cite{hole_cond, polaron-theory} 
In addition, intersite exchange interactions between spins of carriers introduced by the doping and spins of magnetic ions need to be accounted for in case of the formation of small magnetic polarons.  
Inclusion of these interactions in the DMFT calculations would at least necessitate a cluster extension.\cite{Jan_cluster}

\section{Conclusions}\label{Conclusions}

Our neutron powder diffraction, thermodynamic and electrical resistivity measurements show that FeGa$_3$ is an insulator in which iron moments order above room temperature in a complex antiferromagnetic structure that is incommensurate with the nuclear lattice and spin excitations are gapped. Based on many--body calculations within the framework of DMFT we claim that, while the iron atoms in FeGa$_3$ are dominantly in an $S=1$ state, there are strong and temperature--independent charge and spin fluctuations at short time scales, indicating a strongly delocalized character of 3$d$ electrons. Further, the low magnitude of local contributions to the spin susceptibility advocates a predominantly itinerant mechanism for the spin response.
Our combined experimental and computational investigations indicate that FeGa$_3$ is a correlated band insulator with only small effects of many--body renormalizations, in which non--local exchange interactions are presumably responsible for the spin gap of 0.4~eV and the antiferromagnetic order. 

Calculations based on DMFT indicate that dynamical correlation effects in FeGa$_3$ become stronger as a result of hole doping.
Using electrical resistivity and thermodynamic measurements as well as neutron powder diffraction we establish that doping of FeGa$_3$ with Mn and Zn neither affects the long range antiferromagnetic ordering nor leads to an insulator-to-metal transition. 
The effect of doping is very similar for both Mn and Zn. The temperature dependence of the electrical resistivity indicates that localized states are formed in the semiconducting gap.
Magnetization study shows that the introduction of each hole is accompanied by the formation of a spin moment of $S=1/2$, that fluctuates freely at temperatures above $\sim$100~K and decreases gradually with lowering temperature. While these experimental findings suggest the existence of local moments associated with the dopant atoms, the LDA+DMFT calculations argue for an itinerant character of spin excitations that however mimic a Curie--like behavior of the $\chi$($T$) at high temperatures.
At low temperatures, it is tempting to interpret the thermodynamic and transport properties of the Mn-- and Zn--doped FeGa$_3$ in terms of small spin polarons formed by holes that localize into magnetic ``droplets'' due to strong exchange couplings between spins of these free carriers and Fe moments in the antiferromagnetic lattice. Consequently, our study indicates that Fe$_{1-x}$Mn$_x$Ga$_3$ and FeGa$_{3-y}$Zn$_y$ combine features of both itinerant and localized magnetism due to a complex interplay between non--local exchange and correlation effects.

\section{Acknowledgments}\label{Acknowledgments}

Work at Brookhaven National Laboratory (M. G., A. P., and M. C. A.) was carried out under the auspices of the US Department of Energy, Office of Basic Energy Sciences, under Contract No. DE-AC02-98CH1886. Research at Rutgers University (J. M. T. and G. K.) was sponsored by the Department of Defense National Security Science and Engineering Faculty Fellowship via the Air Force Office of Scientific Research. J. M. T. and G. K. were further supported by the NSF DMR 1308141.

\bibliographystyle{apsrev4-1}
\bibliography{main}

\end{document}